\documentclass[preprint,12pt]{article}
\pdfoutput=1
\usepackage[utf8x]{inputenc}      
\usepackage{hyperref}
\usepackage{amsmath,amssymb,mathtools}
\usepackage{slashed}
\usepackage[T1]{fontenc}          
\usepackage{booktabs,tabularx}
\usepackage{graphicx,subfig}
\usepackage{xspace}
\usepackage[usenames]{xcolor}\definecolor{fscolor}{RGB}{44,118,255}
\usepackage{geometry}
\usepackage{todonotes}
\usepackage{listings}
\usepackage[absolute]{textpos}
\usepackage[many]{tcolorbox}
\usepackage{xparse}
\usepackage[font=small,labelfont=bf,format=plain,margin=0.05\textwidth]{caption}
\usepackage{bbm}
\usepackage{tabularx}
\usepackage{cite}
\usepackage{soul}
\usepackage{axodraw2}

\allowdisplaybreaks

\evensidemargin \oddsidemargin
\newcommand{\sbe}{s_\beta}
\newcommand{\cbe}{c_\beta}
\newcommand{\tbe}{t_\beta}
\newcommand{\sbb}{s_\beta^2}
\newcommand{\cbb}{c_\beta^2}
\newcommand{\tbb}{t_\beta^2}

\newcommand{\sz}{s}

\newcommand{\mW}{\ensuremath{M_W}}
\newcommand{\mZ}{\ensuremath{M_Z}}

\newcommand{\vMS}{\ensuremath{v^{\MS, \SM}}}

\newcommand{\vMSSM}{\ensuremath{v^{\DR, \MSSM}}}

\newcommand{\DR}{{\ensuremath{\overline{\text{DR}}}}\xspace}
\newcommand{\MS}{{\ensuremath{\overline{\text{MS}}}}\xspace}
\newcommand{\DRp}{{\ensuremath{\overline{\text{DR}}^\prime}}\xspace}
\newcommand{\OS}{{\text{OS}}\xspace}

\newcommand{\SM}{{\text{SM}}}

\newcommand{\MSSM}{{\text{MSSM}}}

\newcommand{\SUSY}{{\text{SUSY}}}

\newcommand{\FH}{\mbox{{\tt FeynHiggs}}\xspace}

\newcommand{\HS}{\mbox{{\tt HSSUSY}}\xspace}

\newcommand{\Him}{\mbox{{\tt Himalaya}}\xspace}

\newcommand{\Fig}[1]{Fig.~\ref{#1}}

\newcommand{\Sec}[1]{Section~\ref{#1}}
\newcommand{\App}[1]{App.~\ref{#1}}
\newcommand{\Eq}[1]{Eq.~(\ref{#1})}
\newcommand{\Eqs}[2]{Eqs.~(\ref{#1}) and~(\ref{#2})}

\newcommand{\Eqss}[2]{Eqs.~(\ref{#1})-(\ref{#2})}

\renewcommand{\Re}{\text{Re}}

\newcommand{\mh}{m_h}
\newcommand{\mA}{m_A}

\newcommand{\msbote}{m_{\tilde b_1}}
\newcommand{\msbotz}{m_{\tilde b_2}}
\newcommand{\mtL}{m_{\tilde{t}_L}}
\newcommand{\mtR}{m_{\tilde{t}_R}}

\newcommand{\mbR}{m_{\tilde{b}_R}}

\newcommand{\cp}{\ensuremath{{\cal CP}}}

\newcommand{\msusy}{\ensuremath{M_\SUSY}\xspace}

\newcommand{\Li}{\text{Li}_{2}}

\newcommand{\hmg}{\ensuremath{\widehat M_3}\xspace}
\newcommand{\hmu}{\ensuremath{\widehat \mu}\xspace}
\newcommand{\xq}{\ensuremath{\widehat X_q}\xspace}
\newcommand{\xt}{\ensuremath{\widehat X_t}\xspace}

\newcommand{\yq}{\ensuremath{\widehat Y_q}\xspace}
\newcommand{\yt}{\ensuremath{\widehat Y_t}\xspace}

\newcommand{\xb}{\ensuremath{\widehat X_b}\xspace}

\newcommand{\yb}{\ensuremath{\widehat Y_b}\xspace}

\newcommand{\tev}{\,\, \mathrm{TeV}}
\newcommand{\gev}{\,\, \mathrm{GeV}}

\newcommand{\order}[1]{\ensuremath{{\cal O}(#1)}}
\newcommand{\al}{\alpha}
\newcommand{\als}{\al_s}
\newcommand{\alt}{\al_t}
\newcommand{\alb}{\al_b}
\newcommand{\alq}{\al_q}

\newcommand{\giu}{\tilde{g}_{1u}}
\newcommand{\gid}{\tilde{g}_{1d}}
\newcommand{\giiu}{\tilde{g}_{2u}}
\newcommand{\giid}{\tilde{g}_{2d}}

\newcommand{\pMi}{\phi_{M_1}}
\newcommand{\pMii}{\phi_{M_2}}
\newcommand{\pMiii}{\phi_{M_3}}
\newcommand{\pMue}{\phi_{\mu}}
\newcommand{\pXq}{\phi_{X_q}}
\newcommand{\pXt}{\phi_{X_t}}
\newcommand{\pAt}{\phi_{A_t}}
\newcommand{\pYq}{\phi_{Y_q}}
\newcommand{\pYt}{\phi_{Y_t}}
\newcommand{\pXb}{\phi_{X_b}}
\newcommand{\pAb}{\phi_{A_b}}
\newcommand{\pYb}{\phi_{Y_b}}

\begin{document}

\thispagestyle{empty}
\def\thefootnote{\fnsymbol{footnote}}

\begin{flushright}
DESY 20-085
\end{flushright}
\vspace{3em}
\begin{center}
{\Large\bf The light MSSM Higgs boson mass \\[.5em] for large $\tan\beta$ and complex input parameters}
\\
\vspace{3em}
{
Henning Bahl\footnote{email: henning.bahl@desy.de},
Ivan Sobolev\footnote{email: ivan.sobolev@desy.de},
Georg Weiglein\footnote{email: georg.weiglein@desy.de}
}\\[2em]
{\sl Deutsches Elektronen-Synchrotron DESY, Notkestra{\ss}e 85, D-22607 Hamburg, Germany}
\def\thefootnote{\arabic{footnote}}
\setcounter{page}{0}
\setcounter{footnote}{0}
\end{center}
\vspace{2ex}
\begin{abstract}
{}

We discuss various improvements of the prediction for the light MSSM Higgs boson mass in the hybrid framework of the public code \FH, which combines fixed-order and effective field theory results. First, we discuss the resummation of logarithmic contributions proportional to the bottom-Yukawa coupling including two-loop $\Delta_b$ resummation. For large $\tan\beta$, these improvements can lead to large upward shifts of the Higgs mass compared to the existing fixed-order calculations. Second, we improve the implemented EFT calculation by fully taking into account the effect of \cp-violating phases. As a third improvement, we discuss the inclusion of partial N$^3$LL resummation. The presented improvements will be implemented into \FH.

\end{abstract}

\newpage
\tableofcontents
\newpage
\def\thefootnote{\arabic{footnote}}


\section{Introduction}
\label{sec:01_intro}

The discovery of a Higgs boson at the LHC~\cite{Aad:2012tfa,Chatrchyan:2012xdj} was an important step towards the understanding of the fundamental laws of Nature. The properties of the detected particle allow a sensitive test of the predictions of the Standard Model (SM) and of theories of physics beyond the SM (BSM). In particular, in the Minimal Supersymmetric extension of the SM (MSSM) \cite{Nilles:1983ge,Haber:1984rc}, based upon the concept of supersymmetry (SUSY), the mass of the discovered boson is not a free parameter, as in the SM, but is predicted in terms of the model parameters.

While the SM-like Higgs mass in the MSSM is smaller or equal to the mass of the $Z$ boson at the tree-level, large quantum corrections shift it upwards towards the experimentally measured value of $M_h\sim 125\gev$. In order to allow the use of the SM-like Higgs mass as a precision constraint on the MSSM parameter space, the precise determination of these quantum corrections is crucial~\cite{Bahl:2019hmm}.

The quantum corrections can be calculated in different frameworks. In the most direct approach, quantum corrections to the Higgs self-energies are calculated diagrammatically in the full theory (for recent works see~\cite{Borowka:2015ura,Goodsell:2016udb,Passehr:2017ufr,Harlander:2017kuc,Borowka:2018anu,R.:2019ply,Goodsell:2019zfs,Domingo:2020wiy}). This approach has the advantage of capturing all corrections at a specific order in perturbation theory. If the scale of SUSY particles is, however, much larger than the electroweak scale, large logarithms emerge in the fixed-order corrections exacerbating the behaviour of the perturbative expansion. In such situations, effective field theory (EFT) techniques allow the resummation of large logarithmic corrections (for recent works see~\cite{Vega:2015fna,Lee:2015uza,Bagnaschi:2017xid,Bahl:2018jom,Harlander:2018yhj,Bagnaschi:2019esc,Murphy:2019qpm,Bahl:2019wzx}). Without including higher-dimensional operators into the low-energy EFT, terms suppressed by the SUSY scale are, however, missed in this approach.\footnote{An EFT study including the dominant dimension-six operators can be found in \cite{Bagnaschi:2017xid}.} Therefore, the accuracy of the EFT approach can be diminished if one or more SUSY scales are comparable to the electroweak scale. In order to obtain a precise prediction for the SM-like Higgs boson mass for low, intermediary as well as high SUSY scales, both approaches -- the fixed-order and the EFT approach -- can be combined. Such hybrid approaches have been developed in~\cite{Hahn:2013ria,Bahl:2016brp,Athron:2016fuq,Staub:2017jnp,Bahl:2017aev,Athron:2017fvs,Bahl:2018jom,Bahl:2018ykj,Harlander:2019dge,Bahl:2019hmm,Kwasnitza:2020wli}.

In this paper, we focus on the hybrid approach implemented in the publicly available code \FH~\cite{Heinemeyer:1998yj,Heinemeyer:1998np,Hahn:2009zz,Degrassi:2002fi,Frank:2006yh,Hahn:2013ria,Bahl:2016brp,Bahl:2017aev,Bahl:2018qog}. We will discuss various improvements of the incorporated EFT calculation as well as their combination with the implemented fixed-order calculation: resummation of large logarithms proportional to the bottom Yukawa coupling (including two-loop $\Delta_b$ resummation\cite{Noth:2008tw,Noth:2008ths,Noth:2010jy}), extension of the EFT calculation fully taking into account the effects of complex input parameters as well as an inclusion of partial N$^3$LL resummation.

This paper is structured as follows: In \Sec{sec:02_botresum}, we explain how the resummation of logarithms proportional to the bottom Yukawa coupling is incorporated into the hybrid framework. The extension of the EFT calculation for complex input parameters is discussed in \Sec{sec:03_cEFT}. We explain the implementation of partial N$^3$LL resummation in \Sec{sec:04_NNNLL}. Numerical results are presented in \Sec{sec:05_results}. In \Sec{sec:06_conclusions}, we give our conclusions. In the Appendices, we provide more details regarding the calculation of the two-loop threshold corrections (see \App{app:07_2L_thresholds_derivation}), analytic expressions for the threshold corrections to the SM-Higgs self coupling and to the couplings of the Split-SUSY model (see \App{app:08_CPV_thresholds}), more details regarding the dependence of $\Delta_b$ on \cp-violating phases (see \App{app:09_db2L_CPV}), and explicit formulas for the one- and two-loop logarithmic corrections to $M_h$ proportional to the bottom-Yukawa coupling (see \App{app:10_Mh_logs}).


\section{Resummation of logarithmic bottom Yukawa contributions}
\label{sec:02_botresum}

In this Section, we describe the procedure for resumming logarithmic contributions controlled by the bottom Yukawa coupling in our hybrid framework. We first describe the employed fixed-order and EFT calculations separately. Then we discuss their combination. The resummation of the bottom Yukawa coupling for large $\tan\beta$ is discussed in \Sec{sec:02_botresum_deltab}.


\subsection{Fixed-order calculation}
\label{sec:02_FO}

The fixed-order part of the calculation consists of the full one-loop and \order{\alt\als,\alt^2} corrections ($\als = g_3^2/(4\pi)$ with $g_3^2$ being the strong gauge coupling, and $\alt = y_t^2/(4\pi)$ with $y_t$ being the top Yukawa coupling) implemented into \FH~\cite{Heinemeyer:2007aq,Hollik:2014bua,Hollik:2014wea}. In these corrections \cp-violating phases are fully taken into account.

Also, two-loop corrections proportional to the bottom-Yukawa coupling have been calculated using different renormali{\sz}ation schemes for the sbottom sector. In \Sec{sec:05_results}, we will compare the results of two different schemes.


\subsubsection*{Renormali{\sz}ation scheme~1}

The present public version of \FH (version \texttt{2.16.1}) includes the \order{\alb\als,\alb\alt,\alb^2} corrections derived in~\cite{Brignole:2002bz,Dedes:2003km} ($\alb = y_b^2/(4\pi)$ with $y_b$ being the bottom Yukawa coupling). For these corrections the following renormali{\sz}ation scheme is employed: The squark masses, $m_{\tilde q}^2$, and mixing angles $\theta_{\tilde q}$, and the top quark mass are renormali{\sz}ed in the on-shell (OS) scheme,
\begin{eqnarray}
\label{eq:OS_CTs}
\begin{aligned}
& \delta m_{\tilde q_i}^2 = \Re~\Sigma_{\tilde q_i \tilde q_i} (m_{\tilde q_i}^2),
\quad \delta \theta_{\tilde q} = \frac{\Re~\Sigma_{\tilde q_1\tilde q_2}(m_{\tilde q_1}^2)+\Re~\Sigma_{\tilde q_1\tilde q_2}(m_{\tilde q_1}^2)}{2(m_{\tilde q_1}^2 - m_{\tilde q_2}^2)}, \\
& \delta M_t = \frac{M_t}{2}~\Re \left[\Sigma_{t}^{L}(M_t^2)+\Sigma_{t}^{R}(M_t^2)+ 2\Sigma_{t}^{S}(M_t^2)\right],
\end{aligned}
\end{eqnarray}
where $\Sigma_{\tilde q_i \tilde q_i}$ is used to denote the respective scalar self-energy, $\Sigma_{t}^{L},~\Sigma_{t}^{R}$ and $\Sigma_{t}^{S}$ are the coefficients in the Lorentz decomposition of the unrenormali{\sz}ed top-quark self-energy,
\begin{equation}
\label{eq:Lor_dec}
\Sigma_{t}(p) = \slashed{p} P_L \Sigma_{t}^{L}(p^2) + \slashed{p} P_R \Sigma_{t}^{R}(p^2) + M_t \Sigma_{t}^{S}(p^2) + M_t \gamma_{5} \Sigma_{t}^{P}(p^2).
\end{equation}
Additionally, \Re~denotes the real part, and $M_t$ is the OS top-quark mass.

The corrections in~\cite{Brignole:2002bz,Dedes:2003km} have been calculated assuming vanishing \cp-violating phases. Moreover, they have been derived in the large $\tan\beta$ limit which implies that the bottom-quark mass $m_b$ is put to zero if it is not multiplied with $\tan \beta$. In this approach, the soft SUSY-breaking masses, the trilinear couplings of the stop and sbottom sector, as well as the top-quark mass are independent parameters. In contrast, the bottom-quark mass is treated as a dependent quantity and the expression for its counterterm is derived from the equation connecting the bottom-quark mass and the sbottom mixing angle,\footnote{We use a different sign convention for $\mu$ in comparison to~\cite{Brignole:2002bz,Dedes:2003km}.}
\begin{equation}
\label{eq:hb_eq}
s_{2\theta_{\tilde b}} = \frac{2 m_b \mu \tbe}{\msbotz^2 - \msbote^2},
\end{equation}
where we introduced the abbreviations
\begin{align}
s_\gamma = \sin\gamma,\hspace{.5cm}c_\gamma = \cos\gamma,\hspace{.5cm}t_\gamma = \tan\gamma
\end{align}
for a generic angle~$\gamma$. Following the discussion in Refs.\cite{Brignole:2002bz,Dedes:2003km} we omitted terms proportional to $\sim A_b$ in \Eq{eq:hb_eq}, since they are not enhanced by factors of $\tan \beta$.

Therefore, the one-loop counterterm for the bottom-quark mass in this scheme has the following form,
\begin{equation}
  \label{ch3:eq88}
  \delta m_b = m_b \left(\frac{\delta m_{\tilde{b}_2}^2 - \delta m_{\tilde{b}_1}^2}{m_{\tilde{b}_2}^2 - m_{\tilde{b}_1}^2} +
  \frac{\delta s_{2\theta_b}}{s_{2\theta_b}} - \frac{\delta \mu}{\mu} - \delta t_{\beta} \right),
\end{equation}
where $\delta \mu$ is the counterterm for the Higgsino mass parameter $\mu$. The actual bottom quark mass, which is used in the calculation, is then given by
\begin{equation}
  \label{eq:mbPietro}
  \widehat{m}_b = m_b^{\DR,\MSSM}(Q)
  \left.\left(1 +
  \frac{\delta m_{\tilde{b}_2}^2 - \delta m_{\tilde{b}_1}^2}{m_{\tilde{b}_2}^2 - m_{\tilde{b}_1}^2} +
  \frac{\delta s_{2\theta_b}}{s_{2\theta_b}} - \frac{\delta \mu}{\mu} - \delta t_{\beta}
  \right)\right|_{\text{fin}}.
\end{equation}
It can be shown by explicit calculation that the renormali{\sz}ation scale dependence of $m_b^{\DR,\MSSM}(Q)$ is canceled out by the scale dependence contained in the combination of the counterterms $\delta m_{\tilde{b}_2}^2, \delta m_{\tilde{b}_1}^2$ and $\delta s_{2\theta_b}$ in the bracket of \Eq{eq:mbPietro}. Therefore, $\widehat{m}_b$ is scale independent at the one-loop level if the Higgsino mass parameter $\mu$ and $\tbe$ are renormali{\sz}ed in the \DR\footnote{To be more precise, these parameters are defined in the \DRp scheme \cite{Jack:1994rk}. We will, however, not make any distinction between \DR and \DRp schemes throughout this paper except of \Sec{sec:04_NNNLL}.} scheme as assumed throughout this work. In \FH, the associated renormali{\sz}ation scale is by default set equal to $M_t$.

Due to the $SU(2)_L$ gauge symmetry, the bilinear soft SUSY-breaking parameters $m_{\tilde{b}_L}$ and $m_{\tilde{t}_L}$ are equal to each other at the tree level. This relation is broken at the one-loop level. The counterterms for $m_{\tilde{t}_L}$ and $m_{\tilde{b}_L}$ read
\begin{subequations}
  \begin{align}
    \label{ch3:eq89a}
    & \delta m_{\tilde{t}_L}^2 = \cos^2 \theta_{\tilde{t}} \; \delta m_{\tilde{t}_{1}}^2 +  \sin^2 \theta_{\tilde{t}} \; \delta m_{\tilde{t}_{2}}^2 + (m_{\tilde{t}_{2}}^2 - m_{\tilde{t}_{1}}^2) \sin 2 \theta_{\tilde{t}} \; \delta \theta_{\tilde{t}} - 2 \; m_t \; \delta m_{t}, \\
    \label{ch3:eq89b}
    & \delta m_{\tilde{b}_L}^2 = \cos^2 \theta_{\tilde{b}} \; \delta m_{\tilde{b}_{1}}^2 +  \sin^2 \theta_{\tilde{b}} \; \delta m_{\tilde{b}_{2}}^2 + (m_{\tilde{b}_{2}}^2 - m_{\tilde{b}_{1}}^2) \sin 2 \theta_{\tilde{b}} \; \delta \theta_{\tilde{b}} - 2 \; m_b \; \delta m_{b},
  \end{align}
\end{subequations}
and are in general not equal to each other. In the following, we will assume that $m_{\tilde{t}_L}^2$ is given as an input parameter. Then the renormali{\sz}ed soft sbottom mass $m_{\tilde{b}_L}^2$ is given by
\begin{equation}
  \label{ch3:eq90}
  m_{\tilde{b}_L}^2 = m_{\tilde{t}_L}^2 + \delta m_{\tilde{t}_L}^2 - \delta m_{\tilde{b}_L}^2.
\end{equation}
The trilinear soft SUSY-breaking parameter $A_b$ is fixed via the sbottom--sbottom--$A$-boson vertex function (see~\cite{Brignole:2002bz,Dedes:2003km} for more details).


\subsubsection*{Renormali{\sz}ation scheme~2}

For our present study, we, however, do not make use of the \order{\alb\als,\alb\alt,\alb^2} corrections already implemented in \FH. Instead, we employ the \order{\alb\als,\alb\alt,\alb^2} corrections presented in~\cite{Passehr:2017ufr,Borowka:2018anu} (see also \cite{Fritzsche:2013fta,Heinemeyer:2010mm}). These also include terms subleading in $\tan\beta$, allow for easier control of the renormali{\sz}ation scheme and take \cp-violating phases fully into account. They will be part of an upcoming \FH release. For the present work, we evaluate them, however, externally and feed the numerical result back to \FH.\footnote{In practice, we use the \texttt{FHAddSelf} functionality (see \texttt{feynhiggs.de}).} In this scheme the soft SUSY-breaking mass $m_{\tilde{b}_L}$ is not treated as an independent parameter and set equal to $m_{\tilde{t}_L}$. This implies
\begin{equation}
  \label{ch3:eq91}
  \delta m_{\tilde{b}_L}^2 = \delta m_{\tilde{t}_L}^2,
\end{equation}
where $\delta m_{\tilde{t}_L}^2$ is given by \Eq{ch3:eq89a}. The consequence of this relation is that only one of the sbottom masses can be set on-shell. As a matter of convention, the mass of the second sbottom is defined in the on-shell scheme,
\begin{equation}
  \label{ch3:eq92}
  \delta m_{\tilde{b}_{2}}^2 = \Re~\Sigma_{\tilde{b}_2 \tilde{b}_2} (m_{\tilde{b}_{2}}^2).
\end{equation}
We treat the mass of the bottom quark as an independent parameter which is renormali{\sz}ed in the \DR scheme,
\begin{equation}
  \label{ch3:eq93}
  \delta m_b = \frac{m_b}{2}~\Re \left.\left[\Sigma_{b}^{L}(m_b^2)+\Sigma_{b}^{R}(m_b^2)+ 2\Sigma_{b}^{S}(m_b^2)\right]\right|_{\rm div},
\end{equation}
where $\Sigma_{b}^{L},~\Sigma_{b}^{R}$ and $\Sigma_{b}^{S}$ are defined in analogy to \Eq{eq:Lor_dec}. The trilinear soft SUSY-breaking parameter $A_b$ is also defined in the \DR scheme,
  \label{ch3:eq94}
\begin{align}
\delta A_b^{\DR} ={}& \frac{1}{m_b} \left[\left(\delta m_{\tilde{b}_{1}}^{2,\;\DR} - \delta m_{\tilde{b}_{2}}^{2,\;\DR} \right)
    \mathbf{U}_{\tilde{b}_{11}} \mathbf{U}_{\tilde{b}_{12}}^* \right. \notag\\
    & \hspace{0.9cm} \left. + \; \delta m_{\tilde{b}_{12}}^{2,\;\DR} \; \mathbf{U}_{\tilde{b}_{21}} \mathbf{U}_{\tilde{b}_{12}}^* + \delta m_{\tilde{b}_{21}}^{2,\;\DR} \; \mathbf{U}_{\tilde{b}_{11}} \mathbf{U}_{\tilde{b}_{22}}^* \right] \notag\\
  & - \left(A_b - \mu^* \; t_{\beta} \right) \frac{\delta m_b^{\DR}}{m_b} + t_{\beta} \; \delta \mu^{*,\;\DR} + \mu^* \; t_{\beta} \; \delta t_{\beta}^{\DR},
\end{align}
where $\mathbf{U}_{\tilde{b}}$ is the mixing matrix of the sbottom sector, and
\begin{subequations}
  \label{ch3:eq95}
  \begin{align}
    \delta m_{\tilde{b}_{12}}^{2,\;\DR} &= \frac{1}{2} \left( \left. \Sigma^{(1)}_{\tilde{b}_{1} \tilde{b}_{2}} (m_{\tilde{b}_{1}}^2) \right|_{\rm div} +
    \left. \Sigma^{(1)}_{\tilde{b}_{1} \tilde{b}_{2}} (m_{\tilde{b}_{2}}^2) \right|_{\rm div} \right) = (m_{\tilde{b}_{1}}^2 - m_{\tilde{b}_{2}}^2) \delta\theta_{\tilde b}^\DR, \\
    \delta m_{\tilde{b}_{21}}^{2,\;\DR} &= \left(\delta m_{\tilde{b}_{12}}^{2,\;\DR} \right)^*.
  \end{align}
\end{subequations}
Eqs.~\eqref{ch3:eq91}--\eqref{ch3:eq95} fix the renormali{\sz}ation conditions for all parameters of the sector.

In both schemes described above it is assumed that the stop sector is renormali{\sz}ed using the OS scheme. We furthermore implemented a pure \DR renormali{\sz}ation of the stop and sbottom sector as an additional option.


\subsection{EFT calculation}
\label{sec:02_EFT}

We build upon the existing EFT calculation in \FH~\cite{Hahn:2013ria,Bahl:2016brp,Bahl:2017aev,Bahl:2018qog}. At the sfermion mass scale, \msusy, all sfermions as well as the non-SM-like Higgs bosons are integrated out.\footnote{In this paper, we do not consider the \FH implementation incorporating a Two-Higgs-Doublet-Model as EFT below the sfermion scale~\cite{Bahl:2018jom}.} Performing the renormali{\sz}ation-group running to lower scales and passing two additional independent thresholds for electroweakinos (charginos and neutralinos) and the gluino, the SM is recovered as EFT.\footnote{The case of the gluino being much heavier than the rest of the MSSM spectrum requires special care. This case was considered in \cite{Bahl:2019wzx}.} The currently implemented EFT calculation resums leading and next-to-leading (LL and NLL) logarithms as well as next-to-next-to-leading logarithms (NNLL) in the limit of vanishing electroweak gauge couplings. So far, however, all corrections proportional to the bottom Yukawa coupling are set to zero in the EFT calculation.

For the incorporation of the bottom Yukawa contributions, our aim was to reach the same level of accuracy as for the other corrections. For implementing LL and NLL resummation, we include the bottom Yukawa contributions to the one-loop matching condition of the SM Higgs self-coupling, $\lambda$~\cite{Vega:2015fna,Bagnaschi:2017xid}. Also the one- and two-loop RGEs are extended by the RGE of the bottom-Yukawa coupling and by bottom-Yukawa contributions to the RGEs of the other couplings (see e.g.~\cite{Buttazzo:2013uya}).

To achieve resummation at the NNLL level, we derive the \order{\alb\als,\alb\alt,\alb^2} threshold corrections for $\lambda$ making use of the two-loop Higgs self-energy corrections obtained in~\cite{Hollik:2014wea,Passehr:2017ufr,Borowka:2018anu}. Details are given in \App{app:07_2L_thresholds_derivation}. In the limit of vanishing \cp-violating phases, we find agreement with the expressions derived in~\cite{Bagnaschi:2017xid} using the effective-potential approach. We also add the bottom Yukawa contributions to the three-loop SM RGEs in the limit of vanishing electroweak gauge couplings~\cite{Mihaila:2012pz,Bednyakov:2012rb,Bednyakov:2012en,Chetyrkin:2013wya,Bednyakov:2013eba} and to the calculation of the SM \MS vev at the electroweak scale.

For the EFT calculation, all sbottom input parameters are defined in the \DR scheme at the scale \msusy. We choose to define $\tan\beta$ in the \DR scheme at the scale \msusy.


\subsection{Combination in the hybrid approach}
\label{sec:02_Hybrid}

For the combination of the fixed-order and the EFT calculation, we follow the procedure described in~\cite{Hahn:2013ria,Bahl:2016brp,Bahl:2017aev,Bahl:2018ykj}. For the self-energy of the SM-like Higgs boson, the result of the fixed-order calculation, $\widehat\Sigma_{hh}^\text{FO}(p^2)$, and the EFT result, $-2\lambda(M_t)(v_{\MS})^2$ (with $v_{\MS}$ being the SM \MS vev at the scale $M_t$), are summed. Subtraction terms are used to ensure that contributions included in both results are not counted twice,
\begin{align}\label{eq:hybrid_combination}
\widehat\Sigma_{hh}^\text{hybrid}(p^2) = \widehat\Sigma_{hh}^\text{FO}(p^2) - 2\lambda(M_t)\overline{v}^2 - \text{(subtraction terms)}.
\end{align}
In order to ease this combination, we choose to define the sbottom input parameters in the same scheme in the fixed-order and the EFT calculations: We fix them in the \DR scheme at the scale \msusy. Also $\tan\beta$ and $\mu$ are fixed in the \DR scheme at the scale \msusy.

A complication arises through the use of the \DR bottom quark mass in the EFT as well as the fixed-order calculation. After adding both results as shown in \Eq{eq:hybrid_combination}, the Higgs pole masses are determined taking into account the momentum dependence of the fixed-order self-energy. As discussed in detail in~\cite{Bahl:2018ykj}, this momentum dependence arises only from SM-type corrections as well as contributions suppressed by the SUSY scale. In order to match the result of a pure EFT calculation, in which the Higgs pole mass is determined in the SM, we have to ensure that the SM-type corrections are evaluated using only SM quantities. The \DR bottom-quark mass, however, is an MSSM quantity. For this reason, we reparametri{\sz}e the SM bottom-quark contributions to the Higgs self-energies in terms of the SM \MS bottom quark mass at the scale $M_t$.

In our implementation, this is achieved by subtracting the UV-finite $\order{\alb}$ self-energy, containing only the SM contributions and parametri{\sz}ed in terms of the MSSM bottom quark $m_b^{\DR,\MSSM}(\msusy)$, and then adding back the same quantity but parametri{\sz}ed via $m_b^{\MS,\SM}(M_t)$. The explicit expression for the one-loop SM Higgs boson self-energy renormali{\sz}ed in the \MS scheme with the tadpoles renormali{\sz}ed to zero is given by
\begin{equation}
\label{ch5:eq2}
\widetilde{\Sigma}_{hh}^{\MS,\SM}(p^2) = \frac{3 m_b^2}{16 \pi^2 v^2} (p^2 - 4 m_b^2) B_0^{\rm fin}(p^2, m_b^2, m_b^2),
\end{equation}
where the superscript ``fin'' means that in this expression we take only the finite part of the one-loop scalar integral function $B_0(p^2,m_1^2,m_2^2)$, for which we use the definition given in \cite{Denner:1991kt}. Since the SM-like Higgs mass is determined via an iterative solution of the pole equation, the same procedure has to be applied to the derivative of $\widetilde{\Sigma}_{hh}^{~\MS,\SM}(p^2)$ with respect to the external momentum. The $\order{\alb \als,\alt \alb,\alb^2}$ SM self-energies computed in the gaugeless limit are extracted from the code {\tt FlexibleSUSY} \cite{Athron:2014yba,Athron:2016fuq,Athron:2017fvs} and Refs.~\cite{Degrassi:2012ry,Martin:2014cxa}.

In order to allow for an OS definition of the stop input parameters, a conversion of the OS parameters, used in the fixed-order calculation, to the \DR scheme, used in the EFT calculation, is necessary. As argued in \cite{Hahn:2013ria,Bahl:2016brp,Bahl:2017aev,Bahl:2019hmm}, for this conversion only one-loop logarithmic terms should be taken into account. Only in the conversion formula for the stop mixing parameter, $X_t$, large logarithms appear. For the present study, we extend the formula given in~\cite{Bahl:2016brp} by including the bottom Yukawa contributions,
\begin{align}
X_t^\DR(\msusy) &= X_t^\OS\left\{1 + \left[\frac{\als}{\pi} - \frac{3\alt}{16\pi}\left(1 - \frac{\vert X_t \vert^2}{\msusy^2}\right) \right.\right.\nonumber\\
&\left.\left.\hspace{3.15cm} + \frac{3\alb}{16\pi}\left(1 + \frac{\vert X_b \vert^2}{\msusy^2}\right)\right]\ln\frac{\msusy^2}{M_t^2}\right\},
\end{align}
where $X_b$ is the sbottom mixing parameter ($X_b = A_b - \mu^* \tan\beta$).


\subsection{Determination of the MSSM bottom quark mass and Yukawa coupling}
\label{sec:02_botresum_deltab}

Here, we describe how we obtain the \DR bottom quark mass used in the fixed-order calculation as well as the \DR bottom Yukawa coupling used in the EFT calculation. As input, we use the SM \MS bottom Yukawa coupling, $y_b^{\MS,\SM}$, and the SM \MS vev, $v^{\MS,\SM}$, at the scale $M_t$. These are evolved to the SUSY scale. At this scale we determine the MSSM \DR bottom Yukawa coupling, $h_b^{\DR,\MSSM}$ (with $h_b\cbe = y_b$ at the tree level), and the MSSM \DR vev, $v^{\DR,\MSSM}$, by matching the SM to the MSSM,
\begin{align}\label{eq:hb_matching2}
\left(h_b^{\DR,\MSSM}c_\beta\right)(\msusy) &= y_b^{\MS,\SM}(\msusy)\left(1 + \Delta y_b\right), \\
\label{eq:hb_matching3}
\vMSSM(\msusy) &= \vMS(\msusy)\left(1 + \Delta v\right),
\end{align}
where the one-loop expression for $\Delta v$ is given in \Eq{eq:bottom_ths_dv} below. The \DR bottom quark mass is then determined by
\begin{align}
m_b^{\DR,\MSSM}(\msusy) = \left(h_b^{\DR,\MSSM}c_\beta\right)(\msusy) v^{\DR,\MSSM}(\msusy).
\end{align}
It is well-known that the relation between the \DR bottom quark mass and the \SM~\MS bottom Yukawa coupling includes terms proportional to $\tan\beta$. For large $\tan\beta$, the leading $\tan\beta$-enhanced terms can be resummed as described in~\cite{Hempfling:1993kv,Hall:1993gn,Carena:1994bv,Carena:1999py,Brignole:2002bz,Dedes:2003km,Guasch:2003cv,Heinemeyer:2004xw}. Typically, this resummation is written in the form
\begin{align}\label{eq:mbMSSM_def}
m_b^{\DR,\MSSM} = m_b^{\MS,\SM} \frac{1 + \epsilon_b}{|1+\Delta_b|},
\end{align}
where $\Delta_b$ includes $\tan \beta$-enhanced terms, which are not suppressed by powers of $m_b/m_t$. $\epsilon_b$ contains all other terms from the one-loop relation between $m_b^{\DR,\MSSM}$ and $m_b^{\MS,\SM}$. We employ a similar relation for the matching of the bottom Yukawa coupling,\footnote{There are terms proportional to $m_b^2 \tan^2 \beta$ in the one-loop relation between $m_b^{\DR,\MSSM}$ and $m_b^{\MS,\SM}$ as well as in the relation between $v^{\DR,\MSSM}$ and $v^{\MS,\SM}$. These terms are suppressed as $\sim m_b^2 \tan \beta/m_t^2$ compared to the top Yukawa contribution to $\Delta_b$ and, therefore, numerically irrelevant. Therefore, we will refer to the terms in $\epsilon_b$ as ``non-enhanced'' terms in this paper.}
\begin{align}\label{eq:hb_matching}
\left(h_b^{\DR,\MSSM}c_\beta\right)(\msusy) = y_b^{\MS,\SM}(\msusy) \frac{1 + \epsilon_b - \Delta v}{|1 + \Delta_b|}.
\end{align}
A similar procedure for the calculation of the MSSM bottom Yukawa coupling was adopted in~\cite{Bagnaschi:2017xid}. There, however, non-enhanced terms, $\epsilon_b$, and the threshold correction of the vev, $\Delta v$, were included into the definition of $\Delta_b$. In our approach, we separate them to resum only $\tan\beta$ enhanced corrections to the bottom-Yukawa coupling in the same way as in~\cite{Brignole:2002bz,Hofer:2009xb}. This results only in a small numerical difference since the main contribution to $h_b^{\DR,\MSSM}(\msusy)$ comes from $\Delta_b$ (see also the discussion in \Sec{sec:05_results}).

In our implementation, we include full one-loop corrections to $\Delta_b$. The quantity $\Delta v$ is calculated at the one-loop level in the gaugeless limit. In addition, we include the leading two-loop corrections to $\Delta_b$. These two-loop corrections are based on the results from~\cite{Noth:2008tw,Noth:2008ths,Noth:2010jy}.\footnote{Similar results have been obtained in~\cite{Martin:2005ch,Mihaila:2010mp}. Moreover, the authors of \cite{Ghezzi:2017enb} derived subleading two-loop corrections, which are not taken into account in the present work.} We, however, perform an expansion of $\Delta_b$ (at the one- and two-loop level) for large \msusy omitting terms of higher-order in $\order{v^2/\msusy^2}$. In addition, we adapt the renormali{\sz}ation scheme to match our scheme. More precisely, in~\cite{Noth:2008tw,Noth:2010jy} the soft supersymmetry breaking parameters in the stop and sbottom sectors as well as the gluino mass are renormali{\sz}ed on-shell. Moreover, all supersymmetric particles and the top quark are decoupled from the scale dependence of the strong coupling $\als$. This decoupling of the top quark and the on-shell renormali{\sz}ation of the top sector induces large logarithms, $\displaystyle\log(\msusy^2/M_t^2)$, implying that the formulas in~\cite{Noth:2008tw,Noth:2008ths,Noth:2010jy} are not directly applicable in our framework. Since in our case the low-energy model is the SM with possibly light gluinos (and electroweakinos), we do not decouple the top quark and the gluino. Also, to be consistent with the other parts of our EFT calculation we renormali{\sz}e the gluino mass and the stop/sbottom masses in the \DR scheme at the matching scale $Q$.

In the limit of all involved non-SM masses having the same value, we obtain
\begin{align}
& \Delta_b^{2l} = \Delta_b^{2l,\order{\als^2}} + \Delta_b^{2l,\order{\alt \als}}, \\
\label{eq:2LdbQCD}
& \Delta_b^{2l,\order{\als^2}} = \frac{\als(Q)^2 C_F}{12\pi^2} \frac{\mu}{\msusy} t_{\beta} \bigg(2 C_A - C_F + 6 T_R \nonumber\\
& \hspace{5.8cm}-\left(3 C_A - 2 C_F - 9 T_R \right) \log \frac{\msusy^2}{Q^2}\bigg), \\
\label{eq:2LdbEW}
& \Delta_b^{2l,\order{\alt \als}} = -\frac{\als(Q) y_t^2(Q) C_F}{384\pi^3} \frac{A_t}{\msusy} t_{\beta} \left(7 + 10 \log \frac{\msusy^2}{Q^2}\right).
\end{align}
Here, $A_t$ is the stop trilinear coupling ($A_t = X_t + \mu^*/\tan\beta$), $Q$ is the renormali{\sz}ation scale, $C_A = 3, C_F = \frac{4}{3}$ and $T_R = \frac{1}{2}$. Formulas also valid for non-degenerate masses are distributed as ancillary files together with this paper.


\section{EFT calculation for complex input parameters}
\label{sec:03_cEFT}

In the fixed-order approach, the dependence on \cp-violating phases is known at the one- and two-loop level~\cite{Frank:2006yh,Heinemeyer:2007aq,Hollik:2014wea,Hollik:2014bua,Hahn:2015gaa}. In the EFT framework, the phase dependence has so far only been considered in case of a low-energy Two-Higgs-Doublet-Model~\cite{Pilaftsis:1999qt,Carena:2000yi,Carena:2015uoe,Murphy:2019qpm,Bahl:2020mjy}. Here, we work out the dependence on \cp-violating phases in the case of the SM (and the SM plus electroweakinos and/or gluinos) as EFT, for which so far only an interpolation of the result in case of real input parameters has been available~\cite{Bahl:2018qog}.

We first discuss the case of the SM as low-energy EFT. Since the SM includes no phases (apart from the CKM phase, whose effect is negligible for the determination of the Higgs mass), \cp-violating effects in the full MSSM enter only via threshold corrections to real parameters. At the one-loop level, the only contribution to the matching of the Higgs self-coupling with a non-vanishing phase dependence is the electroweakino contribution. It depends on the phases of the bino and wino soft-breaking masses, $\pMi$ and $\pMii$, as well as of the Higgsino mass parameter, $\pMue$ (explicit expressions are listed in \App{app:08_CPV_thresholds}). This implies that at the one-loop level, there is no dependence on the phases of the squark sector (at least if the absolute values of the squark mixing parameters, $|X_q|$, are kept constant).

The phases of the stop and sbottom sector along with the gluino phase, $\pMiii$, however, enter the matching of the Higgs self-coupling at the two-loop level. Based upon the fixed-order results presented in~\cite{Hollik:2014wea,Hollik:2014bua,Passehr:2017ufr,Borowka:2018anu}, we extract the dependence of the two-loop threshold correction on these phases at \order{\alpha_{b,t}\als,\alb^2,\alb\alt,\alt^2} without assumptions on the internal masses (details are given in \App{app:07_2L_thresholds_derivation}). In case of real input parameters, we find full agreement with the results of \cite{Bagnaschi:2014rsa,Vega:2015fna,Bagnaschi:2017xid}. By analy{\sz}ing the obtained expressions, it becomes clear how the expressions derived in \cite{Bagnaschi:2014rsa,Vega:2015fna,Bagnaschi:2017xid} can be generali{\sz}ed to the case of complex input parameters:
\begin{itemize}
\item \order{\alq \als} where $q = {t,b}$:
The expression for zero phases is a polynomial in \xq. To get the expression for non-zero phases every odd power of \xq has to be multiplied by $\cos(\pXq - \pMiii)$, and \xq has to replaced by $|\xq|$.
\item \order{\alq^2} where $q = {t,b}$:
The expression for zero phases is a sum of monomials in the variables \xq and $\displaystyle \yq = \widehat{X}_q + \frac{2 \widehat{\mu}^*}{\sin 2 \beta}$ of one of three types: the monomials which contain only even powers of \xq, the ones which contain only even powers of \yq and the ones which contain both \xq and \yq. The latter contain only even or only odd powers of \xq and \yq at the same time. To get the expression for non-zero phases, every monomial which contains odd powers of \xq and \yq has to be multiplied by $\cos(\pXq - \pYq)$, and every \xq and \yq has to be replaced by $|\xq|$ and $|\yq|$, respectively.
\end{itemize}

The generali{\sz}ation of the \order{\alb\alt} expression from the \cp-conserving case to the \cp-violating case is slightly more complicated since different multiplicative factors arise.

Full explicit expressions in the limit of all sfermions having the same mass are given in \App{app:08_CPV_thresholds_TL}. Fully general expressions can be found in ancillary files distributed alongside this paper.

If the low-energy theory is the SM plus electroweakinos, effective Higgs--Higgsino--gaugino couplings are induced. These are potentially complex-valued. An explicit matching calculation at the one-loop level, however, shows that their phase is zero even if one or more of the electroweakino phases in the MSSM are non-zero. Correspondingly, also the RGEs of the SM plus electroweakinos are not modified in the presence of non-zero phases. The phases, however, enter in the threshold corrections for the bottom and top Yukawa couplings as well as the Higgs self-coupling when integrating out the electroweakinos (full expressions are listed in \App{app:08_CPV_thresholds}).

In addition to the phase dependencies discussed above, also the $\Delta_b$ corrections (see \Sec{sec:02_botresum_deltab}) depend on $\pMue$, $\phi_{M_{1,2,3}}$ and $\pAt$. The phase dependence of the one-loop correction has been derived in~\cite{Williams:2008qpa,Hofer:2009xb, Williams:2011bu}. The phase dependence of the two-loop correction, which we derived based upon the result of~\cite{Noth:2008tw,Noth:2008ths,Noth:2010jy} (see \Sec{sec:02_botresum_deltab}), has, however, been unknown so far. We find that this dependence is the same as for the one-loop result. Namely, \Eq{eq:2LdbQCD} has to be multiplied by $\cos(\pMue + \pMiii)$ and \Eq{eq:2LdbEW} has to be multiplied by $\cos(\pMue + \pAt)$.

This can be understood by looking at the explicit two-loop diagrams (see \App{app:09_db2L_CPV}). They fall into three categories: either a gluon, a gluino or a sbottom quark is added to the one-loop graph. If a gluon is added, the phase dependence of the one-loop graph is obviously not changed, since the two additionally appearing strong gauge couplings do not include a phase dependence. The same is true if a sbottom quark is coupled to the one-loop graph by a four-sfermion vertex. Working in the chiral basis, it is again obvious that this coupling does not induce an additional phase dependence. The case of adding a gluino is slightly more complicated. The two additional gluon-gluino-sbottom couplings do depend on the phase of the gluino mass parameter. Working again in the chiral basis, it is easy to see that one of these two additional couplings is a left-handed coupling and the other one is a right-handed coupling. The dependence on the gluino phase cancels between the left-handed and the right-handed coupling. More details and all relevant two-loop diagrams can be found in \App{app:09_db2L_CPV}.


\section{\texorpdfstring{N$^3$LL}{N3LL} resummation}
\label{sec:04_NNNLL}

Up to now, the EFT calculation implemented in \FH was restricted to full LL and NLL resummation as well as NNLL resummation in the limit of vanishing electroweak gauge couplings. In this Section, we discuss the implementation of N$^3$LL resummation at \order{\alt\als^2} based upon the work presented in~\cite{Harlander:2018yhj}.

The following ingredients are needed in addition to the already implemented corrections for NNLL resummation:
\begin{itemize}
\item SM \order{\alpha_{t}\als^2}~Higgs self-energy corrections,
\item leading QCD corrections to the three-loop RGEs of the Higgs self-coupling, the strong gauge coupling as well as the top Yukawa coupling,
\item \order{\als^3}~extraction of the \MS top Yukawa coupling at the electroweak scale,
\item \order{\alpha_{t}\als^2}~matching condition for the Higgs self-coupling between the SM and the MSSM.
\end{itemize}
The SM \order{\alpha_{t}\als^2} corrections to the Higgs self-energy have been obtained in~\cite{Martin:2007pg,Martin:2014cxa,Martin:2015eia}; the necessary RGEs in~\cite{Martin:2015eia,Chetyrkin:2016ruf}. Formulas for extracting the SM \MS couplings at the three-loop level can be found in~\cite{Buttazzo:2013uya}. The \order{\alpha_{t}\als^2} matching condition of the Higgs self-coupling was computed in~\cite{Harlander:2018yhj} based on the \order{\alt^2\als^2} fixed-order calculation performed in \cite{Harlander:2008ju,Kant:2010tf,Harlander:2017kuc}. The result is implemented in the publicly available code \Him~\cite{Harlander:2017kuc,Harlander:2018yhj}. As discussed in~\cite{Harlander:2018yhj}, this calculation is based on an expansion of three-loop diagrams for certain mass hierarchies. \Him provides an uncertainty estimate for this truncation error.

We implemented all these corrections into the EFT calculation of \FH (the link to \Him has already been implemented for the work presented in~\cite{Bahl:2019hmm}). By default, \Him uses the \DRp scheme~\cite{Jack:1994rk} for the renormali{\sz}ation of the squark input parameters~\cite{Harlander:2018yhj}. Correspondingly, also the input parameters of \FH are defined in the \DRp scheme if N$^3$LL resummation is activated. In case of complex input parameters, we interpolate the \Him result.\footnote{An interpolation in case of a complex $M_3$ is not possible, since the expressions implemented in \Him are not dependent on the sign of $M_3$.}

The inclusion of N$^3$LL resummation in the EFT calculation can also be used within the hybrid approach. In this case we, however, require that also in the fixed-order calculation the parameters entering the three-loop threshold correction are renormali{\sz}ed in the \DRp scheme. The two-loop conversion, that would be necessary between OS parameters used in the fixed-order calculation and \DRp parameters used in the EFT calculation, is beyond the scope of the present paper.\footnote{The necessary two-loop squark self-energy corrections have already been calculated in~\cite{Martin:2005eg,Martin:2006ub}.}


\section{Numerical results}
\label{sec:05_results}

In this Section, we discuss the numerical effects of the various improvements discussed above.


\subsection{Resummation of logarithmic bottom Yukawa contributions}
\label{sec:05_results_botResum}

Here, we investigate the numerical effect of resumming logarithmic contributions proportional to the bottom Yukawa coupling. First, we concentrate on a scenario presented in Ref.~\cite{Bagnaschi:2017xid}. Namely, we assume that all soft SUSY-breaking masses are equal to $\msusy = 1.5\tev$ except the gluino mass which is fixed by $M_3 = 2.5\tev$. The stop mixing parameter is set by $X_t = \sqrt{6}\msusy$, and the trilinear couplings of the third generation fermions are equal to each other, $A_b = A_{\tau} = A_t$. The Higgsino mass parameter, $\mu$, is chosen to be equal to $-1.5\tev$. Due to this choice of the signs of $M_3$, $X_t$ and $\mu$ the MSSM bottom Yukawa coupling is enhanced by the one-loop threshold corrections proportional to the top Yukawa coupling and the strong coupling. As in Ref.~\cite{Bagnaschi:2017xid} all the input parameters listed above and $\tan\beta$ are assumed to be \DR parameters at the scale $\msusy$.\footnote{In the considered scenario the ratio $M_3/\msusy$ equals $5/3$. According to the analysis carried out in \cite{Bahl:2019wzx}, the hierarchy between the gluino and squark masses is not so large that a resummation of the $M_3$-enhanced contributions would be required, see e.g. Fig.~1 of \cite{Bahl:2019wzx}.}

\begin{figure}
\begin{minipage}{.48\textwidth}\centering
\includegraphics[width=\textwidth]{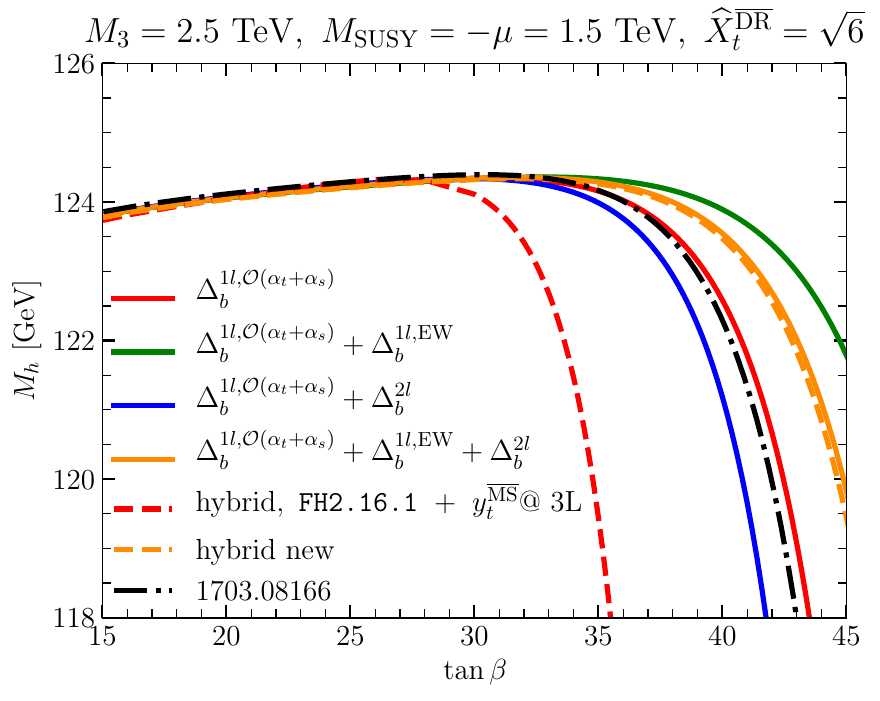}
\end{minipage}
\begin{minipage}{.48\textwidth}\centering
\includegraphics[width=\textwidth]{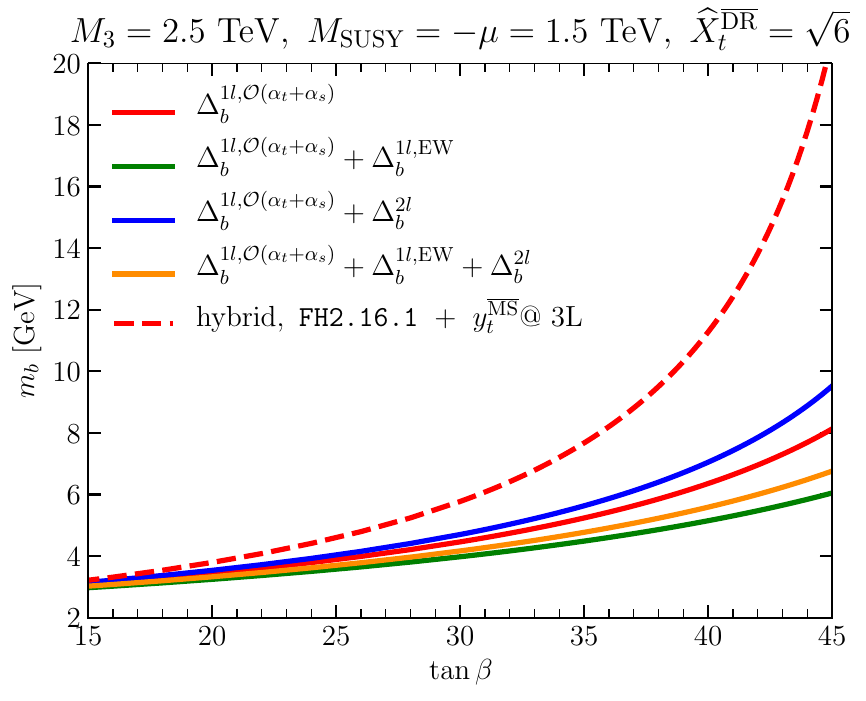}
\end{minipage}
\caption{Predictions for $M_h$ (\textit{left}) and $m_b$ (\textit{right}), which denotes the bottom mass used in the different calculations (see text), as a function of $\tan\beta$ for different accuracy levels in the calculation of $\Delta_b$. For this plot we consider the same MSSM scenario as in Fig.~2 of \cite{Bagnaschi:2017xid}.}
\label{fig:TB_scan}
\end{figure}

In the left panel of~\Fig{fig:TB_scan} we present results for $M_h$ in dependence on $\tan\beta$. In addition to showing results obtained with the calculation presented in this paper, we display results obtained using the most recent public version of \FH (version \texttt{2.16.1}). Moreover, we show the result presented in Fig.~2 of~\cite{Bagnaschi:2017xid} for comparison. This result was obtained in a pure EFT framework using a private code written by the authors of the paper which is formally equivalent to the code \HS~\cite{Athron:2016fuq, Athron:2017fvs} at the discussed order. In the right panel of~\Fig{fig:TB_scan} we show the bottom mass, $m_b$, which is used in the corresponding calculations. In the case of the red dashed curve it is the ``OS'' bottom mass, $\widehat{m}_b$, defined by \Eq{eq:mbPietro} and in case of the blue, red, orange and green solid lines it is $m_b^{\DR,\MSSM}(M_{\rm SUSY})$ given by \Eq{eq:mbMSSM_def}.

In the first step of our numerical analysis, we focus on the various EFT results in the left panel of~\Fig{fig:TB_scan}: the black dot-dashed line corresponds to the result obtained in \cite{Bagnaschi:2017xid} (red solid line in Fig.~2 of~\cite{Bagnaschi:2017xid}). For this curve the full LL and NLL resummation of large logarithms is performed. In addition to that, NNLL logarithms are resummed to all orders in the gaugeless limit (i.e., the electroweak gauge couplings are neglected in the two-loop threshold corrections to $\lambda$). One-loop $\Delta_b$ resummation, including \order{\als,\alt} corrections, is used in the one-loop threshold correction for the bottom Yukawa coupling. The red, blue, green and orange solid lines correspond to the results of our EFT calculation with different approximation levels used in the calculation of the bottom Yukawa threshold correction. We should note here that the results presented in~\cite{Bagnaschi:2017xid} have been obtained using the SM \MS top Yukawa coupling extracted at the ${\rm N^3 LO}$ level while we by default use the NNLO value. For a proper comparison with the results of~\cite{Bagnaschi:2017xid}, we adapted our calculation to use the same level of corrections (see also the discussion in Sections~\ref{sec:04_NNNLL} and~\ref{sec:05_results_NNNLL}). We use this determination of the SM \MS top Yukawa coupling for all curves of~\Fig{fig:TB_scan}.

We observe a very good agreement between our EFT result using only \order{\als,\alt} corrections in the calculation of $\Delta_b$ (solid red curve), which is the same level of accuracy as used in~\cite{Bagnaschi:2017xid}, and the result of~\cite{Bagnaschi:2017xid} (black dot-dashed curve). The absolute difference between the two curves equals $\sim 0.04~{\rm GeV}$ for $t_{\beta} = 15$ and $\sim 0.7~{\rm GeV}$ for $t_{\beta} = 42$ where the curves have a very steep behavior. This difference comes mainly from the determination of the MSSM bottom Yukawa coupling at the scale $M_{\rm SUSY}$. In~\cite{Bagnaschi:2017xid}, the threshold correction for the vacuum expectation value, $\Delta v$, and non-enhanced terms were included in the definition of $\Delta_b$ while we do not include them (see \Eq{eq:hb_matching}). If we include them into $\Delta_b$ as in~\cite{Bagnaschi:2017xid}, the absolute difference between our calculation and the calculation presented in~\cite{Bagnaschi:2017xid} shrinks down even further ($\sim 0.2~{\rm GeV}$ for $t_{\beta} = 42$).

For the green solid curve in the left plot of~\Fig{fig:TB_scan}, we take into account electroweak corrections in the calculation of $\Delta_b$ in addition to the \order{\als,\alt} corrections used for the solid red curve. As a consequence of \Eq{eq:bottom_ths}, this choice leads to a partial cancellation in the calculation of $\Delta_b$ and hence to a suppression of the MSSM bottom mass at the scale $M_{\rm SUSY}$ as one can see on the right panel of~\Fig{fig:TB_scan} showing $m_b^{\DR,\MSSM}(M_{\rm SUSY})$ in dependence of $\tan\beta$. This in turn reduces the downward shift in the Higgs mass by the one-loop threshold corrections to the SM Higgs self-coupling, $\lambda$, that is proportional to the bottom Yukawa coupling.

The blue solid curve in the left plot of~\Fig{fig:TB_scan} shows the prediction for $M_h$ neglecting the electroweak one-loop contributions to $\Delta_b$ but including the leading two-loop QCD corrections to $\Delta_b$. For our parameter choice, these corrections increase the absolute value of $\Delta_b$ by approximately $5\%$. Correspondingly, also the MSSM bottom mass is increased as can be seen in the right plot of~\Fig{fig:TB_scan}. This results in a significant change of the resulting Higgs mass for $\tan \beta \gtrsim 40$ where the dependence on $\tan\beta$ is very pronounced. The orange curves in the left plot of~\Fig{fig:TB_scan} correspond to the inclusion of all corrections to $\Delta_b$ mentioned above. For the considered parameter choice, the electroweak corrections to $\Delta_b$ are roughly three times larger by absolute value than the two-loop corrections to $\Delta_b$. This explains why the orange and the green curves lie quite close to each other.

The orange dashed curve represents the result of the hybrid calculation of $M_h$. Namely, we have merged the fixed-order result with the NNLL EFT calculation (see \Sec{sec:02_botresum}). The orange solid and dashed curves differ essentially by the inclusion of terms which are suppressed by the ratio $v^2/M_{\rm SUSY}^2$ into the hybrid result. Since in our case $M_{\rm SUSY}$ is chosen above the TeV scale, the size of these terms is, as expected, quite small. Therefore, the observed good agreement between the two methods serves as a consistency check of our hybrid calculation.

Finally, the red dashed curve shows the prediction for $M_h$ obtained by \texttt{FeynHiggs-2.16.1} which we ran using the default flags as explained in~\cite{Bahl:2018qog}.\footnote{For reference, here we list the values of these input flags: \texttt{mssmpart} = 4, \texttt{higgsmix} = 2, \texttt{p2approx} = 4, \texttt{looplevel} = 2, \texttt{loglevel} = 3, \texttt{runningMT} = 1, \texttt{botResum} = 1, \texttt{tlCplxApprox} = 0.} As only modification of this version, we have used the ${\rm N^3 LO}$ instead of the NNLO SM \MS top Yukawa coupling to allow for a direct comparison to the result of~\cite{Bagnaschi:2017xid}. We see that the agreement between all the seven curves is quite good for small values of $\tan \beta$, but for $\tan \beta \gtrsim 30$ the red dashed curve shows a steep fall-off, while for the other curves the large downward shift from $b/\tilde{b}$-sector corrections sets in only at higher values of $\tan \beta$. The reason for this behaviour becomes clear when looking at the right panel of~\Fig{fig:TB_scan}: the red dashed curve, which corresponds to $\widehat{m}_b$ defined in \Eq{eq:mbPietro}, increases much more rapidly for rising $\tan\beta$ than the other four lines.\footnote{Note that the red dashed and the red solid line have the same accuracy level of $\Delta_b$.} This expression for the bottom mass is inserted in the leading one-loop fixed order result which gives rise to a large downward shift of $M_h$~\cite{Carena:1995bx, Carena:2000dp, Draper:2013oza},
\begin{equation}
\label{eq:Mh_nonlog}
  (\Delta M_h^2)^{\rm 1-loop, bottom} \simeq - \frac{\widehat{m}_b^4 \tan^4 \beta}{16 \pi^2 v^2}.
\end{equation}
This term grows rapidly in absolute value with increasing $\tan \beta$. A similar effect occurs for all other curves but there the dependence of the bottom mass on $\tan \beta$ is much milder. This is a consequence of our choice of the renormali{\sz}ation scheme. Namely, the bottom mass used in our setup is the \DR MSSM bottom mass calculated at the scale $M_{\rm SUSY}$. All the quantities entering the calculation of $\Delta_b$ and $\epsilon_b$ are also \DR MSSM quantities at this scale. The most important ones are the top Yukawa coupling $\alt$ and the strong Yukawa coupling $\alpha_s$ (see \Eq{eq:bottom_ths} in \App{app:08_CPV_thresholds}). Since their values decrease with increasing scale, \footnote{Note that we have also included the leading two-loop QCD corrections to $\Delta_b$ which reduces its scale dependence.} the $\Delta_b$ correction calculated in our approach is smaller than the corresponding correction in~\texttt{FeynHiggs-2.16.1}. In this way our approach yields more stable results for large values of $\tan \beta$ and for regions of the MSSM parameter space where the signs of the products $\mu M_3$ and $\mu A_t$ are negative.

\medskip

Next, we discuss the numerical effect induced by the resummation of logarithms proportional to the bottom Yukawa coupling. First of all, to have an idea how numerically important the effect is, it is instructive to have a look at the analytic one- and two-loop expressions which one can find in \App{app:10_Mh_logs}. The bottom mass we use in our calculation, even though being potentially enhanced by $\Delta_b$ effects, is the smallest mass taken into account in our EFT calculation. The only way corrections containing the bottom mass may become sizeable is when these terms are additionally proportional to $\tan\beta$. This is the case when $m_b$ is, for example, multiplied by $\widehat{X}_b$,~$\widehat{Y}_t$, $t_{\beta}$ or $1/c_{\beta}$. We only find such enhancements in the two-loop next-to-leading logarithmic contributions when the stop mixing parameter, $\widehat{X}_t$, is renormali{\sz}ed in the OS scheme (see \Eqs{eq:Mh2L_NLL_OS}{eq:Mh2L_NLL_DR}). As a consequence, we expect the effect of the resummation to be small if we renormali{\sz}e $\widehat{X}_t$ in the \DR scheme.

\begin{figure}
\begin{minipage}{.48\textwidth}\centering
\includegraphics[width=\textwidth]{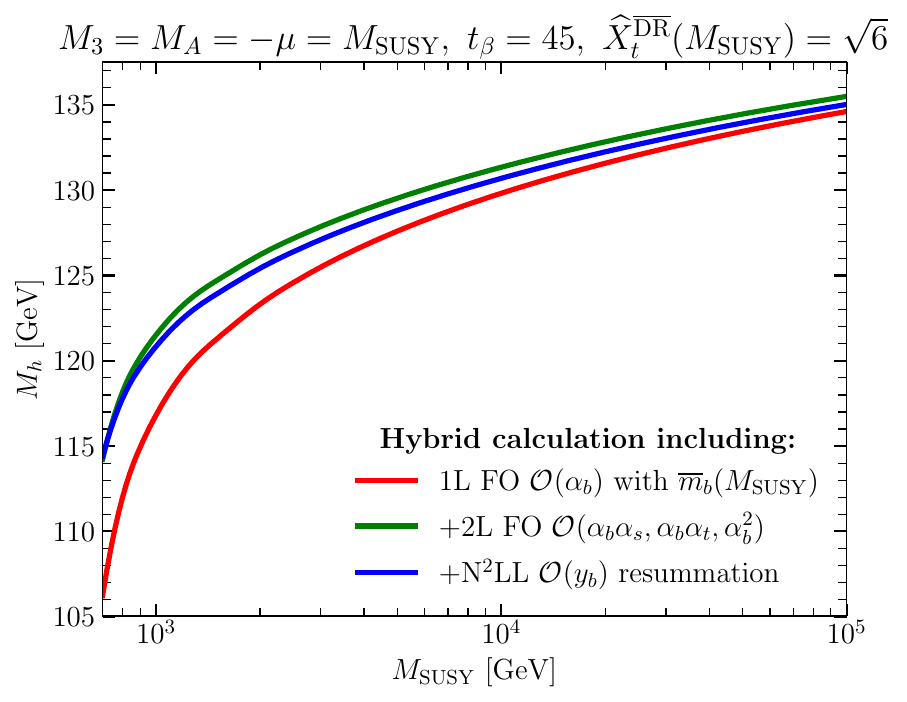}
\end{minipage}
\begin{minipage}{.48\textwidth}\centering
\includegraphics[width=\textwidth]{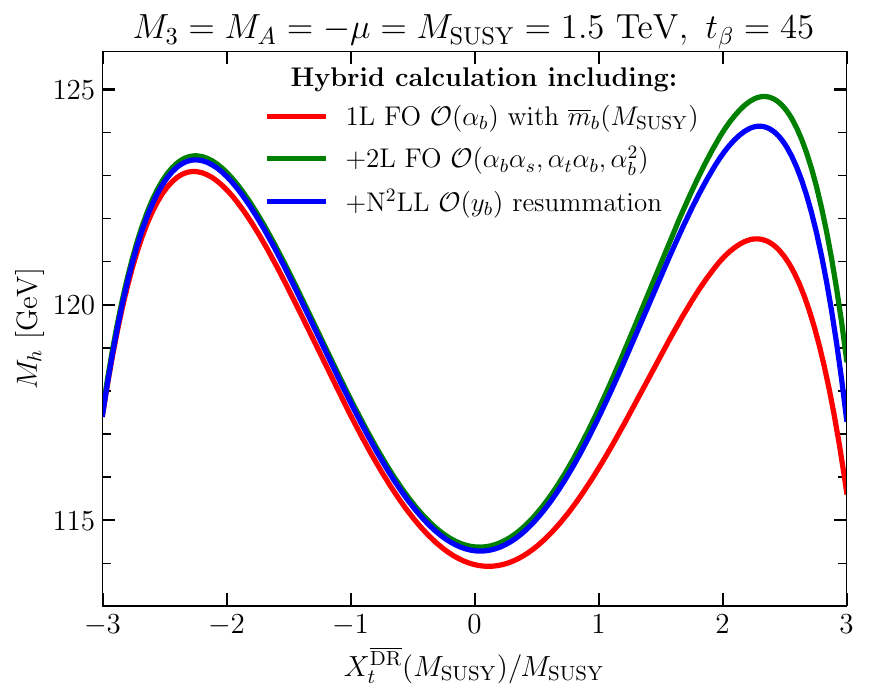}
\end{minipage}
\caption{$M_h$ as a function of \msusy (\textit{left}) and $\xt^\DR(\msusy)$ (\textit{right}). The \textit{red} lines show the prediction of our hybrid calculation including only the one-loop fixed-order \order{\alb} correction. For the \textit{blue} lines, we additionally included the fixed-order \order{\alb\als,\alb\alt,\alb^2} corrections. The \textit{green} lines contain additionally the resummation of logarithms proportional to the bottom Yukawa coupling up to the NNLL level.}
\label{fig:botresumMh_XtDR}
\end{figure}

\begin{figure}
\begin{minipage}{.48\textwidth}\centering
\includegraphics[width=\textwidth]{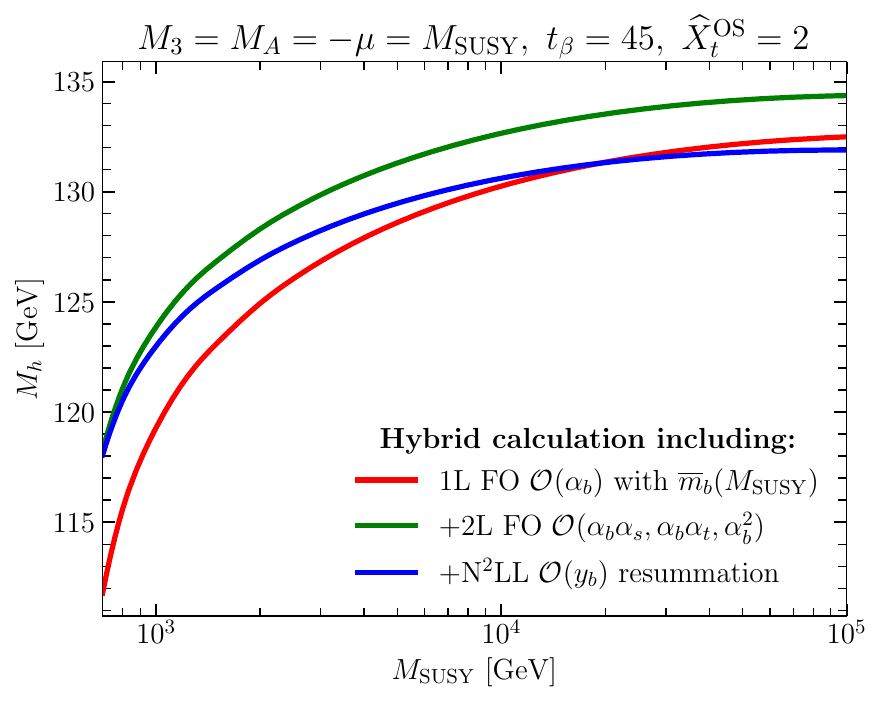}
\end{minipage}
\begin{minipage}{.48\textwidth}\centering
\includegraphics[width=\textwidth]{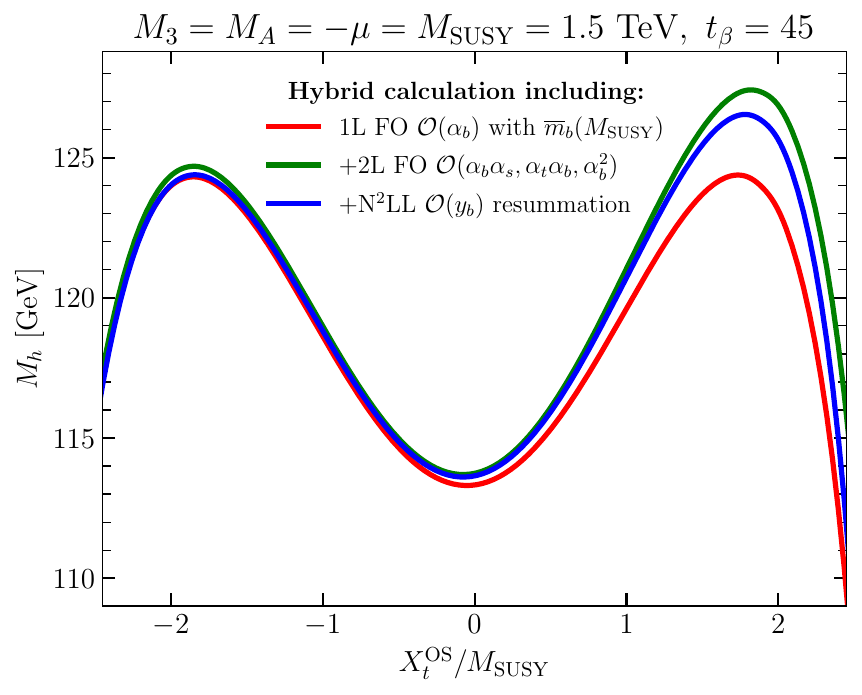}
\end{minipage}
\caption{Same as \Fig{fig:botresumMh_XtDR} but $X_t$ is renormali{\sz}ed in the OS scheme.}
\label{fig:botresumMh_XtOS}
\end{figure}

This qualitative consideration turns out to be reflected in the numerical results as one can see in the left panel of~\Fig{fig:botresumMh_XtDR}. The red curve corresponds to the hybrid result including the effects of the bottom Yukawa coupling only at the one-loop level in the fixed order calculation. The used MSSM \DR bottom mass, $m_b^{\rm \DR, MSSM}(M_{\rm SUSY})$, contains all the corrections discussed above (i.e. the level of accuracy corresponds to the orange curves in~\Fig{fig:TB_scan}). The green curve includes additionally the \order{\alb\als,\alb\alt,\alb^2} fixed-order corrections from~\cite{Passehr:2017ufr, Borowka:2018anu}. Finally, the blue curve also contains the resummation of LL, NLL and NNLL logarithms controlled by the bottom Yukawa coupling. The same color scheme also applies to the right panel of~\Fig{fig:botresumMh_XtDR} and to both plots in~\Fig{fig:botresumMh_XtOS}.  For these plots we have picked a MSSM scenario where all soft-breaking masses and $\mu$ are equal by absolute value to the common mass scale $M_{\rm SUSY}$. Moreover, we set $A_b = 2.5 \msusy$ and $t_{\beta} = 45$. The bino, wino and gluino masses are chosen to be positive, $M_{1,2,3} > 0$,  while the Higgsino mass parameter is negative, $\mu<0$.

For the left plot of \Fig{fig:botresumMh_XtDR} we have chosen $\widehat{X}_t^{\DR}(\msusy) \equiv X_t^{\DR}(\msusy)/\msusy = \sqrt{6}$. First, we note that the green and the blue curves agree very well with each other for low values of $M_{\rm SUSY}$. The difference between the two amounts to only $\sim 0.3~{\rm GeV}$ for $\msusy = 700~{\rm GeV}$. In this region, where the scale separation is relatively small, the resummation of higher-order logarithmic contributions is expected to be subdominant.\footnote{It is worth noting here that the different treatment of the vacuum expectation value in the fixed-order calculation, where $v_{G_F}$ is used, and in the EFT calculation, where $v^{\DR, \rm MSSM}(\msusy)$ is used, induces differences at order $\mathcal{O}(m_b^2 \alpha (\alt + \alb))$ which are beyond the accuracy level of our calculation and hence may be regarded as part of the uncertainties from unknown higher-order corrections. Moreover, the different treatment of the vacuum expectation value in the two-loop threshold corrections and in the respective pieces of the fixed-order calculation, inducing a difference at the three-loop order, contributes to the small shift between the result containing the two-loop fixed order result in the $b/\tilde{b}$-sector and the one including the resummation of higher-order logarithmic contributions. A similar effect was also discussed in~\cite{Bahl:2017aev}.} However, the two curves lie quite close to each other for the whole range of scales, even for $M_{\rm SUSY}$ as high as $10^5~{\rm GeV}$. This is in line with our qualitative analysis above: the logarithms containing bottom Yukawa coupling are numerically small if $\widehat{X}_t^{\DR}(\msusy)$ is used as an input parameter.

On the other hand, a large shift of about $10~{\rm GeV}$ between the red and the green curves in this Figure can be observed for small values of $\msusy$. It decreases for rising $\msusy$ and amounts to about $0.5~{\rm GeV}$ for $\msusy = 10^5~{\rm GeV}$. This result indicates that for this scenario $\Delta_b$ by itself is not a good approximation for the higher-order effects controlled by the bottom Yukawa coupling in the region of small \msusy, since the MSSM bottom mass $m_b^{\DR, \rm MSSM}(\msusy)$ is large in this region due to the large and negative value of $\Delta_b$.\footnote{For example, for $\msusy = 700~{\rm GeV}$ it amounts to $m_b^{\DR, \rm MSSM}(\msusy) \simeq 7.5~{\rm GeV}$.} Thus, in this scenario two-loop fixed-order corrections from the $b/\tilde{b}$-sector that go beyond the $\Delta_b$ contribution are numerically important.\footnote{It is worth noting that due to the parameter choices, which enhance $\Delta_b$, different levels of approximation in $\Delta_b$ yield very different results for $M_{h}$ in this scenario.} With increasing $M_{\rm SUSY}$ the bottom mass $m_b^{\rm \DR, MSSM}(M_{\rm SUSY})$ decreases and the three curves get close to each other.

In the plot on the right panel of~\Fig{fig:botresumMh_XtDR} we fix $M_{\rm SUSY} = 1.5~{\rm TeV}$ and vary $\widehat{X}_t^{\DR}$. One can see that for negative $\widehat{X}_t^{\DR}$ all three curves give roughly the same result. This is due to the fact that the contributions to $\Delta_b$ proportional to the strong coupling and to the top Yukawa coupling partially cancel each other. Correspondingly, the bottom mass does not acquire a significant enhancement for negative $\widehat{X}_t^{\DR}$. The inclusion of the two-loop fixed-order corrections as well as the resummation of the logarithms has only a small effect in this case. On the contrary, for positive $\widehat{X}_t$ the top Yukawa and strong corrections to $\Delta_b$ add up and enhance the bottom mass.\footnote{In particular, in the considered scenario and for $\widehat{X}_t^{\DR} = \sqrt{6}$ the top-Yukawa and strong contributions to the one-loop $\Delta_b$ amount to $\Delta_b^{\order{\alt}} \simeq -0.33$ and $\Delta_b^{\order{\als}} \simeq -0.41$, respectively.} In accordance with \Eq{eq:Mh_nonlog}, this shifts the Higgs mass downwards at the one-loop level. This effect can be seen in the shape of the red curve: while for scenarios where contribution of the $b/\tilde{b}$-sector is numerically small (see e.g. Fig.~9 in \cite{Bahl:2019hmm}) the local maximum for $M_h$ at positive values of $\widehat{X}_t^{\DR}$ is typically several GeV higher than the one at negative values of $\widehat{X}_t^{\DR}$, for the red curve in the right plot of \Fig{fig:botresumMh_XtDR} the maximum at positive values of $\widehat{X}_t^{\DR}$ is about $2~{\rm GeV}$ lower than the one at negative values of $\widehat{X}_t^{\DR}$. Because of the large value of $m_b^{\DR, \rm MSSM}(\msusy)$ the incorporation of the two-loop corrections in the $b/\tilde{b}$-sector (green curve) has a significant effect for $\widehat{X}_t^{\DR} = \sqrt{6}$, giving rise to an upward shift of more than $3~{\rm GeV}$. Since there is only a moderate splitting between \msusy and $M_t$, the effect of the resummation of higher-order logarithmic contributions remains relatively small (blue curve). It amounts to a downward shift of less than $1~{\rm GeV}$ for $\widehat{X}_t^{\DR} = \sqrt{6}$.

\medskip

In \Fig{fig:botresumMh_XtOS}, we renormali{\sz}e the stop sector in the OS scheme. In the left plot $\widehat{X}_t^{\rm OS} = 2$ is chosen. This plot shares the same features at low $M_{\rm SUSY}$ as the corresponding plot in~\Fig{fig:botresumMh_XtDR}. However, for large $M_{\rm SUSY}$ the effect of the resummation becomes more prominent due to the presence of logarithmic terms of $\mathcal{O}(m_b^2 m_t^4)$ that are enhanced by $t_{\beta}^2$ (see \Eq{eq:Mh2L_NLL_OS} below),
\begin{equation}
  (M_h^{\rm 2L,NLL})^2_\text{bot,OS} \simeq 3 \kappa^2 \displaystyle \frac{\overline{m}_b^2 \overline{m}_t^4}{v^4} t_{\beta}^2 \vert \widehat{A}_t \vert^2 (6 - \vert \widehat{A}_t \vert^2) \log \frac{\msusy^2}{M_t^2}.
\end{equation}
In the considered scenario, the resummation gives rise to a downward shift of the Higgs mass, visible as the difference between the blue curve and the green curve, by $\sim 2~{\rm GeV}$ for $M_{\rm SUSY} = 10~{\rm TeV}$ and by $\sim 2.5~{\rm GeV}$ for $M_{\rm SUSY} = 100~{\rm TeV}$. On the right panel of this Figure we show the result of varying $\widehat{X}_t^{\rm OS}$ with fixed $\msusy = 1.5~{\rm TeV}$. As in the case of the $\DR$ stop input parameters the three lines are very close to each other for $\widehat{X}_t^{\rm OS} < 0$. The effect of the inclusion of the two-loop \order{\alb\als,\alb\alt,\alb^2} fixed-order corrections (green curve) and the resummation (blue curve) becomes sizeable in the region $\widehat{X}_t^{\rm OS} \gtrsim 1$. Because of the moderate value of $\msusy = 1.5~{\rm TeV}$ the impact of the higher-order logarithmic contributions is not significantly enhanced compared to the result expressed in terms of $\widehat{X}_t^{\DR}$ shown in the right plot of \Fig{fig:botresumMh_XtDR}.

\medskip

As a final phenomenological application of our improved calculation, we consider the $M_h^{125,\mu-}$ benchmark scenario recently defined in~\cite{Bahl:2020kwe}, accompanying the benchmark scenarios proposed in \cite{Bahl:2018zmf,Bahl:2019ago}. In this scenario the SUSY input parameters are fixed as
\begin{eqnarray}
&M_{Q_3}=M_{U_3}=M_{D_3}=1.5~\text{TeV},\quad
M_{L_3}=M_{E_3}=2~\text{TeV}, \nonumber\\[2mm]
&  \mu=-2~\text{TeV},\,\quad
M_1=1~\text{TeV},\quad M_2=1~\text{TeV},
\quad M_3=2.5~\text{TeV}, \nonumber\\[2mm]
&X_t=2.8~\text{TeV},\quad A_b=A_\tau=A_t\,.
\nonumber
\end{eqnarray}
For the SM parameters the ones recommended by the LHC-HXSWG~\cite{deFlorian:2016spz} are used:
\begin{eqnarray}
&m_t^{\text{pole}}=172.5~\text{GeV},\quad
\alpha_s(\mZ)=0.118,\quad
G_F=1.16637\cdot 10^{-5}~\text{GeV}^{-2},\nonumber\\
&m_b(m_b)=4.18~\text{GeV},\quad
\mZ=91.1876~\text{GeV},\quad
\mW=80.385~\text{GeV}\,.
\nonumber
\end{eqnarray}
The stop SUSY soft-breaking parameters are defined in the OS scheme. In~\cite{Bahl:2020kwe}, also the sbottom trilinear coupling is renormali{\sz}ed in the OS scheme. For better comparison with our previous results, we instead choose to fix $A_b$ and the sbottom masses in the \DR scheme.\footnote{The difference to the corresponding result using the OS scheme for the renormali{\sz}ation of the sbottom sector is very small.} In addition, we define $\tan\beta$ at the scale \msusy instead of at the scale $M_t$, which was used in~\cite{Bahl:2020kwe}.

Note that for this scenario $\mu=-2~\text{TeV}$ is chosen implying relatively large $\Delta_b$ corrections which enhance the cross section times branching ratio for the heavy Higgs bosons decaying to a pair of bottom quarks. In addition, the $\Delta_b$ corrections also affect the prediction for the SM-like Higgs boson, which we will investigate here.

The stop mass scale is equal to $1.5~\rm{TeV}$, so we do not expect the resummation of logarithms controlled by the bottom Yukawa coupling to have a major numerical impact in this case (see discussion above). On the other hand, as we have seen in Figs.\ref{fig:TB_scan}--\ref{fig:botresumMh_XtOS}, large $\Delta_b$ corrections imply that the prediction for $M_h$ can be sensitive to the level of accuracy in the determination of the bottom mass which is used in the fixed-order corrections at the one- and the two-loop level.

\begin{figure}\centering
\begin{minipage}{.68\textwidth}\centering
\includegraphics[width=\textwidth]{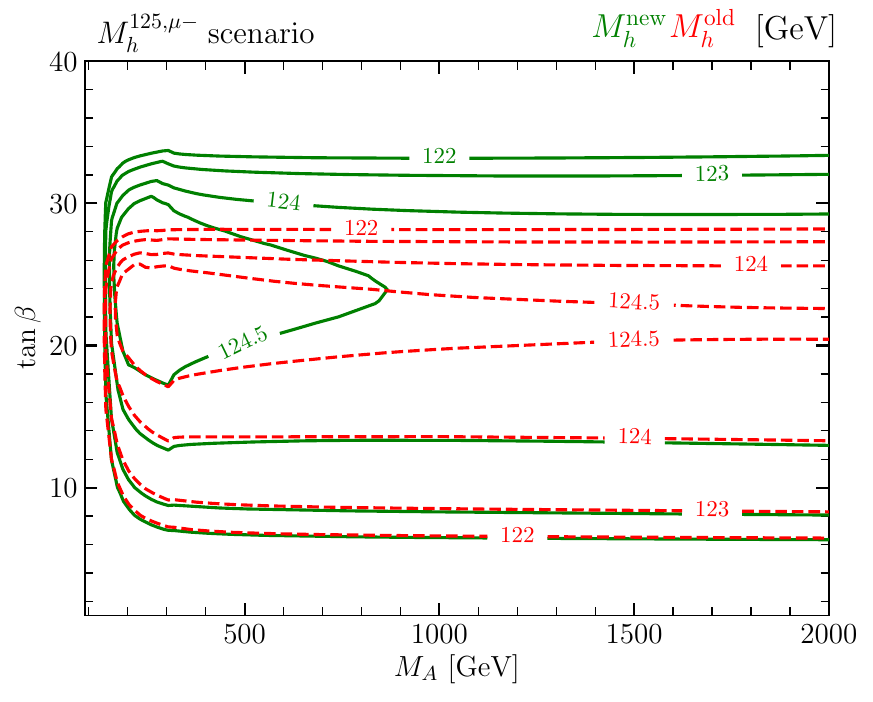}
\end{minipage}
\caption{Predicted contour lines for $M_h$ in the $M_h^{125,\mu-}$ scenario using a calculation including only the leading corrections to $\Delta_b$, corresponding to the one used in~\cite{Bahl:2020kwe} (red dashed lines), and our improved calculation presented in this paper (green solid lines).}
\label{fig:Mh125-mu_scenario}
\end{figure}

In \Fig{fig:Mh125-mu_scenario} we present, in the $(M_A, \tan \beta)$ plane, the contour lines of the SM-like Higgs boson mass ranging from $122~\rm{GeV}$ to $125~\rm{GeV}$.\footnote{Throughout the plane, $M_h < 125$ GeV. Therefore, no 125~GeV contours appear.} We do not consider any of the experimental constraints described in detail in \cite{Bahl:2018zmf,Bahl:2019ago,Bahl:2020kwe} and concentrate only on the prediction for the mass of the lightest Higgs boson of the MSSM. The red dashed and green solid lines correspond to two different computational setups. We calculated the red contours including only the leading one-loop corrections to $\Delta_b$ of \order{\als,\alt} and evaluated the bottom-quark mass according to Renormali{\sz}ation scheme 1 as described in \Sec{sec:02_botresum}. Apart from the different definition of some of the input parameters, as mentioned above, this corresponds to the default settings of \texttt{FeynHiggs-2.16.1}, which was used in~\cite{Bahl:2020kwe} for the analysis of the benchmark scenario. The green lines show the prediction based on the improved calculation described in this paper. In comparison to the red contours, we also include electroweak one-loop as well as the leading two-loop corrections to $\Delta_b$, evaluate the bottom-quark mass at the SUSY scale according to \Eq{eq:mbMSSM_def} and resum logarithms proportional to the bottom-Yukawa coupling.

We notice that in the region of small $\tan \beta$ both calculations agree with each other very well since in this region the corrections from the bottom/sbottom sector are negligible. In this region the Higgs mass grows with increasing $\tan \beta$ mainly due to the growth of the tree-level mass. With a further increase of $\tan \beta$ the Higgs mass starts to decrease due to large $\Delta_b$ corrections and the rapid increase of the \DR bottom mass in the MSSM. This behaviour corresponds to the one that we observed in the left plot of \Fig{fig:TB_scan}. As discussed there, the mass of the SM-like Higgs computed using {\tt FeynHiggs-2.16.1} falls faster with increasing $\tan \beta$ than the mass computed using the calculation presented in the current paper due to the lower accuracy level in the calculation of $\Delta_b$ of the previous result. Consequently, the $\tan \beta$-region in which the SM-like Higgs mass is compatible (taking into account the theoretical uncertainties) with the experimentally measured value is enlarged. The corresponding upper bound on $\tan\beta$ in this scenario is shifted from $\sim 28$ to $\sim 33$.

\subsection{EFT calculation for complex input parameters}
\label{sec:05_results_cEFT}

In this Section, we discuss the numerical effect of including the full phase dependence into the two-loop threshold corrections to the Higgs self-coupling. First, let us briefly review the method used in \FH to handle non-zero phases so far. The treatment of the two-loop corrections in the presence of complex parameters is controlled by the flag \texttt{tlCplxApprox}. When it equals 3, the fixed-order \order{\alt \als,\alt^2} corrections including the full phase dependence are activated and combined with the fixed-order \order{\alb\als,\alb\alt,\alb^2} corrections. Since the implementation of the latter corrections up to now is based on the results of \cite{Brignole:2002bz,Dedes:2003km}, that were obtained for the case of real parameters, an interpolation in the phases is invoked for this part of the two-loop corrections. Specifically, an interpolation is carried out in \texttt{FeynHiggs} when the phases of $\mu,~M_3, X_t$ or $X_b$ are non-zero. The user can choose between interpolation in $A_t$ or $X_t$, and $A_b$ or $X_b$. In the EFT part of the code the interpolation is always carried out in the following way. First, the RGEs are integrated numerically and the subtraction terms are calculated for all possible combinations of $+\vert P \vert$ and $-\vert P \vert$ (where $P \in \{\mu, X_t / A_t, M_3\}$).\footnote{The threshold corrections in \texttt{FH-2.16.1} do not depend on $X_b$ or $A_b$.} After that, linear interpolation is performed on the obtained grid. In this Section, we choose to interpolate $X_t$ in the comparison with our new results when the phase of $X_t$ or $A_t$ is non-zero.

\begin{figure}
\begin{minipage}{.48\textwidth}\centering
\includegraphics[width=\textwidth]{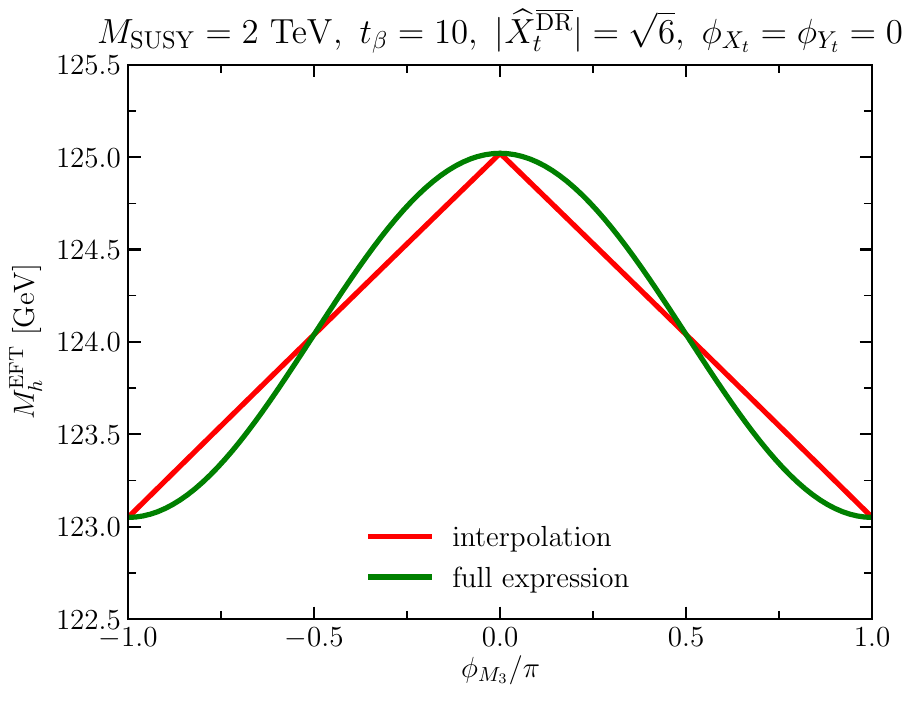}
\end{minipage}
\begin{minipage}{.48\textwidth}\centering
\includegraphics[width=\textwidth]{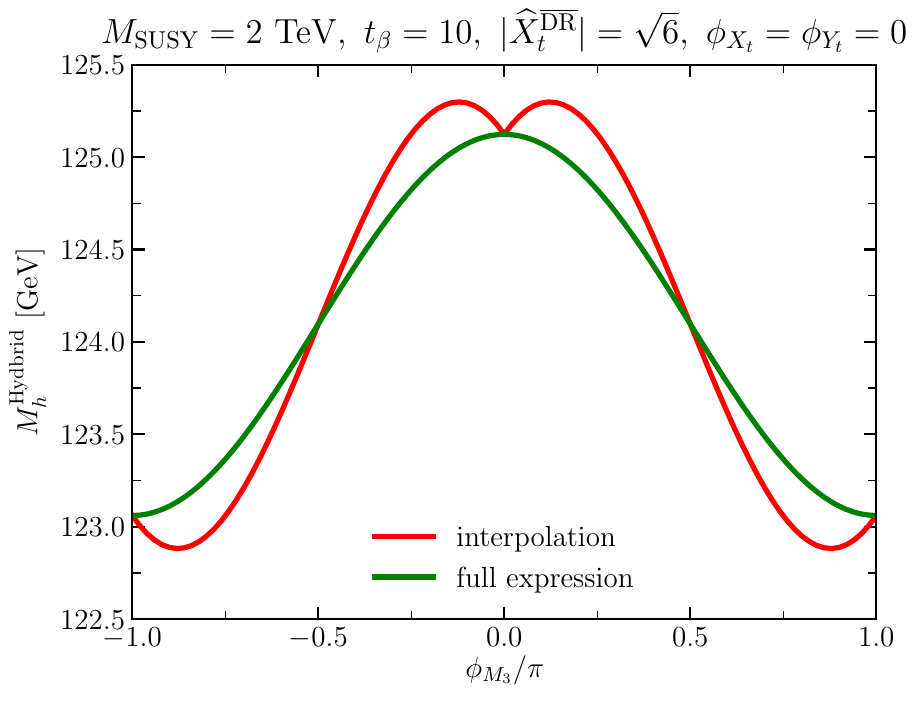}
\end{minipage}
\caption{\textit{Left:} $M_h$ as a function of $\pMiii$ setting $\pXt = \pYt = 0$ calculated using the pure EFT calculation. The results obtained by using an interpolation of the phase and by including the full phase dependence are compared. \textit{Right:} Same as left plot, but the results of the hybrid calculation are shown.}
\label{fig:pM3_scan_pXt0}
\end{figure}

\begin{figure}
\begin{minipage}{.48\textwidth}\centering
\includegraphics[width=\textwidth]{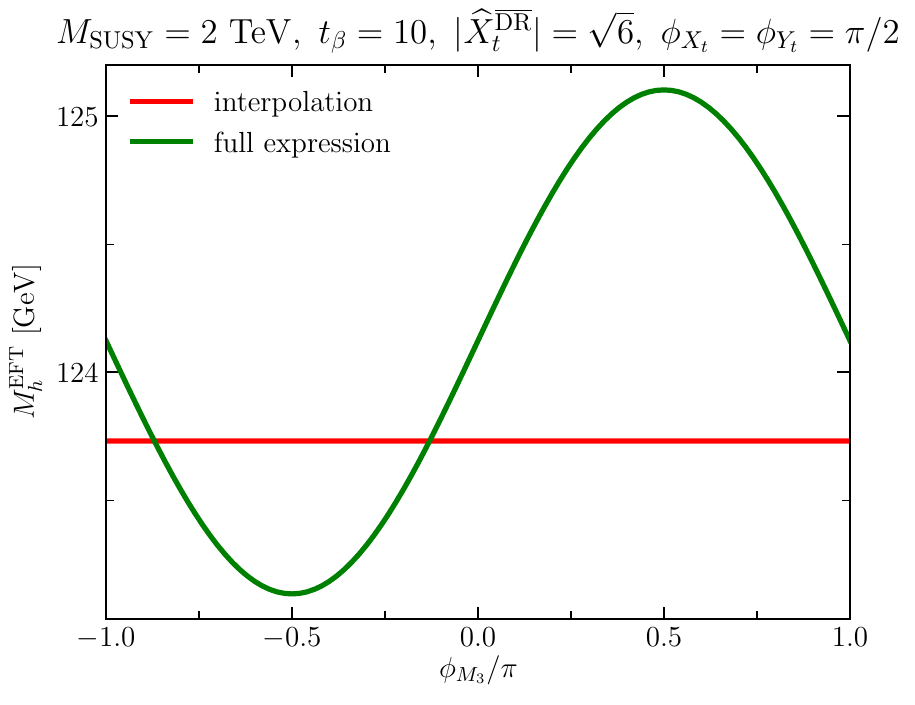}
\end{minipage}
\begin{minipage}{.48\textwidth}\centering
\includegraphics[width=\textwidth]{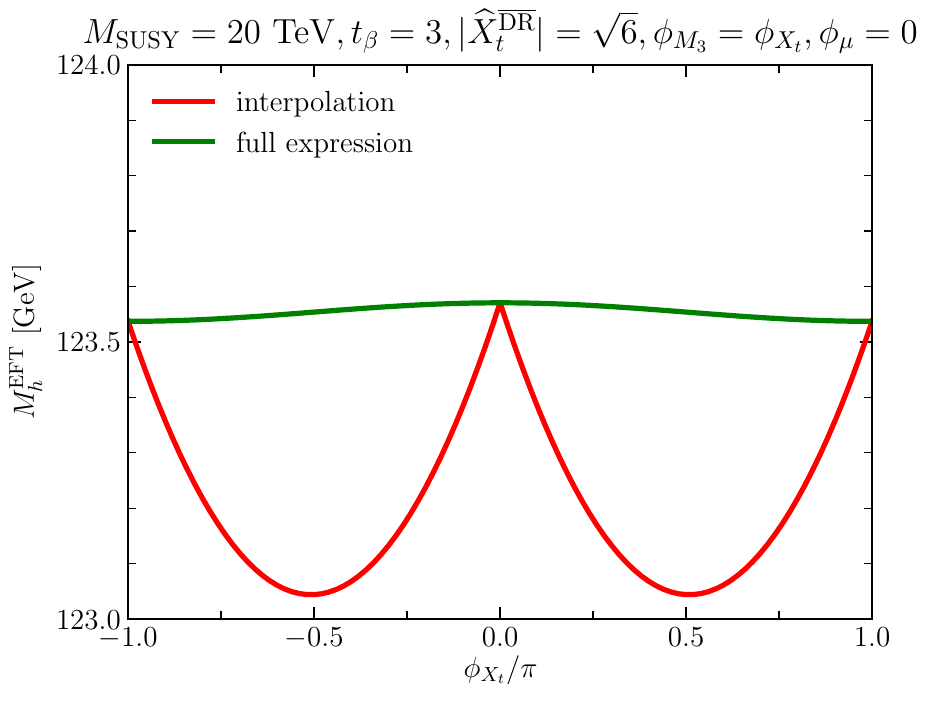}
\end{minipage}
\caption{\textit{Left:} The same as the left plot of~\Fig{fig:pM3_scan_pXt0} but for $\pXt = \pYt = \pi/2$. \textit{Right:} $M_h$ as a function of $\pXt$ setting $\pMiii = \pXt$ and $\pMue = 0$. The results obtained by interpolating the EFT calculation and by including the full phase dependence are compared.}
\label{fig:pM3_scan_pXtPi2}
\end{figure}

The phases of the above-mentioned parameters enter the hybrid calculation via threshold corrections to the Higgs self-coupling and via the subtraction terms. As we mentioned in \Sec{sec:03_cEFT}, both of them depend only on the absolute value $\vert \widehat{X}_t \vert$ at the one-loop level, so the interpolation would give a correct result if only LL and NLL resummation were included and the interpolation was performed in $X_t$. However, the two-loop threshold corrections to the Higgs self-coupling (and hence the two-loop non-logarithmic subtraction terms) do not depend just on the absolute value of $X_t$. For example, the $\mathcal{O}(\alt\als)$ threshold correction also depends on the cosine of the phase difference, $\cos(\pXt - \pMiii)$, and the formula for the $\mathcal{O}(\alt^2)$ threshold correction depends on $\vert \widehat{Y}_t \vert$ and $\cos(\pXt - \pYt)$. In comparison to the full formula, the application of interpolation leads to deviations at the next-to-next-to-leading logarithmic order. The phases also enter the expression for the two-loop threshold corrections of the bottom Yukawa coupling and $\Delta_b$. First, we will, however, concentrate on MSSM scenarios in which the effect of the bottom Yukawa coupling on the Higgs mass is negligible and so we will not include any two-loop corrections of \order{\alb\als,\alb\alt,\alb^2} for the results that are presented in \Fig{fig:pM3_scan_pXt0} and \Fig{fig:pM3_scan_pXtPi2}.

\medskip

In order to test our approach we first consider the same MSSM scenario as in Fig.~3 of~\cite{Bahl:2018qog}: all soft SUSY breaking masses and $\mu$ are equal to the common mass scale $M_{\rm SUSY} = 2~{\rm TeV}$, $\tan \beta = 10$ and $\widehat{X}_t^{\rm \DR}(\msusy) = \sqrt{6}$. We vary the phase of the gluino mass parameter $M_3$ in the interval $[-\pi, +\pi]$ and assume all the other input parameters to be real. In this way, we test the phase dependence of the \order{\alt\als} threshold correction.

In the left plot of~\Fig{fig:pM3_scan_pXt0}, we show the comparison between the pure EFT prediction of \texttt{FeynHiggs-2.16.1} (red line) and our new calculation including the full phase dependence (green line). First, we notice that the two methods give the same answer for $\pMiii = 0,\pm \pi$ which is expected because in these cases $M_3$ is a real parameter. This serves as a cross-check for our implementation. Second, we see that the interpolation in this particular scenario is a fairly good approximation: the absolute difference between the two curves does not exceed $\sim 0.3~{\rm GeV}$. The largest deviations occur for $\pMiii \simeq \pm \frac{\pi}{4}$ and $\pMiii \simeq \pm \frac{3\pi}{4}$. Since the interpolation is only performed in one parameter, $\pMiii$, the resulting curve consists of two straight lines.

In the case of the hybrid calculation (see right plot of~\Fig{fig:pM3_scan_pXt0}), the phase dependence at the two-loop level is fully included in the fixed-order part of the calculation. However, the subtraction terms are interpolated in the same way as the EFT calculation. As a consequence of those phase-dependent contributions in the fixed-order part and the subtraction terms, the curve showing the interpolated hybrid calculation (red) has a different behaviour than the interpolated EFT calculation shown in the left plot. As in the left plot of~\Fig{fig:pM3_scan_pXt0}, the overall difference between the full hybrid and the interpolated hybrid calculation does not exceed  $\sim 0.3~{\rm GeV}$.

\medskip

Next, we proceed with a scenario which is similar to the one described above, but we assume that $\widehat{X}_t$ and $\widehat{Y}_t$ are purely imaginary while keeping the same absolute value for $ \vert \widehat{X}_t \vert = \sqrt{6}$ as before. As one can see in the left plot of~\Fig{fig:pM3_scan_pXtPi2}, where again the result of the pure EFT calculation varying the phase of $M_3$ is shown, the trilinear interpolation procedure results in a straight line that does not depend on $\pMiii$. In the chosen scenario, this line overestimates the result for the full expression for the Higgs mass for $\pMiii/\pi \in [-0.87, -0.13]$ and underestimates it for the other values of $\pMiii$. The absolute difference between the two approaches amounts to $\sim 1.2~{\rm GeV}$ for $\pMiii \simeq \frac{\pi}{2}$. The two results do not agree for $\pMiii = 0,\pm\pi$ since $X_t$ and $Y_t$ are chosen purely imaginary, and therefore an interpolation is also carried out with respect to those phases.

\medskip

As a next step, we investigate the effects of the phase dependence in the $\mathcal{O}(\alt^2)$ threshold correction. To enhance the numerical value of this correction, we choose a low value for $\tan \beta$, namely $\tan \beta = 3$. This choice, however, suppresses the tree-level Higgs mass, so to obtain a predicted value around $125~{\rm GeV}$ we have to choose in this scenario a heavy SUSY scale of $M_{\rm SUSY} = 20~{\rm TeV}$. In order to isolate the effects of the phase dependence in the considered corrections, we fix the phase of the gluino mass parameter to be equal to the phase of $X_t$. As a consequence of this choice, the phase dependence in the $\mathcal{O}(\alt \als)$ threshold correction vanishes. We also choose the Higgsino mass parameter to be positive, $\pMue = 0$.

The EFT prediction, varying the phase $\pXt = \pMiii$, is shown in the right plot of~\Fig{fig:pM3_scan_pXtPi2}. Even though we have chosen a low value of $\tan \beta = 3$ in order to enhance the impact of the \order{\alt^2} threshold correction, the overall phase dependence of the full result (green) is quite small. The difference between the Higgs mass calculated at $\pXt = 0$ and $\pXt = \pi$ is only $\sim 0.05~{\rm GeV}$. Lowering $\tan \beta$ even further (and pushing $\msusy$ higher) does not lead to a stronger phase dependence. The behaviour of the interpolated result (red) is different. As in the case of~\Fig{fig:pM3_scan_pXt0}, we see that the results of both methods coincide for $\pXt = 0,\pm \pi$ since for these three points all parameters are real. For other values of $\pXt$, however, the interpolation procedure underestimates the value of $M_h$ predicted based on the full expression by up to $\sim 0.5$ GeV.

\begin{figure}\centering
\includegraphics[width=0.6\textwidth]{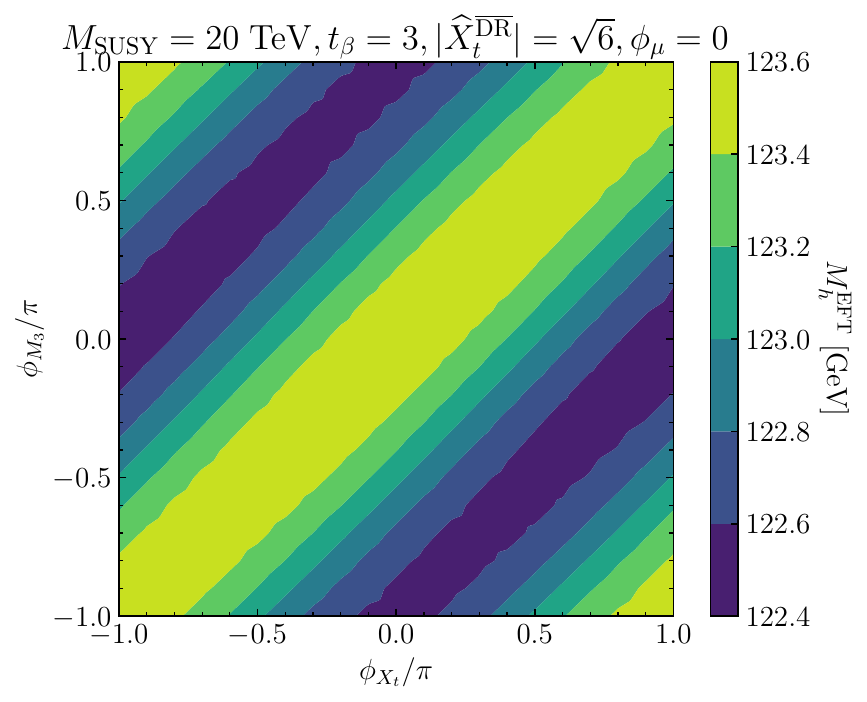}
\caption{Result of the EFT calculation using the full phase dependence. The same scenario as in the right plot of \Fig{fig:pM3_scan_pXtPi2} is used, but $\pXt$ and $\pMiii$ are varied independently.}
\label{fig:pXt_pM3_scan}
\end{figure}

This large deviation can be understood by looking at \Fig{fig:pXt_pM3_scan} showing the $M_h$ prediction of the EFT calculation including the full phase dependence. The same scenario as in the right plot of \Fig{fig:pM3_scan_pXtPi2} is used, but $\pXt$ and $\pMiii$ are varied independently. As visible in the plot, the contours are almost diagonal due to the small phase dependence of the \order{\alt^2} threshold corrections. The parabola-like shape of the interpolated result, as visible for the red curve in the right plot of \Fig{fig:pM3_scan_pXtPi2}, is a consequence of the bilinear interpolation in $\pXt$ and $\pMiii$. For $\pXt = \pMiii > 0$, the Higgs mass values at $(\pXt,\pMiii)=(0,0),(0,\pi),(\pi,0),(\pi,\pi)$ enter the interpolation procedure. For the values $(\pXt,\pMiii)=(0,\pi),(\pi,0)$ the phase dependence of the \order{\alt\als} threshold correction is picked up resulting in the large phase dependence observed for the red curve in the right plot of \Fig{fig:pM3_scan_pXtPi2}. In the considered case, an interpolation in $\pMiii = \pXt$ rather than in $\pMiii$ and $\pXt$ separately would improve the quality of the interpolation.

It should be noted that for the hybrid result the difference between the EFT result incorporating the full phase dependence and the one based on the interpolation, shown in the right plot of \Fig{fig:pM3_scan_pXtPi2}, is further enhanced because of the different treatment of the phase dependence in the fixed-order contribution and the subtraction terms. As a consequence, in this extreme scenario, the incomplete cancellation between the corresponding terms in the fixed-order part and the subtraction terms leads to an artificial enhancement of the deviation that can amount up to $\sim 2$ GeV.

\medskip

\begin{figure}
\begin{minipage}{.48\textwidth}\centering
\includegraphics[width=\textwidth]{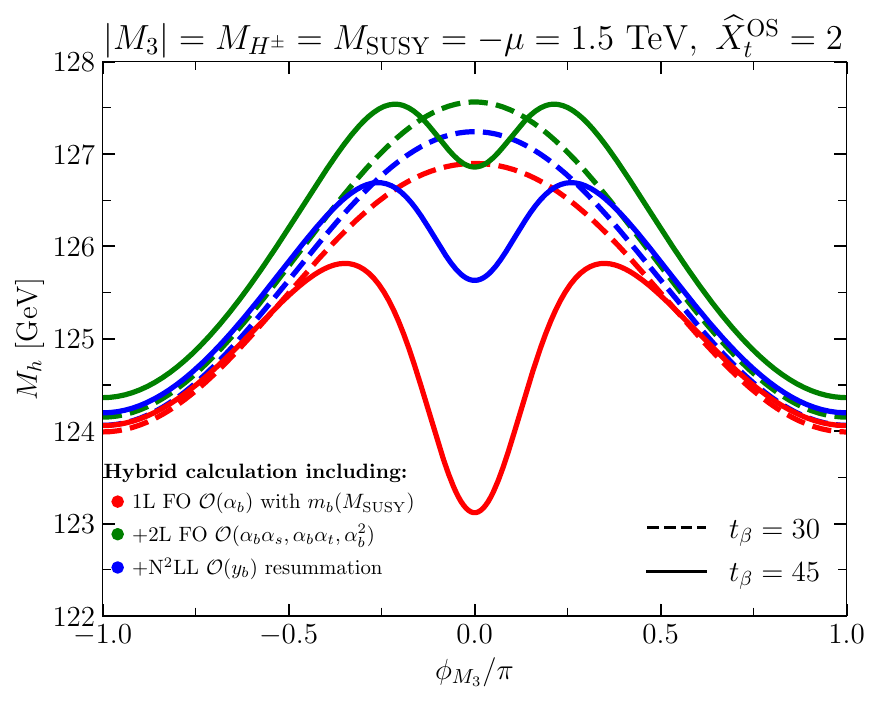}
\end{minipage}
\begin{minipage}{.48\textwidth}\centering
\includegraphics[width=\textwidth]{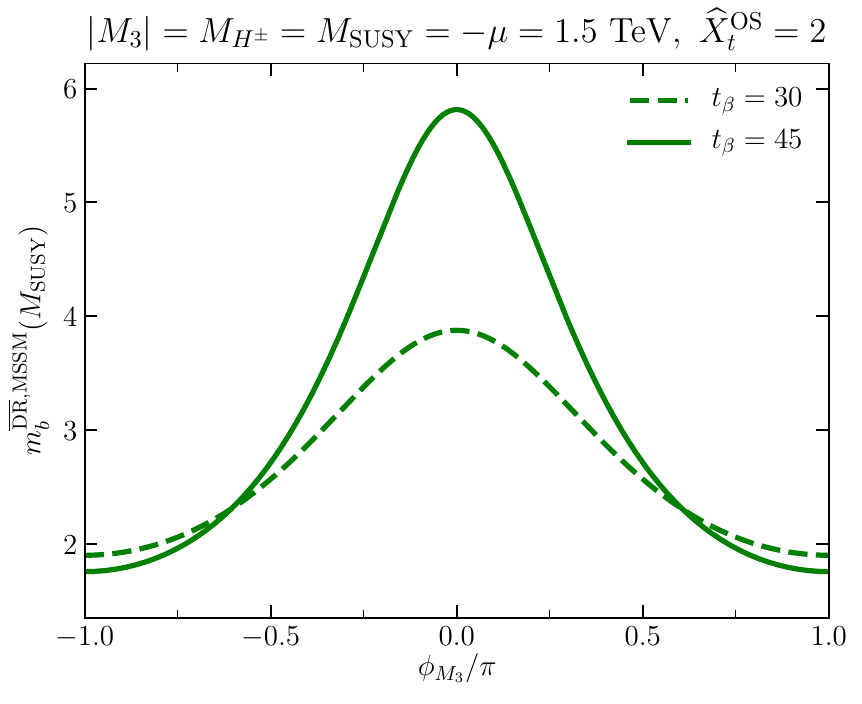}
\end{minipage}
\caption{\textit{Left:} $M_h$ as a function of the phase of the gluino mass $\pMiii$. The red, green and blue colors on this plot mean the same as in~\Fig{fig:botresumMh_XtDR}. The dashed curves correspond to $\tan \beta = 30$ and the solid curves to $\tan \beta = 45$. \textit{Right:} $m_b^{\DR, \rm MSSM}(\msusy)$ as a function of $\pMiii$.}
\label{fig:MhMB_pM3_scan}
\end{figure}

As a final topic in this Section, we analy{\sz}e the interplay between the resummation of the logarithms proportional to the bottom Yukawa coupling and the inclusion of the full phase dependence into the EFT part of our hybrid calculation. As a starting point we go back to the scenario discussed in \Sec{sec:05_results_botResum}. Namely, we consider a single scale scenario, where all soft-breaking masses as well as the mass of the charged Higgs boson\footnote{Since we consider here the \cp-violating case, $M_{H^{\pm}}$ is chosen as an input parameter instead of $M_A$.} are equal to $1.5~{\rm TeV}$, $A_b^{\DR} = 2.5 \msusy$, the Higgsino mass parameter is negative, $\mu = - \msusy$, the bino and wino masses are chosen to be positive, $M_{1,2} > 0$, and $\widehat{X}_t^{\rm OS} = 2$. The phase of the gluino mass parameter is a free parameter, and we vary it in the interval from $-\pi$ to $+\pi$. We examine this scenario for $\tan \beta = 30$ and $\tan \beta = 45$.

The result for $M_h$ as a function of $\pMiii$ is shown in~\Fig{fig:MhMB_pM3_scan}. The colors of the curves on the left panel correspond to the same levels of accuracy as in~\Fig{fig:botresumMh_XtDR}, solid lines correspond to $\tan \beta = 30$, and dashed lines correspond to $\tan \beta = 45$. For $\pMiii = \pm \pi$, the results displayed by all six lines agree with each other within $\sim0.4 \,\,{\rm GeV}$. Here, the strong and the top Yukawa contributions to $\Delta_b$ partially cancel each other, and the MSSM bottom mass does not acquire an enhancement. In fact, for the mentioned points the $\Delta_b$ correction is positive, so that the $\Delta_b$ corrections lead to a suppression of the bottom mass. This is visible in the right panel of~\Fig{fig:MhMB_pM3_scan}, where the solid line (for which $\vert \Delta_b \vert$ is larger) lies below the dashed line for $\pMiii \simeq \pm \pi$.

The red dashed curve resembles the cosine-shape line shown in~\Fig{fig:pM3_scan_pXt0}. This is due to the fact that even for $\pMiii = 0$, where the bottom mass is maximal for $\tan \beta = 30$, it is still too small to have a sizeable effect on $M_h$. Here, the shape of the line can be explained by the phase dependence of the two-loop fixed-order corrections of $\mathcal{O}(\alt\als)$. Adding furthermore the two-loop fixed-order corrections of \order{\alb\als,\alb\alt,\alb^2} (blue dashed line) lifts the prediction for the Higgs mass by $\sim 0.2~\rm{GeV}$ for $\pMiii = \pm \pi$ and by $\sim 0.7~\rm{GeV}$ for $\pMiii = 0$. The inclusion of the resummation of the logarithms proportional to the bottom Yukawa coupling (green dashed line) has a similar numerical effect.

The behaviour as a function of $\pMiii$ is significantly different for $\tan \beta = 45$. The red solid curve starts to grow when $\pMiii$ increases starting from $-\pi$, resembling the red dashed line in shape. However, it reaches a maximum value at $\pMiii \simeq -\frac{\pi}{3}$. This is a consequence of the fact that the $\Delta_b$ correction becomes important in this region, leading to a steep increase of the MSSM bottom mass (see right plot of~\Fig{fig:MhMB_pM3_scan}). Thus, the one-loop corrections involving the bottom mass (see \Eq{eq:Mh_nonlog}) become important, giving rise to a downward shift in $M_h$. At $\pMiii=0$ the bottom mass reaches $\sim 5.8~{\rm GeV}$, and the Higgs mass prediction has a minimum at $\sim 123~{\rm GeV}$. The point $\pMiii = 0$ in this plot corresponds to the point where $\msusy = 1.5~{\rm TeV}$ in the left plot of~\Fig{fig:botresumMh_XtOS}. As in~\Fig{fig:botresumMh_XtOS}, we observe that the inclusion of the two-loop fixed-order corrections controlled by the bottom Yukawa coupling (the difference between the red and the green curves) has a very significant effect. The resummation of higher-order logarithmic contributions (the difference between the blue and the green curves) leads to a downward shift of $\sim 1\gev$ for $\pMiii\simeq \pm\frac{\pi}{3}$ and of $\sim 1.2\gev$ for $\pMiii = 0$. The results displayed in \Fig{fig:MhMB_pM3_scan} demonstrate that the (s)bottom sector contributions can have an important impact on the phase dependence. Similarly to \Fig{fig:botresumMh_XtOS}, we again find that the resummation of logarithms proportional to the bottom-Yukawa coupling amount to an \order{1\gev} effect for large $\tan\beta$ if the OS scheme is used for the renormalization of the stop sector.


\subsection{\texorpdfstring{N$^3$LL}{N3LL} resummation}
\label{sec:05_results_NNNLL}

Here, we study the numerical effects of including N$^3$LL resummation at leading order in the strong gauge coupling (see \Sec{sec:04_NNNLL}) into our hybrid framework. We study a simple single-scale scenario in which all non-SM masses are set to the common scale \msusy. Furthermore, we set all trilinear soft SUSY-breaking couplings, except for $A_t$, to zero. We define the stop parameters in the \DR scheme at the scale \msusy. We set $\tan\beta=10$.

\begin{figure}\centering
\begin{minipage}{.48\textwidth}\centering
\includegraphics[width=\textwidth]{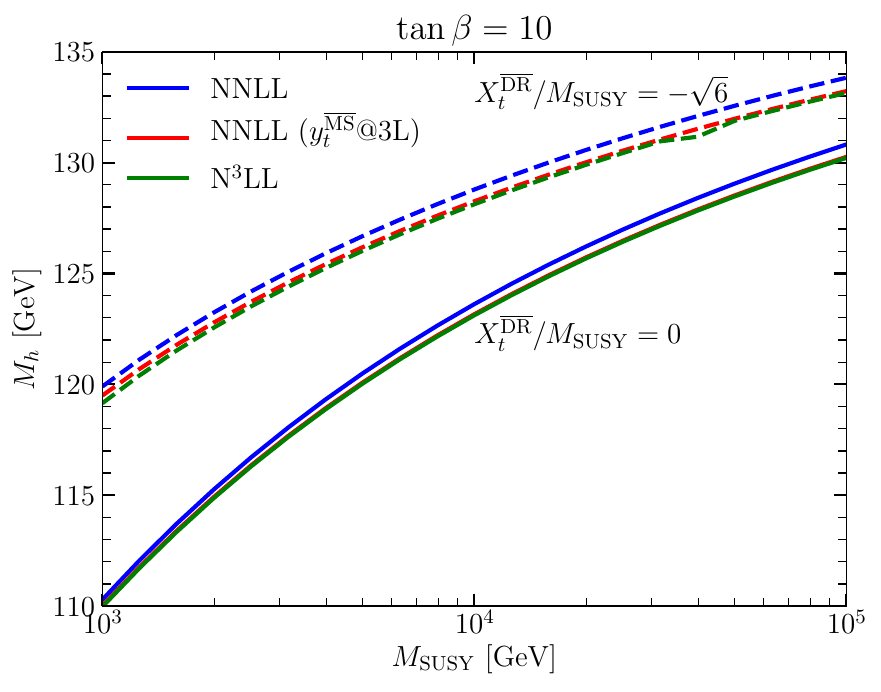}
\end{minipage}
\begin{minipage}{.48\textwidth}\centering
\includegraphics[width=\textwidth]{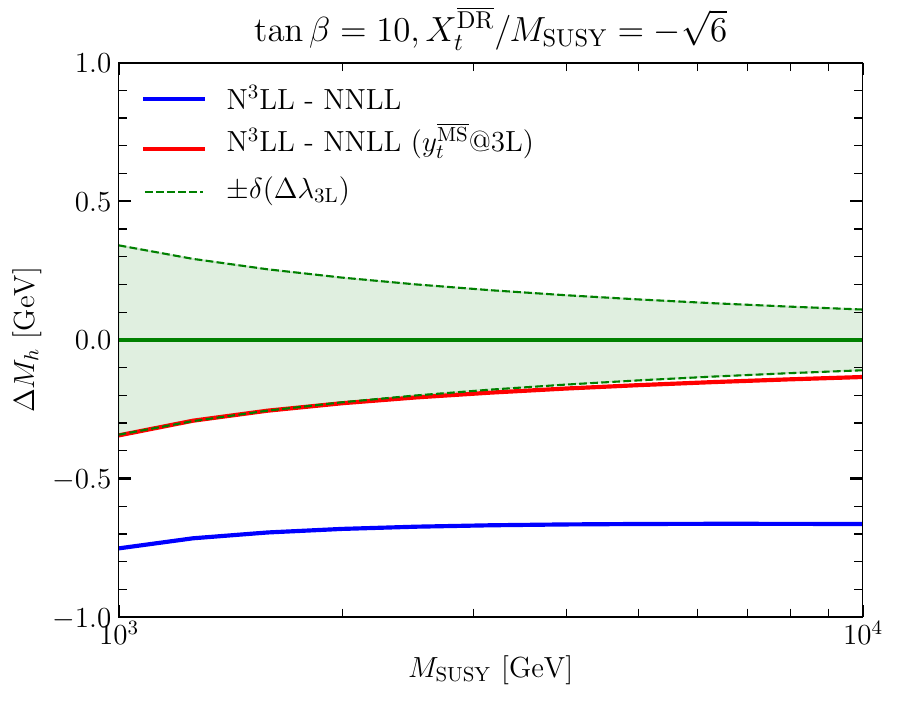}
\end{minipage}
\begin{minipage}{.48\textwidth}\centering
\includegraphics[width=\textwidth]{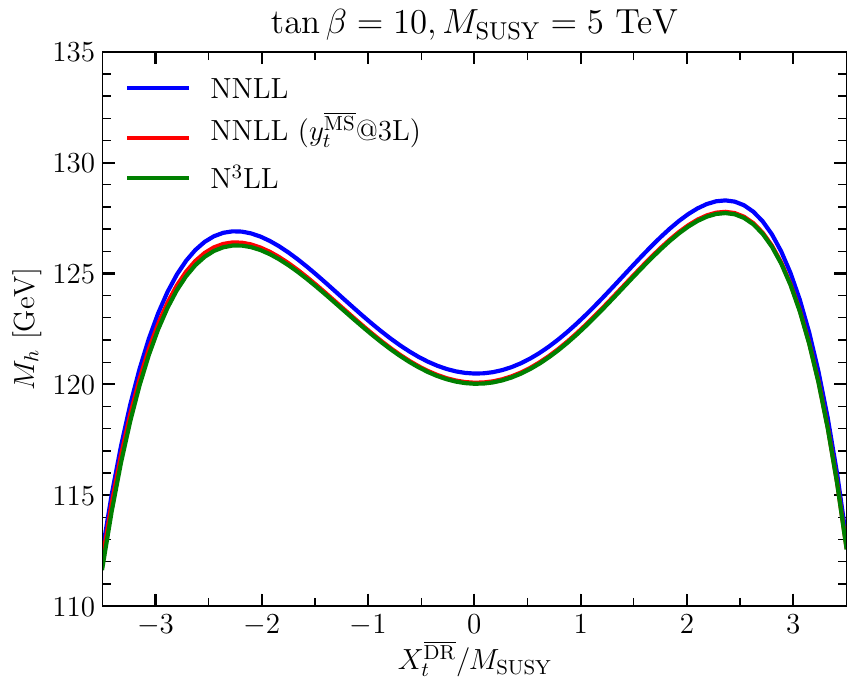}
\end{minipage}
\begin{minipage}{.48\textwidth}\centering
\includegraphics[width=\textwidth]{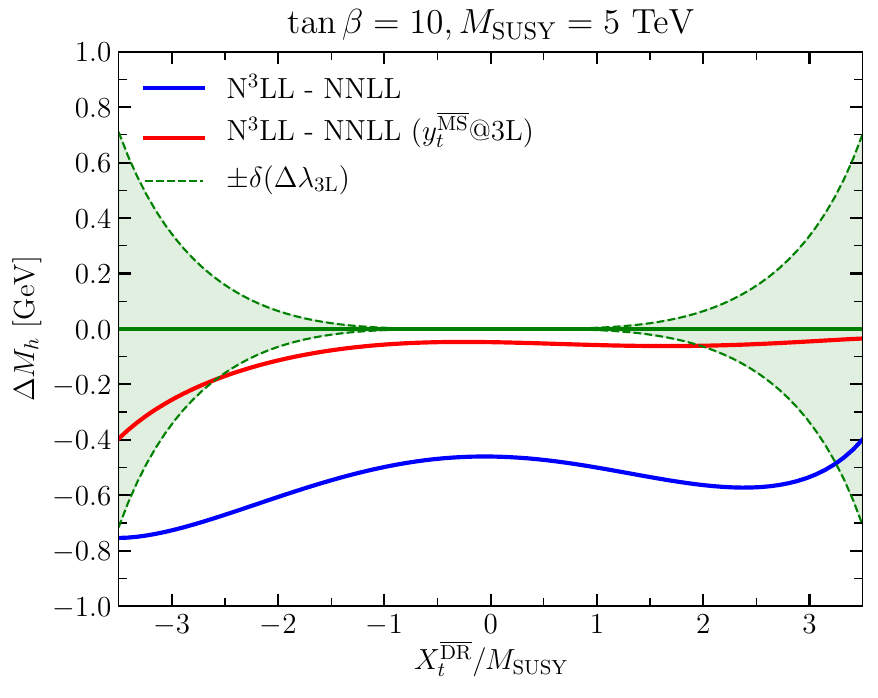}
\end{minipage}
\caption{\textit{Top left}: Prediction for $M_h$ as a function of \msusy for ${\xt=0}$ (solid lines) and ${\xt=-\sqrt{6}}$ (dashed lines). The results using NNLL resummation (blue), NNLL resummation with the SM top Yukawa coupling extracted at the three-loop level (red) and N$^3$LL resummation (green) are compared. \textit{Top right}: Differences of the $M_h$ predictions using N$^3$LL and NNLL resummation (blue line) as well as using N$^3$LL and NNLL resummation with the SM top Yukawa coupling extracted at the three-loop level (red line) as a function of \msusy for ${\xt=-\sqrt{6}}$. In addition, the estimate for the uncertainty associated with the truncation error of the \order{\alt\als^2} Higgs self-coupling threshold correction is shown (green band). \textit{Bottom left}: Same as top left, but $M_h$ is shown as a function of \xt for $\msusy=5\tev$. \textit{Bottom right}: Same as top right, but $\Delta M_h$ is shown as a function of \xt for $\msusy=5\tev$.}
\label{fig:NNNLL}
\end{figure}

In \Fig{fig:NNNLL}, we compare the results obtained using three different accuracy levels to each other: NNLL resummation with the SM top Yukawa coupling extracted at the two-loop level, NNLL resummation with the SM top Yukawa coupling extracted at the three-loop level and N$^3$LL resummation, which also involves the SM top-Yukawa coupling extracted at the three-loop level. The plots in the right part of the Figure display the difference $\Delta M_h$ between the curves in the left panel (see legends of the corresponding plots). In the upper plots, the different results are shown as a function of \msusy. In the upper left plot, the three results (blue, red and green lines) are shown for vanishing stop mixing (solid lines) and for $\xt = -\sqrt{6}$ (dashed lines). For vanishing stop mixing, all three results are in good agreement with each other for low \msusy. If \msusy is raised, there is, however, an increasing difference between the NNLL result (with the two-loop level SM top Yukawa coupling) and the two results involving the three-loop level SM top-Yukawa coupling of up to $\sim 1\gev$ for $\msusy\sim 100\tev$. This shift is almost completely caused by including the three-loop corrections to the extraction of the SM top Yukawa coupling, since the NNLL result with the SM top Yukawa coupling extracted at the three-loop level and the N$^3$LL result are in very good agreement also for $\msusy\sim 100\tev$. Also for $\xt=-\sqrt{6}$, the NNLL result with the SM top Yukawa coupling extracted at the three-loop level and the N$^3$LL result are in good agreement across the considered \msusy range (within $\sim 0.3\gev$). This difference, is displayed by the red curve in the upper right plot of \Fig{fig:NNNLL}. The NNLL result with the SM top Yukawa coupling extracted at the two-loop level deviates from the other two results by $\sim 0.7\gev$, as shown by the blue curve in the top right plot of \Fig{fig:NNNLL}. In this plot, furthermore the estimate of the uncertainty associated with the truncation error in the calculation of the \order{\alt\als^2} threshold correction for the Higgs self-coupling, obtained by including only partially known terms of higher-order in the hierarchy expansion (see~\cite{Harlander:2018yhj} for more details), is shown as a green band. We find that this estimate is of the same size as the shift induced by including the \order{\alt\als^2} threshold correction.

In the lower plots of \Fig{fig:NNNLL}, the same quantities as in the upper plots are shown, but \msusy is set to 5~TeV and \xt is varied. The shifts between the various results are only mildly dependent on \xt (varying \xt leads to shifts of up to 0.4~GeV). This dependence would be stronger for lower \msusy values. The estimate of uncertainty associated with the truncation error, however, shows a strong dependence on \xt. Whereas it is negligible for $-1 \lesssim \xt \lesssim 1$, it increases to up to $0.7\gev$ for $|\xt|\sim 3.5$. As shown by the red curve in the lower right plot of \Fig{fig:NNNLL}, the difference between the NNLL result with the SM top-Yukawa coupling extracted at the three-loop level and the N$^3$LL result is rather small except for large negative values of $\xt$. Where this difference exceeds the level of $0.2\gev$, it is smaller than the estimated uncertainty of the truncation error.

As expected, the results for the N$^3$LL resummation are in very good agreement with the results of~\cite{Harlander:2018yhj}. We observe that the main part of the shift induced by including N$^3$LL resummation is caused by taking into account the three-loop corrections to the extraction of the SM \MS top Yukawa coupling from the measured top mass. The shift caused by including the \order{\alt\als^2} threshold correction for the Higgs self-coupling is smaller and also associated with a rather large uncertainty for large $|\xt|$ values. For small $|\xt|$ values, the shift induced by including the \order{\alt\als^2} threshold correction for the Higgs self-coupling is found to be very small. Therefore, we choose in our implementation to use the result obtained using NNLL resummation with the SM top Yukawa coupling extracted at the three-loop level as default result until the uncertainty in the calculation of the \order{\alt\als^2} threshold correction is further reduced by incorporating additional higher-order contributions.


\section{Conclusions}
\label{sec:06_conclusions}

In this paper, we have presented an improved prediction for the lightest Higgs boson mass in the MSSM in scenarios with large $\tan \beta$, complex input parameters and large $M_{\rm SUSY}$. Our calculation builds on results that are contained in the publicly available code \texttt{FeynHiggs}.

The first improvement concerning scenarios with large $\tan \beta$ includes the change of the renormali{\sz}ation scheme for the bottom mass with respect to the present implementation in \texttt{FeynHiggs}: instead of treating the bottom mass as a derived quantity, in the scheme used in our calculation it is as an independent parameter, renormali{\sz}ed in the \DR scheme in the full MSSM at scale \msusy. The scheme that we have adopted yields numerically more stable results and turned out to be better suited for the combination with the EFT calculation. In the calculation of the \DR bottom mass, we have taken into account higher-order corrections enhanced by $\tan\beta$ by means of a resummation of the quantity $\Delta_b$. We have incorporated full one-loop corrections to $\Delta_b$. Moreover, we have adapted the leading two-loop QCD corrections to $\Delta_b$ obtained in~\cite{Noth:2008tw, Noth:2008ths, Noth:2010jy} such that they are suitable for the framework of our calculation. The inclusion of this correction is formally a three-loop effect. While this correction is numerically not relevant for large parts of the parameter space, it can become sizeable for scenarios with large $\tan \beta$.

Moreover, we have included one- and two-loop threshold corrections to the SM Higgs self-coupling proportional to the bottom Yukawa coupling well as the corresponding RGE contributions up to the three-loop level. This allows resummation up to the next-to-next-to-leading-order. In contrast to the resummation of the logarithms proportional to the top Yukawa coupling or electroweak couplings, here the one- and two-loop leading logarithms are numerically negligible due to the smallness of the bottom mass. However, at the two-loop level for the case where the stop sector is renormali{\sz}ed in the OS scheme the next-to-leading logarithms become parametrically enhanced for large $\tan \beta$. In this case, the resummation can become numerically relevant for large $\tan\beta$ and large $\msusy$.

Secondly, we used the two-loop fixed order results presented in Refs.~\cite{Passehr:2017ufr,Borowka:2018anu} to derive two-loop threshold corrections to the SM Higgs self-coupling for the matching between the SM and the MSSM that are valid for the general case of complex input parameters. This enabled us to perform the EFT calculations for the case of complex parameters. We compared the results including the full phase dependence to the results obtained by the use of the interpolation routine that has been adopted in \texttt{FeynHiggs} up to now. For the pure EFT calculation, we have found the interpolation procedure to perform well in scenarios with only one non-zero phase. In scenarios with more than one non-zero phase, we observed deviations in the prediction for $M_h$ of up to $1~{\rm GeV}$. For the hybrid result the incorporation of the full phase dependence of the EFT part of the calculation yields another important improvement. Up to now the corresponding contributions in the fixed-order result (containing the full phase dependence) and the subtraction terms (based on the interpolated EFT contributions) were treated differently, which could lead to an incomplete cancellation between the two types of contributions. This can lead to numerical deviations of up to $2~{\rm GeV}$ compared to our improved result where the treatment of the phase dependence is the same in all parts of the calculation. We furthermore analysed the interplay between the resummation of the logarithms proportional to the bottom Yukawa coupling and the inclusion of the full phase dependence into the EFT part of the hybrid calculation. We have found that the impact of phase variations on the prediction for $M_h$ can be modified very significantly through the contributions of the $b/\tilde{b}$ sector.

Finally, we combined the publicly available code~\texttt{Himalaya} with~\texttt{FeynHiggs} in order to obtain a prediction for $M_h$ including N$^3$LL resummation at leading order in the strong gauge coupling. A similar analysis was performed in~\cite{Harlander:2019dge}, and we find a very good agreement with the results presented in that paper. The overall effect of the N$^3$LL resummation is $\lesssim 1~{\rm GeV}$, and it only weakly depends on \msusy. We have found that employing the extraction of the SM top Yukawa coupling at the three-loop level within the existing NNLL hybrid calculation yields a result that approximates the N$^3$LL resummation well in view of the remaining theoretical uncertainties of the N$^3$LL contribution.

The improvements described in this paper will be implemented into an upcoming version of the public code~\texttt{FeynHiggs}.


\section*{Acknowledegments}
\sloppy{
We thank Thomas Hahn, Sven Heinemeyer, Sebastian Paßehr and Heidi Rzehak for useful discussions, Sebastian Paßehr for providing two-loop results, Emanuele Bagnaschi and Pietro Slavich for providing expressions for cross-checks as well as Alexander Voigt and Jonas Klappert for help regarding \Him. We acknowledge support by the Deutsche Forschungsgemeinschaft (DFG, German Research Foundation) under Germany‘s Excellence Strategy -- EXC 2121 ``Quantum Universe'' – 390833306.
}


\newpage

\appendix


\section{Derivation of two-loop threshold corrections}
\label{app:07_2L_thresholds_derivation}

In this Appendix, we derive the two-loop threshold corrections for the SM Higgs self-coupling for the matching between the SM and the MSSM based on the fixed-order calculations presented in~\cite{Hollik:2014wea,Passehr:2017ufr,Borowka:2018anu}. We fully take into account the dependence on \cp-violating phases.

The threshold corrections to the quartic coupling $\lambda$ can be obtained via the matching of the four-point vertex function involving the SM Higgs boson as external particle. Here, however, we follow a different approach. Since in the SM the running mass of the lightest Higgs boson is related to its quartic coupling via
\begin{equation}
\overline{m}_h^2 = 2 \lambda v^2,
\end{equation}
the threshold corrections to $\lambda$ can be obtained via the threshold correction to the running Higgs mass $\overline{m}_h$.\footnote{This method is not sufficient to obtain the threshold corrections for all quartic couplings if the EFT below \msusy is the Two-Higgs-Doublet-Model~\cite{Bahl:2020jaq}.} Below we outline the method and derive the general formulas for the one- and two-loop threshold corrections to $\lambda$ in the gaugeless limit. Similar methods can be found in \cite{Vega:2015fna,Kwasnitza:2020wli}.

In the limit $M_A \gg  M_t$, the SM-like Higgs pole mass in the MSSM up to the two-loop level is given by\footnote{In general, only the real part of each term in the sum on the right hand side of \Eq{ch6:eq1} should be considered, since the Higgs self-energies have imaginary parts arising from the contributions of the particles which are lighter than the SM-like Higgs. Since in the MSSM the mass of the SM-like Higgs is close to the electroweak scale, in the usually considered scenarios these imaginary parts arise only from SM particles and, therefore, cancel out in the matching procedure.}
\begin{eqnarray}
  \label{ch6:eq1}
  \begin{aligned}
    & (M_h)_{\MSSM}^2 = m_h^2 - \widehat{\Sigma}_{hh}^{\MSSM,(1)}(\mh^2) - \widehat{\Sigma}_{hh}^{\MSSM,(2)}(\mh^2) \\
    & \hspace{3.5cm} + \widehat{\Sigma}_{hh}^{\MSSM,(1)}(\mh^2)~\widehat{\Sigma}_{hh}^{\MSSM,(1)~\prime}(\mh^2),
  \end{aligned}
\end{eqnarray}
where $m_h$ is the MSSM tree-level mass, and the prime indicates the derivative with respect to the external momentum squared. In the gaugeless limit and the decoupling limit ($M_A \gg  M_t$), $m_h = 0$ can be inserted. All parameters entering the self-energies~$\widehat{\Sigma}_{hh}^{\MSSM,(1)}$~and~$\widehat{\Sigma}_{hh}^{\MSSM,(2)}$~are renormali{\sz}ed in the \DR scheme while the tadpoles are renormali{\sz}ed to zero. The self-energies entering \Eq{ch6:eq1} are assumed to be expanded in the limit $v/\msusy \to 0$.

Below the matching scale $Q$, the effective field theory is the SM. We write the matching condition for the SM running Higgs mass $\overline{m}_h^2$ as a loop expansion,
\begin{equation}
\label{ch6:eq2}
\overline{m}_h^2 = \overline{m}_{h,{\rm tree}}^2 + (\Delta\overline{m}_{h}^{1l})^2 + (\Delta\overline{m}_{h}^{2l})^2 + \ldots,
\end{equation}
where the ellipsis denotes three-loop terms and higher. Since $m_h$ in \Eq{ch6:eq1} equals zero in the considered approximation, we have $\overline{m}_{h,{\rm tree}} = 0$. The pole mass in the SM can then be obtained via the solution of the pole equation
\begin{equation}
\label{ch6:eq3}
M_h^2 = \overline{m}_h^2 - \widetilde{\Sigma}_{hh}^{\MS,\SM}(M_h^2),
\end{equation}
where $\widetilde{\Sigma}_{hh}^{\MS,\SM}$ is the SM Higgs boson self-energy renormali{\sz}ed in the \MS scheme with the tadpoles renormali{\sz}ed to zero. Since the SM is treated as an effective field theory, its parameters are related to the corresponding parameters in the MSSM. This relation can be schematically written as follows,
\begin{equation}
\label{ch6:eq4}
P^{\SM} = P^{\MSSM} + \Delta P,
\end{equation}
where $P$ is a coupling constant, a running quark mass, or the vacuum expectation value. Inserting this relation into the self-energy $\widetilde{\Sigma}_{hh}^{\MS,\SM}(M_h^2)$ induces a shift at one order higher in the loop expansion,
\begin{equation}
\label{ch6:eq5}
\widetilde{\Sigma}_{hh}^{\MS,\SM} = \left.\widetilde{\Sigma}_{hh}^{\SM} \right|_{P^{\SM} \to P^{\MSSM}} + \widehat{\Sigma}_{hh}^{\SM,{\rm shifts}}.
\end{equation}
The first term on the right-hand side of this equation represents the self-energy which has the same analytic form as $\widetilde{\Sigma}_{hh}^{\MS,\SM}$ but with all \MS SM coupling constants and masses being replaced with their \DR MSSM counterparts. Therefore, the two self-energies are equal at the one-loop level (since in the present discussion we neglect electroweak corrections the different regularisation does not lead to a different result). The difference between the two self-energies is encoded in the quantity $\widehat{\Sigma}_{hh}^{\SM,{\rm shifts}}$ which is of two-loop order and higher.

The renormali{\sz}ed self-energy of the SM-like Higgs boson in the full MSSM can be split into parts,
\begin{equation}
\label{ch6:eq7}
\widehat{\Sigma}_{hh}^{\MSSM} = \widehat{\Sigma}_{hh}^{\SM} + \widehat{\Sigma}_{hh}^{\rm n/SM},
\end{equation}
where the SM part contains contributions from the diagrams with only SM particles and the non-SM part (indicated as ``n/SM'') originates from the diagrams with at least one non-SM particle.

At the one-loop level, the following identity holds,
\begin{equation}
\label{ch6:eq8}
\widehat{\Sigma}_{hh}^{\SM,(1)} = \left.\widetilde{\Sigma}_{hh}^{\SM,(1)} \right|_{P^{\SM} \to P^{\MSSM}}.
\end{equation}
where in this equation the symbols ``$\;\widehat{\;}\;$'' and ``$\;\widetilde{\;}\;$'' are used to denote the SM part of the MSSM Higgs self-energy renormali{\sz}ed in the \DR scheme and the SM Higgs self-energy renormali{\sz}ed in the \MS scheme, respectively. This equation means that the SM contributions in the full MSSM self-energy computed in the \DR scheme, $\widehat{\Sigma}_{hh}^{\SM}$, have the same analytic form as the self-energy computed in the SM in the \MS scheme, $\widetilde{\Sigma}_{hh}^{\SM}$. In \Eq{ch6:eq8} the replacement rule on the right-hand side, $P^{\SM} \to P^{\MSSM}$, implies that all \MS SM parameters in the SM self-energy have to be replaced by their \DR MSSM counterparts without additional shifts.\footnote{We assume that the self-energies on the left- and the right-hand sides of \Eq{ch6:eq8} are expressed in terms of quark masses and vacuum expectation values of the SM-like Higgs. In this parametri{\sz}ation, the SM self-energy in the full MSSM does not depend on non-SM parameters.} At the two-loop level, a relation analogous to \Eq{ch6:eq8} holds for the SM-type corrections to the Higgs mass proportional to the Yukawa couplings, i.e.\ the corrections of $\order{\alt^2,\alt \alb,\alb^2}$ without including any parameter shifts in the one-loop self-energy. For the mixed Yukawa-QCD corrections of $\order{\alt \als,\alb \als}$, the two different choices of the regulari{\sz}ation scheme (dimensional regulari{\sz}ation in the case of the SM and dimensional reduction in the case of the MSSM) lead to different expressions for the SM part of the self-energy. This can already be anticipated since the running \MS and \DR quark masses are not equal to each other at the one-loop level (see e.g.~\cite{Carena:2000dp}). By using \texttt{TwoCalc}~\cite{Weiglein:1992hd,Weiglein:1993hd} and the scripts described in~\cite{Hahn:2015gaa} we explicitly checked the following relation,\footnote{\Eq{ch6:eq9} holds only at zero external momenta. At non-zero momenta an additional term, not related to the top-quark mass, appears (see~\cite{Bagnaschi:2019esc} for details).}
\begin{equation}
\label{ch6:eq9}
\widetilde{\Sigma}_{hh}^{\MS,\SM, \order{\alt \als}} + \frac{\partial}{\partial m_t}\widetilde{\Sigma}_{hh}^{\MS,\SM,(1)} \cdot \Delta m_t^{\MS \to \DR} = \widetilde{\Sigma}_{hh}^{\DR,\SM, \order{\alt \als}},
\end{equation}
and the analogous relation for the $\order{\alb \als}$ self-energies. In \Eq{ch6:eq9},\footnote{Here we do not specify the renormali{\sz}ation scheme for the strong coupling and for the top mass since a change of the renormali{\sz}ation scheme is of three-loop order.}
\begin{equation}
\label{ch6:eq10}
\Delta m_t^{\MS \to \DR} = \frac{\als}{3\pi}m_t.
\end{equation}
As explained above, the one-loop reparametri{\sz}ation of the couplings and masses in the one-loop SM self-energy induces shifts at the two-loop order,
\begin{equation}
\label{ch6:eq11}
\widehat{\Sigma}_{hh}^{\SM,{\rm shifts}} = \sum_{P} \frac{\partial}{\partial P} \widetilde{\Sigma}_{hh}^{\MS,\SM,(1)} \Delta^{(1)} P,
\end{equation}
where $P$ are all SM parameters which enter $\widetilde{\Sigma}_{hh}^{\MS,\SM,(1)}$. The one-loop expression for $\widetilde{\Sigma}_{hh}^{\MS,\SM,(1)}$ at zero external momentum of order $\order{\alt,\alb}$ reads
\begin{equation}
\label{ch6:eq12}
\widetilde{\Sigma}_{hh}^{\MS,\SM,(1)} = \sum_{q = t,b} \frac{3 \left(m_q^{\MS,\SM}(Q)\right)^4}{4 \pi^2 \left(\vMS(Q)\right)^2} \log \frac{\left(m_q^{\MS,\SM}(Q)\right)^2}{Q^2}.
\end{equation}
In this expression all masses and the vacuum expectation value (i.e., $m_q^{\MS,\SM}(Q)$ and $\vMS(Q)$) are SM \MS parameters evaluated at the scale $Q$. They are related to the MSSM parameters in the \DR scheme at the scale $Q$ in the following way
\begin{eqnarray}
\label{ch6:eq15}
  \begin{aligned}
    & \vMS(Q) = \vMSSM(Q) \left(1 - \Delta^{(1)} v\right), \\
    & m_{q}^{\MS,\SM}(Q) = m_{q}^{\DR, \MSSM}(Q) - \Delta^{(1)} m_q, \quad q = {t,b},
  \end{aligned}
\end{eqnarray}
where the one-loop shift $\Delta^{(1)} m_q$ contains contributions of BSM particles as well as the transition between the \DR and \MS schemes. This quantity can be computed from the pole mass matching of the bottom and top masses at the one-loop level. The one-loop shift $\Delta^{(1)} v$ includes non-SM $\order{\alt,\alb}$ terms\footnote{Note that the one-loop shift $\Delta^{(1)} v$ in this equation is the same quantity as $\Delta v$ in \Eq{eq:hb_matching3} in \Sec{sec:02_botresum_deltab} for $Q=\msusy$.} while $\Delta^{(1)} m_q$ includes $\order{\alt,\alb,\als}$ corrections. With these definitions and \Eq{ch6:eq8}, the two-loop terms which account for the shifts between the SM and the MSSM quantities acquire the following form,\footnote{For brevity, we will omit arguments of the self-energies in the rest of this section, implying that they always equal zero.}
\begin{equation}
\label{ch6:eq16}
\widehat{\Sigma}_{hh}^{\SM,{\rm shifts}} = -\sum_{q = t,b} \frac{\partial}{\partial m_q} \widehat{\Sigma}_{hh}^{\SM,(1)} \cdot \Delta^{(1)} m_q + 2 \widehat{\Sigma}_{hh}^{\SM,(1)} \; \Delta^{(1)} v.
\end{equation}
Here we have exploited the fact that in the gaugeless limit $\widehat{\Sigma}_{hh}^{\SM,(1)}$ scales as $\propto 1/v^2$. In~\cite{Bahl:2017aev} it was shown that in the heavy SUSY limit the non-SM part of the relative shift in the vacuum expectation value can be expressed via the non-SM part of the Higgs self-energy derivative,
\begin{equation}
\label{ch6:eq17}
\Delta^{(1)} v = -\frac{\widehat{\Sigma}_{hh}^{{\rm n/SM},(1) \prime}(\mh^2)}{2}.
\end{equation}
Using this relation, \Eq{ch6:eq16} can be rewritten as follows,
\begin{equation}
\label{ch6:eq18}
\widehat{\Sigma}_{hh}^{\SM,{\rm shifts}} = -\sum_{q = t,b} \frac{\partial}{\partial m_q} \widehat{\Sigma}_{hh}^{\SM,(1)} \cdot \Delta^{(1)} m_q - \widehat{\Sigma}_{hh}^{\SM,(1)} \widehat{\Sigma}_{hh}^{{\rm n/SM},(1)~\prime}.
\end{equation}
Taking into account \Eq{ch6:eq5}, \Eq{ch6:eq8} and \Eq{ch6:eq18}, the pole equation \Eq{ch6:eq3} can be solved iteratively up to the two-loop level in the considered approximation,
\begin{eqnarray}
  \label{ch6:eq19}
  \begin{aligned}
    (M_h)_{\SM}^2 ={}& (\Delta\overline{m}_{h}^{1l})^2 + (\Delta\overline{m}_{h}^{2l})^2 - \widehat{\Sigma}_{hh}^{\SM,(1)} - \widetilde{\Sigma}_{hh}^{\MS,\SM,(2)} \\
    & -\widehat{\Sigma}_{hh}^{\SM,(1) \prime} \left((\Delta\overline{m}_{h}^{1l})^2 - \widehat{\Sigma}_{hh}^{\SM,(1)} \right)
    + \sum_{q = t,b} \frac{\partial}{\partial m_q} \widehat{\Sigma}_{hh}^{\SM,(1)} \cdot \Delta^{(1)} m_q \\
    & +\widehat{\Sigma}_{hh}^{\SM,(1)} \widehat{\Sigma}_{hh}^{{\rm n/SM},(1) \prime}.
  \end{aligned}
\end{eqnarray}
At the matching scale $Q$, the predictions for the physical Higgs mass, $M_h$, in the SM and the MSSM have to be equal order by order,
\begin{equation}
\label{ch6:eq20}
(M_h)_{\SM}^2 = (M_h)_{\rm MSSM}^2.
\end{equation}
By equating the one-loop pieces in \Eq{ch6:eq20} and taking into account \Eq{ch6:eq1} and \Eq{ch6:eq19} we get,
\begin{equation}
\label{ch6:eq21}
(\Delta\overline{m}_{h}^{1l})^2 = -\widehat{\Sigma}_{hh}^{\MSSM,(1)} + \widehat{\Sigma}_{hh}^{\SM,(1)} = - \widehat{\Sigma}_{hh}^{{\rm n/SM},(1)}.
\end{equation}
After inserting this one-loop solution back into \Eq{ch6:eq19}, we arrive at
\begin{eqnarray}
  \label{ch6:eq22}
  \begin{aligned}
    (M_h)_{\SM}^2 ={}&  (\Delta\overline{m}_{h}^{2l})^2 - \widehat{\Sigma}_{hh}^{\MSSM,(1)} - \widetilde{\Sigma}_{hh}^{\MS,\SM,(2)} \\
    & + \widehat{\Sigma}_{hh}^{\SM,(1) \prime} \widehat{\Sigma}_{hh}^{\MSSM,(1)}
    + \sum_{q = t,b} \frac{\partial}{\partial m_q} \widehat{\Sigma}_{hh}^{\SM,(1)} \cdot \Delta^{(1)} m_q \\
    & +\widehat{\Sigma}_{hh}^{\SM,(1)} \widehat{\Sigma}_{hh}^{{\rm n/SM},(1) \prime}.
  \end{aligned}
\end{eqnarray}
By equating the two-loop pieces in \Eq{ch6:eq20} and expanding the one-loop self-energies of the Higgs boson in the full MSSM according to \Eq{ch6:eq7}, we get
\begin{eqnarray}
  \label{ch6:eq23}
  \begin{aligned}
    (\Delta\overline{m}_{h}^{2l})^2 ={}& - \widehat{\Sigma}_{hh}^{\MSSM,(2)} + \widetilde{\Sigma}_{hh}^{\MS,\SM,(2)} \\
    & - \sum_{q = t,b} \frac{\partial}{\partial m_q} \widehat{\Sigma}_{hh}^{\SM,(1)} \cdot \Delta^{(1)} m_q
     +\widehat{\Sigma}_{hh}^{{\rm n/SM},(1)} \widehat{\Sigma}_{hh}^{{\rm n/SM},(1) \prime}.
  \end{aligned}
\end{eqnarray}
The running Higgs-boson mass $\overline{m}_h^2$ can be related to the threshold corrections to the quartic coupling $\lambda$ via the relation
\begin{equation}
\label{ch6:eq24}
\overline{m}_h^2 = 2 \Delta \lambda(Q) (\vMS)^2(Q).
\end{equation}
To express the one- and two-loop corrections in terms of the MSSM coupling constants we have to perform the shift of the vacuum expectation value in \Eq{ch6:eq24},
\begin{equation}
\label{ch6:eq25}
\overline{m}_h^2 = 2 \Delta \lambda(Q) (\vMSSM)^2(Q) \left(1 + \widehat{\Sigma}_{hh}^{{\rm n/SM},(1)'}\right).
\end{equation}
By solving this equation at the one- and two-loop levels, we obtain the expressions for the matching coefficients for the quartic coupling,
\begin{subequations}
  \begin{align}
    \label{ch6:eq26a}
    & \Delta \lambda^{1l} = -\frac{\widehat{\Sigma}_{hh}^{{\rm n/SM},(1)}}{2 (\vMSSM)^2(Q)}, \\
    \label{ch6:eq26b}
    & \Delta \lambda^{2l} = -\frac{1}{2 (\vMSSM)^2(Q)} \Bigl(\widehat{\Sigma}_{hh}^{\MSSM,(2)} - \widetilde{\Sigma}_{hh}^{\MS,\SM,(2)} - 2\;\widehat{\Sigma}_{hh}^{{\rm n/SM},(1)}\widehat{\Sigma}_{hh}^{{\rm n/SM},(1)\prime} \notag \\
    & \hspace{5.1cm}  + \sum_{q = t,b} \frac{\partial}{\partial m_q} \widehat{\Sigma}_{hh}^{\SM,(1)} \cdot \Delta^{(1)} m_q \Bigr).
  \end{align}
\end{subequations}
As already mentioned, in the expressions above all couplings are \DR MSSM couplings at the scale $Q$. Another option is to parametri{\sz}e these threshold corrections in terms of the \MS SM couplings at $Q$. In this work, we will use the \MS top-Yukawa coupling in the SM and the \DR MSSM bottom-Yukawa coupling to parametri{\sz}e the one- and two-loop threshold corrections. To express the two-loop threshold corrections in terms of the SM \MS top-Yukawa coupling we have to reparametri{\sz}e the top mass and the vacuum expectation value in the one-loop $\order{\alt}$ threshold correction. This generates the following two-loop terms,
\begin{align}
  \Delta \lambda \bigg|_{h_t^{\MSSM} \to y_t^{\SM}} = -\frac{1}{2(\vMS)^2} \Bigl(&\frac{\partial}{\partial m_t} \widehat{\Sigma}_{hh}^{{\rm n/SM},\order{\alt}} \cdot \Delta^{(1)} m_t \\ \nonumber
  & + 2\; \widehat{\Sigma}_{hh}^{{\rm n/SM},\order{\alt}} \; \widehat{\Sigma}_{hh}^{{\rm n/SM},(1) \prime} \Bigr),
\end{align}
which has to be added to \Eq{ch6:eq26b}.

We have evaluated the non-SM part of the one-loop Higgs boson self-energy with the help of \texttt{FeynArts}~\cite{Kublbeck:1990xc,Eck:1992ms,Hahn:2000kx} and \texttt{FormCalc}~\cite{Hahn:1998yk} and then expanded in the limit $\mtL, \mtR, \mbR \gg m_t, m_b$. The explicit expressions for them (and their derivatives) in the gaugeless limit and for the case $\mtL = \mtR = \mbR \gg m_t, m_b$ read
\begin{align}
  & \widehat{\Sigma}_{hh}^{{\rm n/SM},(1)} = -\sum_{q = t,b} \frac{3 m_q^4}{4 \pi^2 v^2} \left(\log \frac{\msusy^2}{Q^2} + \vert \xq \vert^2 - \frac{\vert \xq \vert^4}{12}\right), \\
  & \widehat{\Sigma}_{hh}^{{\rm n/SM},(1) \prime} = \sum_{q = t,b} \frac{m_q^2}{32 \pi^2 v^2} \; \vert \xq \vert^2.
\end{align}
From these expressions and \Eq{ch6:eq26a} it is clear how the one-loop threshold corrections to $\lambda$ computed in \cite{Bagnaschi:2014rsa,Bagnaschi:2017xid} can be generali{\sz}ed to the case of the MSSM with complex parameters. These corrections are polynomials in the squark mixing parameter $\xq$. To obtain the expression in the MSSM with complex parameters \xq has to replaced by $|\xq|$.

The two-loop self-energies were taken from~\cite{Hollik:2014wea,Passehr:2017ufr,Borowka:2018anu} and expanded in the limit
\[
\mtL, \mtR, \mbR, \mA, \vert \mu \vert, \vert M_3 \vert \gg m_t, m_b
\]
without any additional assumptions on the internal masses and the phases of $X_t, X_b, \mu$ and $M_3$. The two-loop SM self-energies in the \MS scheme were taken from~\cite{Degrassi:2012ry,Martin:2014cxa} and extracted from the code \texttt{FlexibleSUSY}~\cite{Athron:2014yba,Athron:2016fuq,Athron:2017fvs}. Finally, the one-loop shifts $\Delta^{(1)} m_t$ and $\Delta^{(1)} m_b$ have been computed using \texttt{FeynArts} and \texttt{FormCalc}. The resulting two-loop formulas for the threshold corrections are presented in \App{app:08_CPV_thresholds_TL}.


\section{Threshold corrections for the case of non-vanishing \texorpdfstring{\cp}{CP}-violating phases}
\label{app:08_CPV_thresholds}

\subsection{One-loop threshold corrections}
\label{app:08_CPV_thresholds_OL}

If the sfermions and heavy Higgses are integrated out from the MSSM, effective Higgs--gaugino--Higgsino couplings, $\tilde{g}_{1u,1d,2u,2d}$, are generated (for their exact definition see e.g.~\cite{Bagnaschi:2014rsa}). In principle, they can be complex. An explicit calculation of their matching conditions at the SUSY scale $\msusy = \sqrt{\mtL \mtR}$, however, shows that they remain real if the sfermions and heavy Higgses are integrated out. All other couplings of the EFT below the SUSY scale are also real-valued.\footnote{The CKM phase is neglected here.}

The only exception are the mass parameters of the EWinos for the case of light EWinos. The phases of these parameters become relevant if the EWinos are integrated out at the EWino mass scale, $M_\chi = \sqrt{\vert M_2 \vert \vert \mu \vert}$, and the SM is recovered as EFT.\footnote{Here, we assume that the absolute values of the EWino mass parameters $M_1,M_2$ and $\mu$ are of the same order of magnitude.}

The threshold corrections of the top and bottom Yukawa couplings originate only from the corrections to the external Higgs leg. They read
\begin{align}
y_t^\SM(M_\chi) ={}& y_t^\text{SM+EWinos}(M_\chi)\Big(1 + \kappa \Delta_{\text{WFR}}\Big),\\
y_b^\SM(M_\chi) ={}& y_b^\text{SM+EWinos}(M_\chi)\Big(1 + \kappa \Delta_{\text{WFR}}\Big),\\
\Delta_{\text{WFR}} ={}& \frac{1}{12} \bigg[2 \giu\gid\cos(\pMi+\pMue)f(r_1) \nonumber\\
&\hspace{1.1cm} + (\giu^2 + \gid^2)\bigg(g(r_1)+3\ln\frac{|\mu|^2}{M_\chi^2}\bigg) \nonumber\\
&\hspace{1.1cm} + 6 \giiu\giid\cos(\pMii+\pMue)f(r_2) \nonumber\\
&\hspace{1.1cm} + 3 (\giiu^2 + \giid^2)\bigg(g(r_2)+3\ln\frac{|\mu|^2}{M_\chi^2}\bigg)
\bigg],
\end{align}
with
\begin{align}
r_1 = \left|\frac{M_1}{\mu}\right|, \hspace{1cm} r_2 = \left|\frac{M_2}{\mu}\right|.
\end{align}
In the expressions above, $\kappa = 1/(4\pi)^2$ is used to indicate the loop order, and $\Delta_{\text{WFR}}$ are the one-loop corrections to the external Higgs legs. Setting the phases to zero, we recover the result presented in Ref.\cite{Bagnaschi:2014rsa}. The loop functions $f$ and $g$ are defined in the Appendix of Ref.\cite{Bagnaschi:2014rsa}.

Similarly, also the matching condition of the Higgs self-coupling is modified,
\begin{align}
\lambda^\SM(M_\chi) = \lambda^\text{SM+EWinos}(M_\chi) + \Delta\lambda
\end{align}
with
\begin{align}
\label{eq:EWino_th}
(4\pi)^2\Delta\lambda ={}& \frac{1}{2}\Big[2\lambda\big(\giu^2+\gid^2+3\giid^2+3\giiu^2\big) -\giu^4-\gid^4-5\giiu^2-5\giid^2 \nonumber\\
&\hspace{.5cm}- 4\giu\gid\giiu\giid - 2\big(\giu^2+\giid^2\big)\big(\gid^2+\giiu^2\big)\Big]\ln\frac{|\mu|^2}{M_\chi^2} \nonumber\\
&-\frac{7}{12}\big(\giu^4+\gid^4\big)f_1(r_1)-\frac{9}{4}f_2(r_2)\big(\giiu^4+\giid^4\big) \nonumber\\
&+\frac{1}{6}\giu^2\gid^2\Big[2\cos(2\pMi+2\pMue\big)h_1(r_1)-11 h_2(r_1)\Big] \nonumber\\
&+\frac{1}{2}\giiu^2\giid^2\Big[2\cos(2\pMii+2\pMue\big)h_1(r_2)-9 h_3(r_2)\Big] \nonumber\\
&+\frac{1}{3}\giu\gid\giiu\giid\Big[\cos(\pMi+\pMii+2\pMue\big)h_4(r_1,r_2)\nonumber\\
&\hspace{3.1cm}-4\cos(\pMi-\pMii)\frac{r_1 r_2}{r_1+r_2}f_8(r_1,r_2)-7 h_5(r_1,r_2)\Big] \nonumber\\
&-\frac{1}{3}\big(\giu^2\giiu^2+\gid^2\giid^2\big)\Big[2\cos(\pMi-\pMii)\frac{r_1 r_2}{r_1+r_2}f_8(r_1,r_2) + \frac{5}{2}h_6(r_1,r_2)\Big]\nonumber\\
&+\frac{1}{6}\big(\giu^2\giid^2+\gid^2\giiu^2\big)\Big[\cos(\pMi+\pMii+2\pMue)h_4(r_1,r_2) - \frac{4}{r_1+r_2}f_8(r_1,r_2))\Big] \nonumber\\
&-\frac{4}{3}\big(\giu\giiu+\gid\giid\big)\big(\giu\giid+\gid\giiu\big)\Big[\frac{r_1}{r_1+r_2}\cos(\pMi+\pMue) \nonumber\\
&\hspace{7.0cm}+\frac{r_2}{r_1+r_2}\cos(\pMii+\pMue)\Big]f_8(r_1,r_2)\nonumber\\
&+\frac{2}{3}\giu\gid\cos(\pMi+\pMue)\Big[\lambda-2\big(\giu^2+\gid^2\big)\Big]f(r_1)\nonumber\\
&+2\giiu\giid\cos(\pMii+\pMue)\Big[\lambda-2\big(\giiu^2+\giid^2\big)\Big]f(r_2)\nonumber\\
&+\frac{1}{3}\lambda\big(\giu^2+\gid^2\big)g(r_1)+\lambda\big(\giiu^2+\giid^2\big)g(r_2).
\end{align}
The loop functions $f_i$ are defined in the Appendix of Ref.\cite{Bagnaschi:2014rsa}. The loop functions $h_i$ are defined by
\begin{subequations}
\begin{align}
h_1(r) &= -\frac{6r^2}{(1-r^2)^3}\big[2-2r^2 + (1+r^2)\ln r^2\big], \\
h_2(r) &= \frac{6}{11(1-r^2)^3}\big[2 + 3 r^2 - 4 r^4 - r^6 + r^2 (4 + 5 r^2 - r^4) \ln r^2\big], \\
h_3(r) &= \frac{2}{9(1-r^2)^3}\big[6 + 7 r^2 - 8 r^4 - 5 r^6 + r^2 (12 + 13 r^2 - r^4) \ln r^2\big], \\
h_4(r_1,r_2) &= -\frac{6 r_1 r_2}{(1-r_1^2)^2 (1-r_2^2)^2 (r_1^2 - r_2^2)}\big[r_1^2 (1-r_2^2)^2 \ln r_1^2 + (1 - r_1^2)(1-r_2^2)(r_1^2-r_2^2) \nonumber\\
&\hspace{5.8cm}-(1 - r_1^2)^2 r_2^2\ln r_2^2\big], \\
h_5(r_1,r_2) &= \frac{6}{7(1-r_1^2)^2 (1-r_2^2)^2 (r_1^2 - r_2^2)}\big[
-r_1^6 (1-r_2^2)^2-r_2^2(1-r_2^4)-r_1^4 r_2^2(1-r_2^4) \nonumber\\
&\hspace{5.7cm}+r_1^2(1+r_2^4-2 r_2^6)
+r_1^4(1+r_1^2)(1-r_2^2)^2\ln r_1^2 \nonumber\\
&\hspace{5.7cm}- (1-r_1^2)^2(1+r_2^2)r_2^4\ln r_2^2\big], \\
h_6(r_1,r_2) &= \frac{6}{5(1-r_1^2)^2 (1-r_2^2)^2 (r_1^2 - r_2^2)}\big[-(1-r_1^2)(1-r_2^2)(r_2^4-r_1^2 r_2^4 - r_1^4 + r_1^4 r_2^2)\nonumber\\
&\hspace{5.7cm}  + (1-r_2^2)^2 r_1^6 \ln r_1^2 - (1-r_1^2)^2 r_2^6 \ln r_2^2\big].
\end{align}
\end{subequations}
In the limit of $r,r_1,r_2\rightarrow 1$ all of the loop functions $h_i$ approach 1. Setting all phases to zero, we again recover the expression given in Ref.~\cite{Bagnaschi:2014rsa}.

The corresponding expressions for the EWino contribution to the matching between the SM and the MSSM can be obtained by replacing the effective Higgs--Higgsino--gaugino couplings $\tilde{g}_{1u,1d,2u,2d}$ in \Eq{eq:EWino_th} using their tree-level matching conditions\cite{Giudice:2004tc,Giudice:2011cg},

\begin{subequations}
\begin{align}
  \label{eq:SplitSUSY_matching}
  & \gid = g' \cos \beta, \\
  & \giid = g \cos \beta, \\
  & \giu = g' \sin \beta, \\
  & \giiu = g \sin \beta.
\end{align}
\end{subequations}
The expressions for $\Delta_b$, $\epsilon_b$ and $\Delta v$ entering the one-loop threshold correction of the bottom Yukawa coupling (see \Eq{eq:hb_matching}) read
\begin{align}
\label{eq:bottom_ths}
(4\pi)^2\Delta_b ={}& C_F g_3^2 t_{\beta} \cos(\pMiii + \pMue) \left \vert \frac{\mu}{M_3} \right \vert \widetilde{F}_9 \left(\frac{M_{Q_3}}{\vert M_3 \vert}, \frac{M_{D_3}}{\vert M_3 \vert}\right)  \nonumber\\
& + \frac{1}{2} y_t^2t_{\beta} \cos(\pAt + \pMue) \left \vert \frac{A_t}{\mu} \right \vert \widetilde{F}_9 \left(\frac{M_{Q_3}}{|\mu|},\frac{M_{U_3}}{|\mu|} \right) \nonumber\\
&  -\frac{3}{4} g^2 t_{\beta} \cos(\pMii + \pMue) \left|\frac{M_2}{\mu}\right|  \widetilde{F}_{9}\left(\frac{M_{Q_3}}{|\mu|}, \left|\frac{M_2}{\mu}\right| \right) \nonumber\\
&- \frac{g'^2}{6} t_{\beta} \cos(\pMi + \pMue)
\Bigg[
\frac{1}{3}\left|\frac{\mu}{M_1}\right| \widetilde{F}_{9} \left(\frac{M_{Q_3}}{|M_1|},\frac{M_{D_3}}{|M_1|} \right) \nonumber \\
& \hspace{4.2cm} + \frac{1}{2}\left|\frac{M_1}{\mu}\right| \widetilde{F}_{9} \left(\frac{M_{Q_3}}{|\mu|}, \left|\frac{M_1}{\mu}\right| \right) \nonumber \\
& \hspace{4.2cm} + \left|\frac{M_1}{\mu}\right| \widetilde{F}_{9} \left(\frac{M_{D_3}}{|\mu|}, \left|\frac{M_1}{\mu}\right| \right) \Bigg],\\
(4\pi)^2\epsilon_b =&{}  - C_F g_3^2 \Bigg[1 + \log \frac{\vert M_3 \vert^2}{Q^2} + \widetilde{F}_6 \left(\frac{M_{Q_3}}{\vert M_3 \vert}\right) + \widetilde{F}_6 \left(\frac{M_{D_3}}{\vert M_3 \vert}\right) \nonumber\\
&\hspace{1.7cm} - \left|\frac{A_b}{M_3} \right|\cos(\pAb - \pMiii) \widetilde{F}_9 \left(\frac{M_{Q_3}}{\vert M_3 \vert}, \frac{M_{D_3}}{\vert M_3 \vert}\right) \Bigg] \nonumber \\
& - \frac{y_b^2}{\cbb} \Bigg[\frac{3}{4} \log \frac{|\mu|^2}{Q^2} + \frac{3}{8} \sbb \Big(2 \log \frac{M_A^2}{Q^2} - 1 \Big)
+ \widetilde{F}_{6} \left(\frac{M_{Q_3}}{|\mu|}\right) + \frac{1}{2} \widetilde{F}_{6} \left(\frac{M_{D_3}}{|\mu|}\right) \Bigg] \nonumber \\
& - \frac{y_t^2}{\sbb} \Bigg[\frac{1}{4} \log \frac{|\mu|^2}{Q^2} + \frac{1}{8} \cbb \Big(2 \log \frac{M_A^2}{Q^2} - 1 \Big)
+ \sbb \Big(\log \frac{M_A^2}{Q^2} - 1 \Big) + \frac{1}{2} \widetilde{F}_{6} \left(\frac{M_{U_3}}{|\mu|}\right) \nonumber \\
&\hspace{1.3cm}+ \frac{1}{2} \widetilde{F}_9 \left(\frac{M_{Q_3}}{|\mu|},\frac{M_{U_3}}{|\mu|} \right) \left(\left|\frac{A_t}{\mu}\right| \cos(\pAt + \pMue) \sbe \cbe - 1 \right)
\Bigg] \nonumber \\
& - g^2 \Bigg[\frac{3}{8} \log \frac{|M_2|^2}{Q^2} - \frac{3}{2} \log \frac{|\mu|^2}{Q^2} + \frac{3}{4} \widetilde{F}_{6} \left(\frac{M_{Q_3}}{|M_2|}\right) - \frac{3}{4} \widetilde{F}_{8} \left(\frac{M_{Q_3}}{|\mu|},\left|\frac{M_2}{\mu}\right|\right)
\Bigg]\nonumber \\
& - g'^2 \Bigg[\frac{5}{72} \log \frac{|M_1|^2}{Q^2}- \frac{1}{2} \log \frac{|\mu|^2}{Q^2} + \frac{1}{36}\widetilde{F}_{6}\left(\frac{M_{Q_3}}{|M_1|}\right) + \frac{1}{9}\widetilde{F}_{6}\left(\frac{M_{D_3}}{|M_1|}\right) \nonumber\\
&\hspace{1.2cm} - \frac{1}{12} \widetilde{F}_{8} \left(\frac{M_{Q_3}}{|\mu|},\left|\frac{M_1}{\mu} \right|\right) - \frac{1}{6} \widetilde{F}_{8}\left(\frac{M_{D_3}}{|\mu|},\left|\frac{M_1}{\mu}\right|\right) \nonumber\\
&\hspace{1.2cm}+ \frac{1}{18}\left|\frac{A_b}{M_1}\right|\cos(\pAb-\pMi)\widetilde{F}_9\left(\frac{M_{Q_3}}{|M_1|},\frac{M_{D_3}}{|M_1|}\right)
\Bigg], \\
(4\pi)^2\Delta v ={}&  -\frac{y_t^2}{4} \frac{\vert X_t \vert^2}{M_{Q_3} M_{U_3}} \widetilde{F}_5 \left(\frac{M_{Q_3}}{M_{U_3}}\right) - \frac{y_b^2}{4} \frac{\vert X_b \vert^2}{M_{Q_3} M_{D_3}} \widetilde{F}_5 \left(\frac{M_{Q_3}}{M_{D_3}} \right). \label{eq:bottom_ths_dv}
\end{align}
The functions $\widetilde{F}_{5,6,8,9}(x)$ are defined in Appendix A of Ref.\cite{Bagnaschi:2014rsa}. $Q$ is the renormali{\sz}ation scale, which we set equal to \msusy. We neglect electroweak contributions to $\Delta v$.

In addition, we give here the result for the one-loop threshold correction for the top Yukawa coupling, which can be used to reexpress the two-loop threshold corrections of the Higgs self-coupling in terms of the MSSM top Yukawa coupling (see \Sec{app:08_CPV_thresholds_TL}). It is given by
\begin{align}
y_t^\SM(Q) ={}&  h_t^\MSSM s_\beta \left(1 + \Delta h_t\right),\\
(4\pi)^2\Delta h_t ={}& \frac{4}{3}g_3^2\left[1 + \ln\frac{|M_3|^2}{Q^2} + \widetilde{F}_6\left(\frac{M_{Q_3}}{|M_3|}\right) + \widetilde{F}_6\left(\frac{M_{U_3}}{|M_3|}\right) \right.\nonumber\\
&\left.\hspace{.9cm} - \left|\frac{X_t}{M_3}\right|\cos(\pMiii-\pXt) \widetilde{F}_9 \left(\frac{M_{Q_3}}{|M_3|},\frac{M_{U_3}}{|M_3|} \right) \right] \nonumber\\
& + \frac{y_t^2}{\sbb} \left[\frac{3}{4}\ln\frac{|\mu|^2}{Q^2} + \frac{3}{8}\cbb\left(2\ln\frac{M_A^2}{Q^2}-1\right) - \frac{1}{4}\sbb|\widetilde{X}_t|\widetilde{F}_5\left(\frac{M_{Q_3}}{M_{U_3}}\right)  \right.\nonumber\\
&\left.\hspace{1.1cm} + \widetilde{F}_6\left(\frac{M_{Q_3}}{|\mu|}\right) + \frac{1}{2}\widetilde{F}_6\left(\frac{M_{U_3}}{|\mu|}\right)\right] \nonumber\\
& + \frac{y_b^2}{\cbb}\left[\frac{1}{4}\ln\frac{|\mu|^2}{Q^2} + \frac{3}{8}\sbb\left(2\ln\frac{M_A^2}{Q^2}-1\right) - \frac{1}{4}\cbb|\widetilde{X}_b|\widetilde{F}_5\left(\frac{M_{Q_3}}{M_{D_3}}\right)  \right.\nonumber\\
&\left.\hspace{1.1cm} + \frac{1}{2}\widetilde{F}_6\left(\frac{M_{D_3}}{|\mu|}\right) + \frac{1}{2}\frac{|X_b|}{|\mu|\tbe}\cos(\pMue+\pXb)\widetilde{F}_9\left(\frac{M_{Q_3}}{|\mu|},\frac{M_{D_3}}{|\mu|}\right)\right] \nonumber\\
& + g^2\left[\frac{3}{8}\ln\frac{|M_2|^2}{Q^2} - \frac{3}{2}\ln\frac{|\mu|^2}{Q^2} + \frac{3}{4}\widetilde{F}_6\left(\frac{M_{Q_3}}{|M_2|}\right) - \frac{3}{4}\widetilde{F}_8\left(\frac{M_{Q_3}}{|\mu|},\frac{|M_2|}{|\mu|}\right) \right.\nonumber\\
&\left.\hspace{1.1cm} - \frac{3}{4\tbe}\frac{|M_2|}{|\mu|}\cos(\pMii + \pMue)\widetilde{F}_9\left(\frac{M_{Q_3}}{|\mu|},\frac{|M_2|}{|\mu|}\right) - \frac{3}{8}\right] \nonumber\\
& + g'^2\left[\frac{17}{12}\ln\frac{|M_1|^2}{Q^2} - \frac{1}{2}\ln\frac{|\mu|^2}{Q^2} + \frac{1}{36}\widetilde{F}_6\left(\frac{M_{Q_3}}{|M_1|}\right) + \frac{4}{9}\widetilde{F}_6\left(\frac{M_{U_3}}{|M_1|}\right) \right.\nonumber\\
&\left.\hspace{1.1cm} + \frac{1}{12}\widetilde{F}_8\left(\frac{M_{Q_3}}{|\mu|},\frac{|M_1|}{|\mu|}\right) - \frac{1}{3}\widetilde{F}_8\left(\frac{M_{U_3}}{|\mu|},\frac{|M_1|}{|\mu|}\right)\right.\nonumber\\
&\left.\hspace{1.1cm}- \frac{1}{9}\frac{|X_t|}{|M_1|}\cos(\pMi-\pXt)\widetilde{F}_9\left(\frac{M_{Q_3}}{|M_1|},\frac{M_{U_3}}{|M_1|}\right) \right.\nonumber\\
&\left.\hspace{1.1cm} + \frac{1}{3\tbe}\frac{|M_1|}{|\mu|}\cos(\pMi + \pMue)\left(\frac{1}{4}\widetilde{F}_9\left(\frac{M_{Q_3}}{|\mu|},\frac{|M_1|}{|\mu|}\right) - \widetilde{F}_9\left(\frac{M_{U_3}}{|\mu|},\frac{|M_1|}{|\mu|}\right)\right)\right.\nonumber\\
&\left.\hspace{1.1cm} - \frac{1}{72}\right]
\end{align}
Setting all phases to zero, we again recover the expression given in Ref.\cite{Bagnaschi:2014rsa}.


\subsection{Two-loop threshold corrections}
\label{app:08_CPV_thresholds_TL}

Here, we list the two-loop threshold corrections to the Higgs self-coupling in the limit where all involved non-SM particles except for EWinos and gluinos have the same mass,
\begin{subequations}
\begin{align}
\label{eq:2Lthatatat}
(4\pi)^4(\Delta\lambda)_{\alt^2} &= - y_t^6 \bigg\{
\frac{3 |\xt|^6}{2} + \frac{1}{\tbb} \Big(\cos(\pXt - \pYt) \big( 12 (3 + 16 K) |\xt| \nonumber\\
& \hspace{0.8cm} - 12 (1 + 4 K) |\xt|^3 \big) |\yt| + 3 (3 + 16 K) |\yt|^2 \Big) \nonumber\\
& \hspace{0.8cm} - |\xt|^2 \Big(2 (7 + 36 K) \frac{|\yt|^2}{\tbb} - \frac{3}{2\sbb} \big(7 + 24 K \nonumber\\
& \hspace{0.8cm} - 3 (5 - 8 K) c_{2\beta} + (32 |\hmu|^2 - 12 |\hmu|^4) f_{2}(|\hmu|) \big) \Big) \nonumber\\
& \hspace{0.8cm} + |\xt|^4 \left(\frac{|\yt|^2}{4\tbb} (19 + 96 K) - \frac{3}{8\sbb} \Big(23 - 25 c_{2\beta} + (16 |\hmu|^2 - 8 |\hmu|^4) \widetilde{f}_{2}(|\hmu|) \Big) \right) \nonumber\\
& \hspace{0.8cm} + \frac{3}{4\sbb} \Big(21 + 120 K + 32 |\hmu|^2 + 2 \pi^2 - (13 - 120 K - 2 \pi^2) c_{2\beta}  \nonumber\\
& \hspace{0.8cm} + \big(-32 + 36 |\hmu|^2 + 8 |\hmu|^4 \big) \widetilde{f}_{2}(|\hmu|) + 16 \widetilde{f}_{3}(|\hmu|) \Big)
\bigg\},\\
\label{eq:2Lthasatat}
(4\pi)^4(\Delta\lambda)_{\alt\als} & = g_3^2 y_t^4\bigg\{
\frac{4}{3} \Big(-12 |\xt|^2 \big(\widetilde{f}_{1}(|\hmg|) - (1 - 7 |\hmg|^2 + 2 |\hmg|^4) \widetilde{f}_{2}(|\hmg|) \big) \nonumber\\
& + |\xt|^4 \big(\widetilde{f}_{1}(|\hmg|) - (1 - 9 |\hmg|^2 + 4 |\hmg|^4) \widetilde{f}_{2}(|\hmg|) \big) \nonumber\\
& - 6 \big(3 + 4 |\hmg|^4 \widetilde{f}_{2}(|\hmg|) - |\hmg|^2 \big(6 \widetilde{f}_{2}(|\hmg|) - 8 \widetilde{f}_{4}(|\hmg|) \big) \big) \Big) \nonumber\\
& + \frac{16}{3} \cos(\pXt - \pMiii) |\hmg| |\xt| \Big( \big(2 (8 - |\hmg|^2) |\xt|^2 - |\xt|^4 \big) \widetilde{f}_{2}(|\hmg|) \nonumber\\
& - 12 \big(1 - \widetilde{f}_{4}(|\hmg|) - |\hmg|^2 \widetilde{f}_{5}(|\hmg|) \big) \Big)
\bigg\}, \\
\label{eq:2Lthasabab}
(4\pi)^4(\Delta\lambda)_{\alb\als} &= -\frac{8}{3}g_3^2 y_b^4\bigg\{
9 + 4 |\xb|^3 |\hmg| \cos(\pXb - \pMiii) (-2 + |\hmg|^2) \widetilde{f}_{2}(|\hmg|) \nonumber \\
& - 12 |\xb|^2 \big(1 + |\hmg|^2 (-2 + |\hmg|^2) \widetilde{f}_{2}(|\hmg|) \big) \nonumber \\
& + |\xb|^4 \big(1 + |\hmg|^2 (-3 + 2 |\hmg|^2) \widetilde{f}_{2}(|\hmg|) \big) \nonumber \\
& + 24 |\xb| |\hmg| \cos(\pXb - \pMiii) \big(1 + \widetilde{f}_{1}(|\hmg|) - 2 \widetilde{f}_{4}(|\hmg|) \big) \nonumber \\
& + 6 |\hmg|^2 \big( (-3 + 2 |\hmg|^2) \widetilde{f}_{2}(|\hmg|) + 4 \widetilde{f}_{4}(|\hmg|) \big) \bigg\},\\
\label{eq:2Lthabatat}
(4\pi)^4(\Delta\lambda)_{\alt\alb} &= y_t^4 y_b^2
\bigg\{
-\frac{3}{2} + 36 K + 3 \pi^2 + \frac{24 K}{\tbb} |\yt|^2 + \frac{1}{\tbb}|\xb|^2 |\yt|^2 + \frac{1}{\tbb}|\xt|^2 |\yt|^2 \nonumber \\
& \hspace{0.8cm} + \frac{2}{9} |\xt| |\yt| |\xb| |\yb| \cos(\pXt + \pXb - \pYt - \pYb) (18 + (1 + 24 K) |\xt|^2) \nonumber \\
& \hspace{0.8cm} - \frac{1}{6 \tbb} (11 + 48 K) |\xb|^2 |\xt|^2 |\yt|^2 + \frac{48 K}{\cbb} \cos(\pXb - \pXt) |\xb| |\xt| \nonumber \\
& \hspace{0.8cm} + \frac{2}{\sbb} |\yt| |\xb| \cos(\pYt - \pXb) (24 K + |\xt|^2) + \frac{24}{\sbb} (3 K - {\Li}(1-|\hmu|^2) + \widetilde{f}_{1}(|\hmu|)) \nonumber \\
& \hspace{0.8cm} + 2 \cos(\pXt - \pYt) |\xt| |\yt| \Bigl(\frac{|\xb|^2}{\tbb} + 6 \Bigl(1 + 4 K \Bigl(-3 + \frac{1}{\sbb} \Bigr) + 2 \widetilde{f}_{1}(|\hmu|) \Bigr)  \nonumber \\
& \hspace{0.8cm} + 2 |\xt|^2 \Bigl(-\frac{1}{3} + 4 K - \widetilde{f}_{1}(|\hmu|) - \widetilde{f}_{2}(|\hmu|) \Bigr) \Bigr) \nonumber \\
& \hspace{0.8cm} + |\xb|^2 |\xt|^4 \Bigl(\frac{1}{2} - 2 \widetilde{f}_{2}(|\hmu|) \Bigr) + \frac{3}{2} |\xb|^2 |\xt|^2 \Bigl(-2 + \frac{1}{\sbb} + 16 \widetilde{f}_{2}(|\hmu|) \Bigr) \nonumber \\
& \hspace{0.8cm} + 2 \cos(\pXb - \pYb) |\xb| |\yb| (24 K + |\xt|^2 (1 + (-12 + |\xt|^2) \widetilde{f}_{2}(|\hmu|) ) - 24 \widetilde{f}_{4}(|\hmu|)) \nonumber \\
& \hspace{0.8cm} + 3 |\xb|^2 (5 + 8 K - \frac{1}{\sbb} (1 + 8 K + 2 |\hmu|^2 \widetilde{f}_{2}(|\hmu|)) + 16 \widetilde{f}_{4}(|\hmu|) ) \nonumber \\
& \hspace{0.8cm} + \frac{4}{3\cbb} \cos(\pXt - \pYb) |\xt| |\yb| (-9 + 2 |\xt|^2 + 12 K (-6 + |\xt|^2) ) \nonumber \\
& \hspace{0.8cm} + \frac{1}{\cbb} \Bigl(36 K + \big(-6 + \pi^2 - |\hmu|^2 (-15 + 6|\hmu|^2 + \pi^2) \big) \widetilde{f}_{2}(|\hmu|) \nonumber \\
& \hspace{0.8cm} - (18 + \pi^2) \widetilde{f}_{4}(|\hmu|) + \big(12 + |\hmu|^2 (-6 + \pi^2) \big) \widetilde{f}_{5}(|\hmu|) \Bigr) \nonumber \\
& \hspace{0.8cm} + 3 |\xt|^2 \Bigl(-2 (5 + 8 K + 4 \widetilde{f}_{1}(|\hmu|)) - \frac{1}{\sbb} (1 + 8 K + 2 |\hmu|^2 \widetilde{f}_{2}(|\hmu|)) \nonumber \\
& \hspace{0.8cm} - \frac{2}{\cbb} \big(-1 + 4 K + |\hmu|^2 \big(1 + \widetilde{f}_{1}(|\hmu|) + \widetilde{f}_{2}(|\hmu|) \big) \big) \Bigr) \nonumber \\
& \hspace{0.8cm} + |\xt|^4 \Big(-\frac{3}{4} \Big(1 + \frac{1}{\cbb} \Big) + \big(1 + \widetilde{f}_{1}(|\hmu|) + \widetilde{f}_{2}(|\hmu|) \big) \Big(4 + \frac{|\hmu|^2}{\cbb} \Big) \Big) \nonumber \\
& \hspace{0.8cm} -3 (1 + 8 K) |\yb|^2 \tbb + 4 (1 + 6 K) |\xt|^2 |\yb|^2 \tbb - \frac{1}{36} (35 + 192 K) |\xt|^4 |\yb|^2 \tbb
\bigg\} \nonumber \\
& +y_t^2 y_b^4
\bigg\{
|\xb|^4 |\xt|^2  - \frac{3}{\tbb}(1 + 8 K) |\yt|^2 + \frac{4}{\tbb} (1 + 6 K) |\xb|^2 |\yt|^2 \nonumber\\
& \hspace{0.8cm} - \frac{1}{36\tbb} (35 + 192 K) |\xb|^4 |\yt|^2 \nonumber \\
& \hspace{0.8cm} + \frac{2}{9} \cos(\pXt + \pXb - \pYt - \pYb) |\xb| |\xt| |\yb| |\yt| (18 + (1 + 24 K) |\xb|^2) \nonumber \\
& \hspace{0.8cm} + \frac{4}{3\sbb} \cos(\pXb - \pYt) |\xb| |\yt| (-9 + 2 |\xb|^2 + 12 K (-6 + |\xb|^2) )  \nonumber \\
& \hspace{0.8cm} + |\xb|^4 \Big(1 - 4 (-2 + |\hmu|^2) \widetilde{f}_{2}(|\hmu|) - \frac{1}{4\sbb} \big(1 - 6 |\hmu|^2 \widetilde{f}_{2}(|\hmu|) + 4 |\hmu|^4 \widetilde{f}_{2}(|\hmu|) \big) \Big) \nonumber \\
& \hspace{0.8cm} + 2 \cos(\pXt - \pYt) |\xt| |\yt| \big(24 K + |\xb|^2 - 24 \widetilde{f}_{4}(|\hmu|) \big) \nonumber \\
& \hspace{0.8cm} + \frac{48 K}{\cbb} \cos(\pXt - \pXb) |\xb| |\xt| + \frac{2}{\cbb} \cos(\pXt - \pYb) (24 K + |\xb|^2) |\xt| |\yb| \nonumber \\
& \hspace{0.8cm} + \frac{3}{2} |\xb|^2 |\xt|^2 \Big(-6 + \frac{1}{\cbb} \Big) - \frac{1}{6} (11 + 48 K) |\xb|^2 |\xt|^2 |\yb|^2 \tbb \nonumber \\
& \hspace{0.8cm} + \frac{1}{2} \Bigl(-3 + 72 K + 6 \pi^2 + \frac{2}{\sbb} \Big(36 K + \pi^2 - 3 \Big(-6 + \big(8 - 11 |\hmu|^2 + 2 |\hmu|^4 \big) \widetilde{f}_{2}(|\hmu|) \nonumber \\
& \hspace{0.8cm} + (-4 + 8 |\hmu|^2) \widetilde{f}_{4}(|\hmu|) \Big) \Big) + \frac{8}{\cbb} \big(-1 + 3 K - \Li(1 - |\hmu|^2) + (1 - |\hmu|^2) \widetilde{f}_{2}(|\hmu|) \big) \Bigr) \nonumber \\
& \hspace{0.8cm} + 3 |\xt|^2 \Big(5 + 8 K + 16 \widetilde{f}_{3}(|\hmu|) - \frac{1}{\cbb} \big(1 + 8 K + 2 |\hmu|^2 \widetilde{f}_{2}(|\hmu|) \big) \Big) \nonumber \\
& \hspace{0.8cm} + 3 |\xb|^2 \Bigl(9 + 16 K - 8 (-1 + |\hmu|^2) \widetilde{f}_{2}(|\hmu|) + \frac{1}{\cbb} \Big(1 + 2 |\hmu|^4 \widetilde{f}_{2}(|\hmu|) \nonumber \\
& \hspace{0.8cm} + \frac{2}{\sbb} \big(4 K + |\hmu|^2 \widetilde{f}_{2}(|\hmu|) - |\hmu|^4 \widetilde{f}_{2}(|\hmu|) \big) \Big) \Bigr) + (24 K + |\xb|^2 + |\xt|^2) |\yb|^2 \tbb \nonumber \\
& \hspace{0.8cm} + \frac{2}{3} \cos(\pXb - \pYb) |\xb| |\yb| \Bigl(2 |\xb|^2 \big(2 + 12 K + 3 (-2 + |\hmu|^2) \widetilde{f}_{2}(|\hmu|) \big) \nonumber \\
& \hspace{0.8cm} + 18 \Big(1 + 2 \widetilde{f}_{1}(|\hmu|) + 4 K \Big(-3 + \frac{1}{\cbb} \Big) \Big) + 3 |\xt|^2 \tbb \Bigl)
\bigg\}, \\
\label{eq:2Lthababab}
(4\pi)^4(\Delta\lambda)_{\alb^2} &= y_b^6 \bigg\{
12 \cos(\pXb - \pYb) |\xb| |\yb| \tbb (4 K (|\xb|^2-4)+|\xb|^2-3) \nonumber \\
& \hspace{1.1cm} + \frac{1}{4} \Bigl( \frac{3}{\cbb} \Big(48 K (5-2 |\xb|^2)+2 |\hmu|^2 \Big(\widetilde{f}_{2}(|\hmu|) \big( (3-2 |\hmu|^2) |\xb|^4 \nonumber \\
& \hspace{1.1cm} + 4 (3 |\hmu|^2-5) |\xb|^2+4 |\hmu|^2+18 \big)+16 \Big)-32 \widetilde{f}_{2}(|\hmu|) \nonumber \\
& \hspace{1.1cm} + 16 \widetilde{f}_{3}(|\hmu|)+|\xb|^4-8 |\xb|^2+4 \pi^2+8 \Big) \nonumber \\
& \hspace{1.1cm} - |\yb|^2 \tbb (|\xb|^2 - 2)  \big( 96 K (|\xb|^2 - 1) + 19 |\xb|^2 - 18 \big) \nonumber \\
& \hspace{1.1cm} + 2 \big(72 K (2 |\xb|^2-5)-2 |\xb|^2 (|\xb|^2-6)^2-6 \pi^2+39 \big) \Bigr)
\bigg\}.
\end{align}
\end{subequations}
The listed expressions are for the case where the one-loop \order{\alt,\alb} corrections are expressed in terms of the SM \MS top Yukawa coupling, $y_t$, and the MSSM \DR bottom Yukawa coupling, $y_b$. Expressions to translate them to the MSSM \DR top Yukawa coupling and the SM \MS bottom Yukawa coupling, respectively, are provided in \Sec{app:08_CPV_thresholds_OL}.

In the \Eqss{eq:2Lthatatat}{eq:2Lthababab}, $\widetilde{f}_{1,2,3,4,5}(x)$ are non-singular functions of $\displaystyle \hmu = \mu/\msusy$ or $\displaystyle \hmg = M_3/\msusy$. The functions $\widetilde{f}_{1,2,3}(x)$ are the same as the functions $f_{1,2,3}(x)$ from Ref.\cite{Espinosa:2000df},
\begin{subequations}
\begin{align}
  & \widetilde{f}_{1}(x) = \frac{x^2 \log x^2}{1-x^2}, \\
  & \widetilde{f}_{2}(x) = \frac{1}{1-x^2} \left[1 + \frac{x^2}{1-x^2} \log x^2 \right], \\
  & \widetilde{f}_{3}(x) = \frac{(-1+2x^2+2x^4)}{(1-x^2)^2} \left[\log x^2 \log(1-x^2) + \Li(x^2) - \frac{\pi^2}{6} - x^2 \log x^2 \right], \\
  & \widetilde{f}_{4}(x) = \frac{x^2 (\log x^2 + \Li(1-x^2))}{(1-x^2)^2}, \\
  & \widetilde{f}_{5}(x) = \frac{x^2 \log x^2 + \Li(1-x^2)}{(1-x^2)^2}.
\end{align}
\end{subequations}
with $\widetilde{f}_{1}(0) = 0,~\widetilde{f}_{2}(0) = 1,~\widetilde{f}_{3}(0) = \frac{\pi^2}{6},~\widetilde{f}_{4}(0) = 0,~\widetilde{f}_{5}(0) = \frac{\pi^2}{6}$ and $\widetilde{f}_{1}(1) = -1,~\widetilde{f}_{2}(1) = \frac{1}{2},~\widetilde{f}_{3}(1) = -\frac{9}{4},~\widetilde{f}_{4}(1) = -\frac{1}{4},~\widetilde{f}_{5}(1) = \frac{3}{4}$. We use a different notation for them than in Ref.\cite{Espinosa:2000df}, since we have already used the notation $f_{i}(x)$ for the functions in the threshold correction to the quartic coupling above in \Eq{eq:EWino_th}.\footnote{We note that these functions are not independent from each other. For example, using the identities for the Spence function $\Li(x^2)$  one can show that $\widetilde{f}_{3}(x) = (1-2x^2-2x^4) \widetilde{f}_{5}(x)$. However, we decided to stick to the notations of Ref.\cite{Espinosa:2000df}. So, we expressed our result in terms of $\widetilde{f}_{1,2,3}$ and added two more functions for better readability of the results.}

The constant $K$ is
\begin{align}
K = - \frac{1}{\sqrt{3}} \int_{0}^{\pi/6} dx \log(2\cos x) \sim -0.1953256.
\end{align}
Fully general expressions for the two-loop threshold corrections to the SM Higgs self-coupling can be found in ancillary files distributed alongside this paper.


\section{Dependence of \texorpdfstring{$\Delta_b^{2l}$}{2L~Deltab} on \texorpdfstring{\cp}{CP}-violating phases}
\label{app:09_db2L_CPV}

Before deriving the phase dependence of the two-loop correction to $\Delta_b$, we first consider the one-loop correction. The \order{\als,\alt} one-loop $\Delta_b$ correction in the heavy SUSY limit can be obtained by evaluating the diagrams in \Fig{sec6.2:fig1}.\footnote{All diagrams in this Section were produced with \texttt{Axodraw} \cite{Collins:2016aya}.}

\begin{figure}
\centering
\fontsize{12}{12}
\input{plots/db1Ldiag}
\caption{Diagrams contributing to $\Delta_b^{\order{\als,\alt}}$ at the leading order in the chiral basis.}
\label{sec6.2:fig1}
\end{figure}

These diagrams include the incoming and the outgoing $b$ quarks with different chirality. The diagrams which involve quarks with the same chirality are subleading with respect to their powers of $\tan \beta$ and do not contribute to $\Delta_b$ \cite{Hofer:2009xb}. The diagrams are drawn using the chiral basis in which the ``left'' and the ``right'' squarks propagate and the off-diagonal mass term is interpreted as an additional interaction (denoted as $\bigotimes$) which flips the chirality quantum number of the squark. In the limit $\msusy \gg v$ only the diagrams with a single mass insertion contribute. This can be seen in the following way. The left diagram in \Fig{sec6.2:fig1} is proportional to
\begin{equation}
\label{ch6:eq30}
\propto \als \;  m_b \;  \mu \; t_{\beta} \; M_3 \; C_0(0, 0, 0, \mtL^2, \mtR^2, \vert M_3 \vert^2),
\end{equation}
where $C_0$ is a Passarino-Veltman function corresponding to the scalar vertex function with three external legs. If all soft SUSY-breaking masses are equal to \msusy, the expression in \Eq{ch6:eq30} reduces to
\begin{equation}
\label{ch6:eq31}
\propto \als \; m_b \; t_{\beta} \frac{\mu~M_3}{\msusy^2} = \als \; m_b \; t_{\beta} \frac{\mu}{\msusy}.
\end{equation}
The diagram with two mass insertions, which is only possible if the two external quarks have the same chirality, is proportional to
\begin{align}
\label{ch6:eq32}
& \propto \als \;  (m_b \;  \mu \; t_{\beta})^2 \; M_3 \; D_0(0, 0, 0, 0, 0, 0, \mtL^2, \mtR^2, \mtL^2, \vert M_3 \vert^2) = \nonumber\\
& = \als \;  (m_b \;  \mu \; t_{\beta})^2 \; \msusy \; D_0(0, 0, 0, 0, 0, 0, \msusy^2, \msusy^2, \msusy^2, \msusy^2),
\end{align}
where $D_0$ is a Passarino-Veltman function corresponding to the scalar vertex function with four external legs. If all soft SUSY-breaking masses are equal to \msusy this diagrams behaves like
\begin{equation}
\label{ch6:eq33}
\propto \als \; m_b \; t_{\beta} \frac{\mu~M_3}{\msusy^2} \times \frac{m_b~\mu~t_{\beta}}{\msusy^2} = \als \; m_b \; t_{\beta} \frac{\mu}{\msusy} \times \frac{m_b~\mu~t_{\beta}}{\msusy^2}.
\end{equation}
We see that it is suppressed by an additional factor $m_b/\msusy$ compared to the diagram with one insertion and is therefore subleading. This is consistent with the fact that only diagrams with external legs of different chirality contribute to $\Delta_b$. Clearly, diagrams with more insertions will be suppressed by additional factors of $m_b/\msusy$. Following similar arguments, one can show that diagrams similar to the right diagram in \Fig{sec6.2:fig1} with more than one mass insertion are suppressed by powers of $m_t/\msusy$.

The same kind of argument applies to higher-order corrections to $\Delta_b$ \cite{Noth:2008ths} as can be proven by using the Kinoshita-Lee-Nauenberg theorem \cite{Kinoshita:1962ur,Lee:1964is}. Namely, the diagrams contributing to the two-loop quantity $\Delta_b$ of order $\order{\als^2, \als\alt}$ contain only one mass insertion.

The phases of the complex parameters $A_t, \mu$ and $M_3$ enter the diagrams of \Fig{sec6.2:fig1} through the mass insertion and through $(b_L \widetilde{b}_L^* \widetilde{g})$, and $(b_L \widetilde{t}_R^* \widetilde{H}_{2}^+)$ vertices. In particular, the mass insertion in the left diagram in \Fig{sec6.2:fig1} yields the phase factor $\propto e^{+i \pMue}$ while both vertices contain $e^{+i\frac{\pMiii}{2}}$. The overall diagram is then $\propto e^{i (\pMiii + \pMue)}$. The result for the analogous diagram with incoming $b_R$ and outgoing $b_L$ leads to the phase factor $\propto e^{-i (\pMiii + \pMue)}$. The overall phase dependence is then $\propto \cos(\pMiii + \pMue)$ (see \Eq{eq:bottom_ths} in \App{app:08_CPV_thresholds_OL}). The mass insertion in the right diagram in \Fig{sec6.2:fig1} gives the phase factor $\propto e^{+i \pAt}$, while the $(b_L \widetilde{t}_R^* \widetilde{H}_{2}^+)$ and $(\bar{b}_R \widetilde{t}_L \widetilde{H}_{2}^-)$ vertices contain the entries of the chargino mixing matrices $\mathbf{V}_{22}^*$ and $\mathbf{U}_{22}^*$. In the gaugeless limit, they are proportional to the phase factors $\propto e^{+i\frac{\pMue}{2}}$. The overall diagram (together with its complex conjugated) is $\propto \cos(\pAt + \pMue)$ (see \Eq{eq:bottom_ths} in \App{app:08_CPV_thresholds_OL}).

\medskip

\begin{figure}
\centering
\fontsize{12}{12}
\input{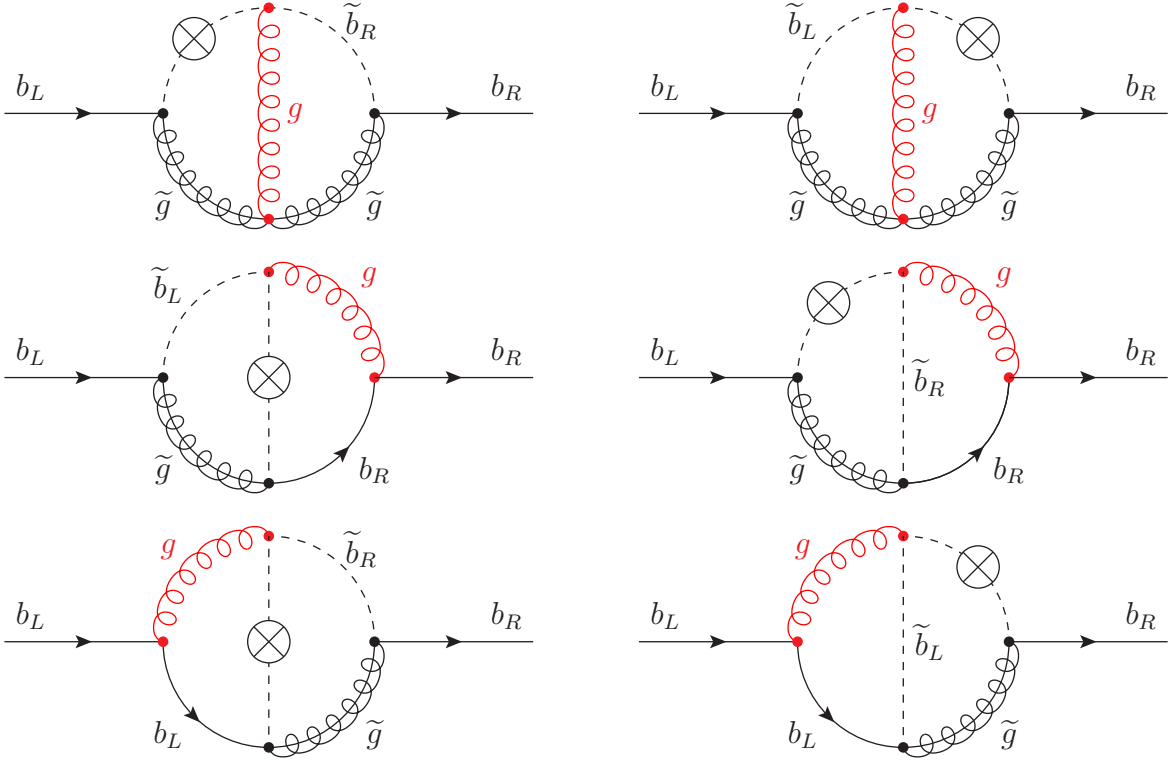}
\caption{The first class of two-loop diagrams contributing to $\Delta_b^{2l,\order{\als^2}}$: a gluon is added to the $\order{\als}$ one-loop graph.}
\label{sec6.2:2Lfig1}
\end{figure}

\begin{figure}
\centering
\fontsize{12}{12}
\input{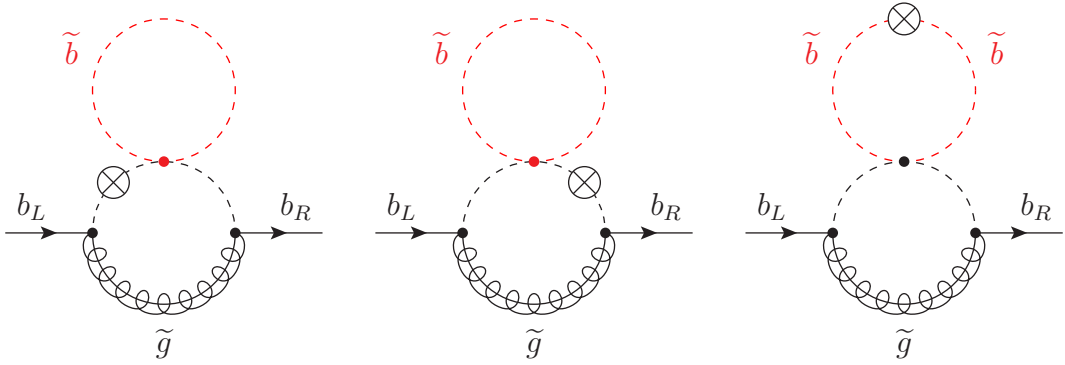}
\caption{The second class of two-loop diagrams contributing to $\Delta_b^{2l,\order{\als^2}}$: a sbottom is added to the $\order{\als}$ one-loop graph.}
\label{sec6.2:2Lfig2}
\end{figure}

\begin{figure}
\centering
\fontsize{12}{12}
\input{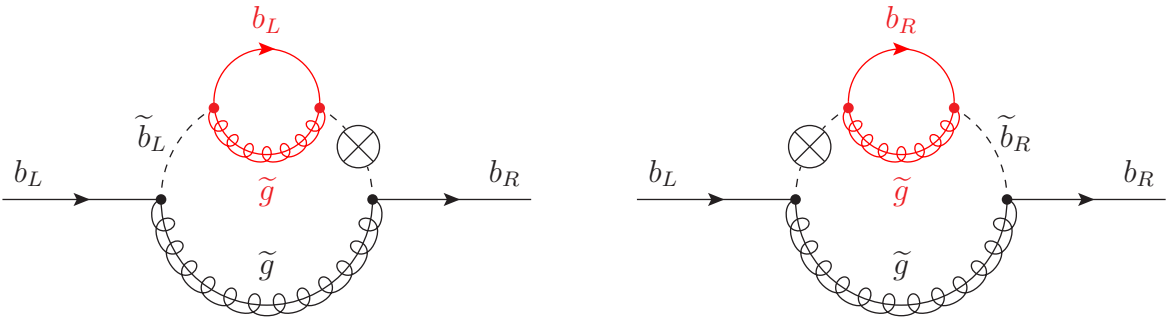}
\caption{The third class of two-loop diagrams contributing to $\Delta_b^{2l,\order{\als^2}}$: a gluino is added to the $\order{\als}$ one-loop graph.}
\label{sec6.3:2Lfig3}
\end{figure}

The two-loop diagrams contributing to the quantity $\Delta_b$ at $\order{\als^2}$ can be split into three categories: either a gluon, a sbottom or a gluino is added to the one-loop $\order{\als}$ graph. Examples of the corresponding diagrams are depicted in Figs.~\ref{sec6.2:2Lfig1}-\ref{sec6.3:2Lfig3}. The particles which are added to the one-loop graph are highlighted with red color.\footnote{The rightmost diagram in \Fig{sec6.2:2Lfig2} cannot be reduced to the left diagram in \Fig{sec6.2:fig1}. This fact, however, does not change our arguments that we present here.}

Following the argumentats given at the end of \Sec{sec:03_cEFT}, we can conclude that in the case of the MSSM with complex parameters the two-loop $\Delta_b$ of $\order{\als^2}$ given in \Eq{eq:2LdbQCD} has to be multiplied by $\cos(\pMue + \pMiii)$. The same reasoning can be applied to the $\order{\als\alt}$ $\Delta_b$ corrections: \Eq{eq:2LdbEW} has to be multiplied by $\cos(\pMue + \pAt)$.\footnote{Corresponding two-loop diagrams can be found in Appendix C of \cite{Noth:2008ths}.}


\section{Leading and next-to-leading logarithms}
\label{app:10_Mh_logs}

In this part of the Appendix we present analytic expressions for the leading (LL) and next-to-leading (NLL) logarithms proportional to the bottom Yukawa coupling, which appear at the one and two-loop order in the calculation of the lightest Higgs boson mass in the MSSM. They are derived in the special case of
\begin{equation}
  M_A = \vert M_3 \vert = \vert \mu \vert = M_{Q_3} = M_{U_3} = M_{D_3} = M_{\rm SUSY}.
\end{equation}
In the following expressions, $\kappa = 1/(4\pi)^2$, $L$ is the logarithm of the ratio of $M_{\rm SUSY}$ and $M_t$,
\begin{equation}
L = \log \displaystyle \frac{M_{\rm SUSY}^2}{M_t^2},
\end{equation}
$\overline{m}_t$ is \MS SM top mass, $\overline{m}_t \equiv \overline{m}_t^{\MS, {\rm SM}}(m_t)$, $\overline{m}_b$ is the \DR MSSM bottom mass at the scale $M_{\rm SUSY}$ defined in \Eq{eq:mbMSSM_def}, $v$ is the SM Higgs vacuum expectation value $v \equiv v_{G_F} = (2\sqrt{2} G_F)^{-1/2} \simeq 174~{\rm GeV}$. The ratio of the vacuum expectation values of the two Higgs doublets, $\tan \beta$, is renormali{\sz}ed in the \DR scheme at the scale $M_{\rm SUSY}$. $g_3$ is the strong gauge coupling, $g_3^2 = 4 \pi \als$. We do not specify the renormali{\sz}ation prescription for it, since it appears in the expressions for $M_h$ starting at the two-loop level. Therefore, a change of the renormali{\sz}ation scheme for $g_3$ is a three-loop effect. The parameter $\widehat{X}_b$ is renormali{\sz}ed in the \DR scheme at the scale $M_{\rm SUSY}$, while $\widehat{X}_t$ is fixed either in the OS scheme or in the \DR scheme at the scale $M_{\rm SUSY}$.

The logarithmic terms can be derived in one of the two following ways. The first method is an approximate perturbative solution of the renormali{\sz}ation group equation (RGE) for the quartic coupling $\lambda$
\begin{equation}
\frac{d\lambda}{d \log Q^2} = \kappa \beta_{\lambda}^{(1)}(Q) + \kappa^2 \beta_{\lambda}^{(2)}(Q) + \ldots,
\end{equation}
where $Q$ is a renormali{\sz}ation scale, $\beta_{\lambda}^{(1)}$, $\beta_{\lambda}^{(2)}$ are the one- and two-loop contributions to the beta function, and the ellipsis encodes higher-order terms. Truncating the result at the second order, one obtains~\cite{Draper:2013oza,Draper:2016pys}
\begin{equation}
  \label{eq:lammt}
  \lambda(M_t) = \lambda(M_{\rm SUSY}) - \beta_{\lambda}^{(1)}(\msusy) \kappa L + \frac{1}{2} \beta_{\lambda}^{(1,1)}(\msusy) \kappa L^2 - \beta_{\lambda}^{(2)}(\msusy) \kappa^2 L + \ldots,
\end{equation}
where $\beta_{\lambda}^{(1,1)}(Q) =\displaystyle d \beta_{\lambda}^{(1)}/d \log Q^2$. The running Higgs mass at the scale $M_t$ is obtained from~\Eq{eq:lammt} by mulitplying the result by $2v_{\MS}^2$,
\begin{equation}
  (M_h^{\MS, {\rm SM}})^2 = 2 \lambda(M_t) v_{\MS}^2.
\end{equation}
The pole Higgs mass is then calculated from the pole equation~\cite{Bahl:2017aev}
\begin{equation}
\label{eq:MhEFT}
(M_h)^2_{\rm EFT} = 2 \lambda(M_t) v_{\MS}^2 - \widetilde{\Sigma}_{hh}^{\rm SM}(m_h^2) - \widetilde{\Sigma}_{hh}^{\rm SM'}(m_h^2) \left(2 \lambda(M_t) v_{\MS}^2 - \widetilde{\Sigma}_{hh}^{\rm SM}(m_h^2) - m_h^2\right) + \ldots.
\end{equation}
The second method is an iterative solution of the Higgs boson pole mass equation in the MSSM which in the decoupling limit $M_A \gg M_Z$ reads~\cite{Bahl:2017aev}
\begin{equation}
\label{eq:MhFO}
  (M_h)^2_{\rm FO} = m_h^2 - \widehat{\Sigma}_{hh}^{\rm MSSM}(m_h^2) + \widehat{\Sigma}_{hh}^{\rm MSSM}(m_h^2) \widehat{\Sigma}_{hh}^{\rm MSSM'}(m_h^2) + \ldots,
\end{equation}
where the prime stands for the derivative of the self-energy with respect to the momentum squared, and $m_h^2$ is the lightest Higgs boson mass at the tree level. We have checked analytically that the logarithmic terms are the same in $(M_h)^2_{\rm FO}$ and $(M_h)^2_{\rm EFT}$. This is an important cross-check of our calculation.\footnote{Both methods have to give the same answer also for the non-logarithmic terms in leading order of the expansion in $v/\msusy$. Imposing this condition we have derived the two-loop threshold conditions listed in \App{app:08_CPV_thresholds}.}

At the one-loop level, we obtain the following contribution proportional to the bottom quark mass,
\begin{equation}
\label{eq:Mh1L}
(M_h^{\rm 1L})^2_\text{bot} = 6 \kappa \displaystyle \frac{\overline{m}_b^2}{v^2} \left(2\overline{m}_b^2 - c_{2\beta}^2 m_Z^2 \right) L
\end{equation}
At the two-loop level the leading logarithmic terms read
\begin{equation}
\label{eq:Mh2L_LL}
(M_h^{\rm 2L,LL})^2_\text{bot} = - 18 \kappa^2 \displaystyle \frac{\overline{m}_b^2}{v^4} \left(\overline{m}_b^4 -\overline{m}_b^2\overline{m}_t^2 + \overline{m}_t^4 \right) L^2 + 96 \kappa^2 \frac{g_3^2 \overline{m}_b^4}{v^2} L^2.
\end{equation}
If $\widehat{X}_t$ is renormali{\sz}ed in the OS scheme, the sub-leading logarithms read
\begin{eqnarray}
\label{eq:Mh2L_NLL_OS}
\begin{aligned}
(M_h^{\rm 2L,NLL})^2_\text{bot,OS} = & -3 \kappa^2 \displaystyle \frac{\overline{m}_b^2 \overline{m}_t^4}{v^4} \left(-18 + 6 \vert \widehat{X}_t \vert^2 - \vert \widehat{X}_b \vert^2 \vert \widehat{X}_t \vert^2
    (6 - \vert \widehat{X}_t \vert^2) + 24 \log \displaystyle \frac{\overline{m}_t}{\overline{m}_b} \right) L \\
  & -6 \kappa^2 \displaystyle \frac{\overline{m}_b^4 \overline{m}_t^2}{v^4} \left( 1 + 4 \vert \widehat{X}_t \vert^2 - 4 \vert \widehat{X}_t \vert \vert \widehat{Y}_t\vert \cos(\pXt - \pYt) + \frac{1}{s_{\beta}^2} \right) L \\
  & -18 \kappa^2 \displaystyle \frac{\overline{m}_b^6}{v^4} \left(-1 + 12 \log \displaystyle \frac{\overline{m}_t}{\overline{m}_b} + \frac{1}{c_{\beta}^2} \right) L \\
  & -32 \kappa^2 \frac{g_3^2 \overline{m}_b^4}{v^2} \left(-1 - 12 \log \displaystyle \frac{\overline{m}_t}{\overline{m}_b} + 2 \vert \widehat{X}_b \vert \cos(\pXb - \pMiii) \right) L.
\end{aligned}
\end{eqnarray}
If $\widehat{X}_t$ is renormali{\sz}ed in the \DR scheme at the scale \msusy, we obtain
\begin{eqnarray}
\label{eq:Mh2L_NLL_DR}
\begin{aligned}
(M_h^{\rm 2L,NLL})^2_\text{bot,\DR} = & -3 \kappa^2 \displaystyle \frac{\overline{m}_b^2 \overline{m}_t^4}{v^4} \left(-18 + 6 \vert \widehat{X}_t \vert^2 + 24 \log \displaystyle \frac{\overline{m}_t}{\overline{m}_b} \right) L \\
  & -6 \kappa^2 \displaystyle \frac{\overline{m}_b^4 \overline{m}_t^2}{v^4} \left( 1 + 4 \vert \widehat{X}_t \vert^2 - 4 \vert \widehat{X}_t \vert \vert \widehat{Y}_t\vert \cos(\pXt - \pYt) + \frac{1}{s_{\beta}^2} \right) L \\
  & -18 \kappa^2 \displaystyle \frac{\overline{m}_b^6}{v^4} \left(-1 + 12 \log \displaystyle \frac{\overline{m}_t}{\overline{m}_b} + \frac{1}{c_{\beta}^2} \right) L \\
  & -32 \kappa^2 \frac{g_3^2 \overline{m}_b^4}{v^2} \left(-1 - 12 \log \displaystyle \frac{\overline{m}_t}{\overline{m}_b} + 2 \vert \widehat{X}_b \vert \cos(\pXb - \pMiii) \right) L.
\end{aligned}
\end{eqnarray}
We see that the choice of the renormali{\sz}ation scheme for $\widehat{X}_t$ affects only the logarithmic terms proportional to $\overline{m}_b^2 \overline{m}_t^4$.



\newpage

\bibliographystyle{JHEP.bst}
\bibliography{bibliography}{}

\end{document}